\documentclass[longauth]{aa} 

\usepackage{graphicx}
\usepackage{pdflscape}
\usepackage{dcolumn, siunitx}
\newcolumntype{d}[1]{D{.}{.}{#1}}
\newcolumntype{C}{D>{\;\to\;}{2.0}}
\usepackage{amsmath}
\usepackage{subfig}
\PassOptionsToPackage{hyphens}{url}
\usepackage[dvipsnames]{xcolor}
\usepackage[varg]{txfonts}
\usepackage{multirow}
\usepackage{placeins}
\graphicspath{{./Figures/}}
\usepackage{twoopt}
\usepackage{natbib}
\bibpunct{(}{)}{;}{a}{}{,} 
\usepackage[breaklinks=true]{hyperref} 
\makeatletter
\newcommandtwoopt{\citeads}[3][][]{\href{https://ui.adsabs.harvard.edu/\#abs/#3}%
{\def\hyper@linkstart##1##2{}%
\let\hyper@linkend\@empty\citealp[#1][#2]{#3}}}
\newcommandtwoopt{\citepads}[3][][]{\href{https://ui.adsabs.harvard.edu/\#abs/#3}%
{\def\hyper@linkstart##1##2{}%
\let\hyper@linkend\@empty\citep[#1][#2]{#3}}}
\newcommandtwoopt{\citetads}[3][][]{\href{https://ui.adsabs.harvard.edu/\#abs/#3}%
{\def\hyper@linkstart##1##2{}%
\let\hyper@linkend\@empty\citet[#1][#2]{#3}}}
\newcommandtwoopt{\citeyearads}[3][][]%
{\href{https://ui.adsabs.harvard.edu/\#abs/#3}
{\def\hyper@linkstart##1##2{}%
\let\hyper@linkend\@empty\citeyear[#1][#2]{#3}}}
\makeatother
\hypersetup{colorlinks = true, allcolors={cyan}} 
\newcommand{\cdbox}[1]{%
  {\color{blue}%
    \dbox{\color{black}#1}}%
}
\usepackage{dashbox,framed,color,ocg-p}
\newcommand{\W}{$\mathrm{W3(H_2O)}$}

\newcommand{\mc}{$\mathrm{CH_3CN}$}
\newcommand{\mck}[1]{$\mathrm{CH_3CN}\,(12_#1-11_#1)$}
\newcommand{\mckr}[2]{$\mathrm{CH_3CN}\,(12_K-11_K)\,K = #1-#2$}
\newcommand{\mciso}{$\mathrm{C{H_3}^{13}CN}$}

\newcommand{\mcisokr}[2]{$\mathrm{C{H_3}^{13}CN}\,(12_K-11_K)\,K = #1-#2$}

\newcommand{\eg}    {e.g.}
\newcommand{\ie}    {i.e.}
\newcommand{\jpb}   {Jy~beam$^{-1}$}
\newcommand{\jpp}   {Jy~pixel$^{-1}$}
\newcommand{\kms}   {km~s$^{-1}$}
\newcommand{\lo}    {$L_{\sun}$}
\newcommand{\mo}    {$M_{\sun}$}

\newcommand{\tq}{Toomre~$Q$}

\newcommand{\uv}{\textit{uv}}
\newcommand{\str}[1]{\renewcommand{\arraystretch}{#1}}
\usepackage{orcidlink}

\begin{document} 

   \title{Kinematics and stability of high-mass protostellar disk candidates at sub-arcsecond resolution\thanks{Based on observations from an IRAM large program L14AB. IRAM is supported by INSU/CNRS (France), MPG (Germany), and IGN (Spain).}}

   \subtitle{Insights from the IRAM NOEMA large program CORE}

	\author{A.~Ahmadi \orcidlink{0000-0003-4037-5248}\,\inst{\ref{inst:leiden},\ref{inst:mpia}}
	\and H.~Beuther\inst{\ref{inst:mpia}}
	\and F.~Bosco\inst{\ref{inst:mpia}}
	\and C.~Gieser\inst{\ref{inst:mpe}}
	\and S.~Suri\inst{\ref{inst:vienna}}
	\and J.~C.~Mottram\inst{\ref{inst:mpia}}
    \and R.~Kuiper\inst{\ref{inst:duisburg}}
    \and T.~Henning\inst{\ref{inst:mpia}}
    \and {\'A}.~S{\'a}nchez-Monge\inst{\ref{inst:barcelona1},\ref{inst:barcelona2},\ref{inst:madrid},\ref{inst:cologne}}
    \and H.~Linz\inst{\ref{inst:mpia}}
    \and R.~E.~Pudritz\inst{\ref{inst:mcmaster}}
    \and D.~Semenov\inst{\ref{inst:mpia},\ref{inst:lmu}}
    \and J.~M.~Winters\inst{\ref{inst:iram}}
	\and T.~M\"oller\inst{\ref{inst:cologne}}
    \and M.~T.~Beltr{\'a}n\inst{\ref{inst:arcetri}}
    \and T.~Csengeri\inst{\ref{inst:bordeaux}}
    \and R.~Galv{\'a}n-Madrid\inst{\ref{inst:irya}}
    \and K.~G.~Johnston\inst{\ref{inst:leeds}}
    \and E.~Keto\inst{\ref{inst:cfa}}
    \and P.~D.~Klaassen\inst{\ref{inst:ukatc}}
    \and S.~Leurini\inst{\ref{inst:cagliari}}
    \and S.~N.~Longmore\inst{\ref{inst:liverpool}}
    \and S.~L.~Lumsden\inst{\ref{inst:leeds}}
    \and L.~T.~Maud\inst{\ref{inst:eso}}
    \and L.~Moscadelli\inst{\ref{inst:arcetri}}
    \and A.~Palau\inst{\ref{inst:irya}}
    \and T.~Peters\inst{\ref{inst:mpa}}
    \and S.~E.~Ragan\inst{\ref{inst:cardiff}}
    \and J.~S.~Urquhart\inst{\ref{inst:kent}}
    \and Q.~Zhang\inst{\ref{inst:cfa}}
    \and H.~Zinnecker\inst{\ref{inst:chile}} 
    }

    \institute{Leiden Observatory, Leiden University, PO Box 9513, 2300 RA Leiden, The Netherlands,  \email{aahmadi@strw.leidenuniv.nl} \label{inst:leiden}
    \and Max-Planck-Institut f\"ur Astronomie, K\"onigstuhl 17, 69117 Heidelberg, Germany \label{inst:mpia}
    \and Max Planck Institute for Extraterrestrial Physics, Gie{\ss}enbachstra{\ss}e 1, 85749 Garching bei M\"unchen, Germany\label{inst:mpe}
    \and Department of Astrophysics, University of Vienna, Türkenschanzstra{\ss}e 17 (Sternwarte) 1180 Wien, Austria \label{inst:vienna}
	\and Faculty of Physics, University of Duisburg-Essen, Lotharstra{\ss}e 1, D-47057 Duisburg, Germany \label{inst:duisburg}
	\and Institut de Ci\`encies de l'Espai (ICE, CSIC), Can Magrans s/n, E-08193, Bellaterra, Barcelona, Spain \label{inst:barcelona1}
	\and Institut d'Estudis Espacials de Catalunya (IEEC), Barcelona, Spain \label{inst:barcelona2}
	\and Observatorio Astron\'omico Nacional (OAN, IGN), Calle Alfonso XII 3, 28014, Madrid, Spain \label{inst:madrid}
	\and I.\ Physikalisches Institut der Universit\"at zu K\"oln, Z\"ulpicher Str.\ 77, 50937, K\"oln, Germany \label{inst:cologne}
	\and Dept. of Physics and Astronomy, McMaster University, 1280 Main Street West, Hamilton, ON L8S 4K1, Canada \label{inst:mcmaster}
	\and Department of Chemistry, Ludwig Maximilian University, Butenandtstr. 5-13, 81377 Munich, Germany \label{inst:lmu}
	\and IRAM, 300 rue de la Piscine, Domaine Universitaire, F-38406 Saint Martin d'H{\`e}res, France \label{inst:iram}
	\and INAF - Osservatorio Astrofisico di Arcetri, Largo E. Fermi 5, 50125 Firenze, Italy \label{inst:arcetri}
	\and Laboratoire d'astrophysique de Bordeaux, Univ. Bordeaux, CNRS, B18N, all\'ee Geoffroy Saint-Hilaire, 33615 Pessac, France \label{inst:bordeaux}
	\and Instituto de Radioastronom\'ia y Astrof\'isica, Universidad Nacional Aut\'onoma de M\'exico, Antigua Carretera a P\'atzcuaro 8701, Ex-Hda. San Jos\'e de la Huerta, 58089, Morelia, Michoac\'an, M\'exico  \label{inst:irya}
	\and School of Physics and Astronomy, The University of Leeds, Sir William Henry Bragg Building, Woodhouse Lane, Leeds, LS2 9JT, UK \label{inst:leeds}
	\and Center for Astrophysics | Harvard \& Smithsonian, 60 Garden St., Cambridge, MA 02420, USA \label{inst:cfa}
	\and UK Astronomy Technology Centre, Royal Observatory Edinburgh, Blackford Hill, Edinbugh EH9 3HJ, UK \label{inst:ukatc}
	\and INAF - Osservatorio Astronomico di Cagliari, via della Scienza 5, 09047, Selargius (CA), Italy \label{inst:cagliari}
	\and Astrophysics Research Institute, Liverpool John Moores University, 146 Brownlow Hill, Liverpool L3 5RF, UK \label{inst:liverpool}
	\and European Southern Observatory, Karl-Schwarzschild-Str. 2, 85748, Garching bei M\"unchen, Germany \label{inst:eso}
	\and Max-Planck-Institut f\"ur Astrophysik, Karl-Schwarzschild-Str. 1, 85748 Garching, Germany \label{inst:mpa}
	\and School of Physics and Astronomy, Cardiff University, Queen's Buildings, The Parade, Cardiff, CF24 3AA, UK\label{inst:cardiff}
    \and Centre for Astrophysics and Planetary Science, University of Kent, Canterbury, CT2 7NH, UK \label{inst:kent}
    \and Universidad Autonoma de Chile, Nucleo de Astroquimica y Astrofisica, 425 Avda Pedro de Valdivia, Providencia, Santiago de Chile, Chile \label{inst:chile}
    }
       
   \authorrunning{A. Ahmadi et al.}
   \titlerunning{Kinematics and stability of high-mass protostellar disk candidates}
   \date{Version: April 28, 2023}
 
  \abstract
   {The fragmentation mode of high-mass molecular clumps and the accretion processes that form the most massive stars ($M\gtrsim 8$~\mo) are still not well understood. A growing number of case studies have found massive young stellar objects (MYSOs) to harbour disk-like structures, painting a picture that the formation of high-mass stars may proceed through disk accretion, similar to that of lower mass stars. However, the properties of such structures have yet to be uniformly and systematically characterised. Massive disks are prone to fragmentation via gravitational instabilities due to high gas densities and accretion rates. Therefore, it is important to study the stability of such disks in order to put into context the role of disk fragmentation in setting the final stellar mass distribution in high-mass star forming regions. }
   {The aim of this work is to uniformly study the kinematic properties of a large sample of MYSOs and characterise the stability of possible circumstellar disks against gravitational fragmentation.}
   {We have undertaken a large observational program (CORE) making use of interferometric observations from the Northern Extended Millimetre Array (NOEMA) for a sample of 20 luminous ($L>10^4$~\lo) protostellar objects in the 1.37 mm wavelength regime in both continuum and spectral line emission, reaching ~0.4\arcsec\ resolution (800~au at 2~kpc).}
   {We present the gas kinematics of the full sample and detect dense gas emission surrounding 15 regions within the CORE sample. Using the dense gas tracer \mc, we find velocity gradients across 13 cores perpendicular to the directions of bipolar molecular outflows, making them excellent disk candidates. The extent of the \mc\ emission tracing the disk candidates varies from $1800-8500$~au. Analysing the free-fall to rotational timescales, we find that the sources are rotationally supported. The rotation profiles of some disk candidates are well described by differential rotation while for others the profiles are poorly resolved. Fitting the velocity profiles with a Keplerian model, we find protostellar masses in the range of $\sim10-25$~\mo. Modelling the level population of \mckr{0}{6} lines we present temperature maps and find median temperature in the range 70--210 K with a diversity in distributions. Radial profiles of the specific angular momentum ($j$) for the best disk candidates span a range of 1--2 orders of magnitude, on average $\sim10^{-3}$~\kms\,pc, and follow $j \propto r^{1.7}$, consistent with a poorly resolved rotating and infalling envelope/disk model. Studying the Toomre stability of the disk candidates, we find almost all (11 out of 13) disk candidates to be prone to fragmentation due to gravitational instabilities at the scales probed by our observations, as a result of their high disk to stellar mass ratio. In particular, disks with masses greater than $\sim10-20\%$ of the mass of their host (proto)stars are Toomre unstable, and more luminous YSOs tend to have disks that are more massive compared to their host star and hence more prone to fragmentation.}
   {In this work, we show that most disk structures around high-mass YSOs are prone to disk fragmentation early in their formation due to their high disk to stellar mass ratio. This impacts the accretion evolution of high-mass protostars which will have significant implications for the formation of the most massive stars.}

   \maketitle


\section{Introduction}

There is a growing consensus that the formation of high-mass stars ($M\gtrsim 8$~\mo) proceeds through disk accretion, similar to that of lower mass stars. This has been shown through numerical simulations (\eg~\citeads{2002ApJ...569..846Y}; \citeads{2009Sci...323..754K}; \citeads{2010ApJ...711.1017P}; \citeads{2010ApJ...722.1556K}, \citeyearads{2011ApJ...732...20K}; \citeads{2013ApJ...772...61K}; \citeads{2016ApJ...823...28K}) and is supported by ubiquitous observations of molecular outflows in such regions (\eg~\citeads{2000A&A...364..613H}; \citeads{2001ApJ...552L.167Z}, \citeyearads{2005ApJ...625..864Z}; \citeads{2002A&A...387..931B}; \citeads{2009A&A...504..127F}; \citeads{2011A&A...530A..12L}; \citeads{2014prpl.conf..451F}; \citeads{2015MNRAS.453..645M}). Direct detection of disks around high-mass protostars has been difficult for many reasons. Firstly, high-mass star-forming regions are rare compared to their lower mass counterparts (\citeads{2001MNRAS.322..231K}; \citeads{2003PASP..115..763C}), and are typically located at further distances, requiring observations at high angular resolution to study them in great detail. High-mass stars evolve much more rapidly than low-mass stars and in their short formation phase are deeply embedded within their natal molecular cloud. Furthermore, a significant fraction of OB-type stars are found in binary or multiple systems (\citeads{2007ApJ...655..484A}; see review by \citeads{2022arXiv220310066O}). Spectroscopic studies have found $>70\%$ of main sequence O-type stars in close binary pairs, attributing this outcome to their formation process rather than tidal capture or other scenarios (\citeads{2012MNRAS.424.1925C}; \citeads{2012Sci...337..444S}). A pilot survey dedicated to the study of multiplicity in massive young stellar objects (YSOs) using $K$-band observations has found a binary fraction of at least 50\% \citepads{2019MNRAS.484..226P}. High angular resolution observations are therefore especially critical for disentangling contributions from individual protostars due to this clustered mode of formation of high-mass stars (see review by \citeads{2018ARA&A..56...41M}). 

One of the first systematic searches for disks around intermediate- and high-mass YSOs at high angular resolution (tens of milli-arcsecond) was performed using the mid-infrared instrument (MIDI) on the Very Large Telescope Interferometer (VLTI) in the \textit{N}~band ($8-13\,\mu m$) \citepads{2013A&A...558A..24B}. Fitting the visibilities with 2-D geometric models, they derived size, orientation, and elongation of the circumstellar material for a sub-sample of 20 objects. They found that almost all objects showed deviations from spherical symmetry. In general, the compact mid-IR emission is elongated along the direction of the circumstellar disk and perpendicular to the outflow (e.g. S255 IRS3: \citeads{2018RAA....18...93Z}), but sometimes the illumination footprints of outflow cones can dominate the thermal emission at mid-IR and therefore the elongated nature of the intensity distribution at mid-IR for some sources may be due to outflows rather than the embedded disks \citepads{2017ApJ...843...33D}, hence the geometric models of the MIDI data are in principle susceptible to both disks and outflow cavity walls \citepads{2013A&A...558A..24B}.

Since the bulk of disk material is colder than what the near- and mid-IR regimes can trace, (sub)millimetre observations at high angular resolutions are necessary to obtain a full picture, especially of the outermost regions of the disk and to study the properties of the transition between the infalling envelope and the disk. A review by \citetads{2016A&ARv..24....6B} summarising the evidence for disks around intermediate- to high-mass stars by compiling multi-wavelength observations from literature, confirmed that early-B to late-O type (proto)stars may indeed form through a disk-mediated accretion mechanism. However, the existence of disks around early O-type stars has been elusive. \citetads{2017A&A...602A..59C} employed the Atacama Large Millimeter/submillimeter Array (ALMA) to obtain 0.2\arcsec\ resolution data towards six high-luminosity YSOs and found signatures of Keplerian-like rotation of gas around half of the sample, demonstrating that observing this short-lived early stage of high-mass star formation, when disks may be deeply embedded within their surrounding envelopes, is difficult. 

It is important to target a large sample of high-mass YSOs and uniformly search for and characterise the properties of a statistically significant sample of disk candidates in order to determine whether disk-mediated accretion is in fact the standard mechanism for high-mass star formation. In recent years, a growing number of case studies have found disk-like structures around intermediate- to high-mass (proto)stars (\citeads{2010Natur.466..339K}; \citeads{2015ApJ...813L..19J}, \citeyearads{2020A&A...634L..11J}; \citeads{2015MNRAS.447.1826Z}, \citeyearads{2019ApJ...872..176Z}; \citeads{2016MNRAS.462.4386I}, \citeyearads{2018ApJ...869L..24I}; \citeads{2016ApJ...823..125C}; \citeads{2017A&A...602A..59C}; \citeads{2018A&A...617A..89C}; \citeads{2018A&A...618A..46A}; \citeads{2018A&A...620A..31M}, \citeyearads{2019A&A...627L...6M}; \citeads{2019A&A...623A..77S}; \citeads{2019A&A...622A.206M}; \citeads{2020ApJ...904...77G}; \citeads{2020ApJ...900L...2T}). If disk-mediated accretion is the standard mechanism for high-mass star formation, then a closely related question that needs investigation is whether or not the existence of close binary or multiple systems is the product of disk fragmentation. Massive disks are prone to fragmentation via gravitational instabilities due to high gas densities and accretion rates (\citeads{2006MNRAS.373.1563K}; \citeads{2010ApJ...708.1585K}; review by \citeads{2016ARA&A..54..271K}). Recent numerical simulations have shown that binary/multiple systems with separations on the scales of hundreds of au can be formed as a result of disk fragmentation (\citeads{2018MNRAS.473.3615M}; \citeads{2020A&A...644A..41O}), confirming speculations based on earlier, lower-resolution simulations that identified fragmentation as a crucial process in accretion flows around high-mass stars but that did not resolve the disk scale \citepads{2010ApJ...725..134P}. Observational studies have also started to resolve arc-, ring- or spiral-like substructures that presumably develop as a result of gravitational instabilities in disks and inner envelopes (\citeads{2019A&A...627L...6M}; \citeads{2019ApJ...872..176Z}; \citeads{2020A&A...634L..11J}). Furthermore, a recent study by \citetads{2021A&A...645L..10R} has shown that velocity dispersion of massive stars in young clusters increases as they age, suggesting that binaries are formed at larger separations and migrate inwards as a result of interactions with a remnant accretion disk or with other YSOs in the system. Therefore, both disk fragmentation and disk-mediated inward migration of fragments can be responsible for the observed close binary or multiple systems. Disk fragmentation may also be an important mechanism in seeding the initial binary and multiple systems before the inward migration occurs.

We have observed 20 high-mass YSOs at the highest angular resolution currently possible for 1.3~mm observations in the northern sky (0.4\arcsec, 800~au at 2~kpc) using the Northern Extended Millimetre Array (NOEMA) interferometer and the IRAM 30-m telescope (PI: Henrik Beuther). One of the objectives of this large program, CORE\footnote{\url{http://www.mpia.de/core}}, was to investigate the fragmentation of cloud cores via gravitational Jeans instability \citepads{2018A&A...617A.100B}. We found a diversity in fragmentation of pc-size clumps into cores with several thousand au sizes, whereby some regions were found to have many cores, while other regions contain a very low number of cores. We attributed this diversity in fragmentation to a variety in initial density structures and/or different initial magnetic field strengths. 

The aim of the work presented in this paper is to study the small-scale kinematics of the CORE sample to search for and characterise disks around high-mass YSOs. In particular, we study the stability of disk candidates against gravitational instabilities in order to understand the importance of disk fragmentation in setting the final stellar mass distribution of a given region. Other results from the CORE survey include various case studies of individual regions (\citeads{2018A&A...618A..46A}; \citeads{2019A&A...627A..68C}; \citeads{2019A&A...629A..10B}; \citeads{2019A&A...631A.142G}; \citeads{2020A&A...636A.118M}; \citeads{2021A&A...647A.114M}; \citeads{2021A&A...655A..84S}) as well as a chemical analysis of the early phase of high-mass star formation based on the full sample \citepads{2021A&A...648A..66G}. 

The paper is organised as follows. In Sect.~\ref{s: source_selection} we describe how the CORE sample was selected followed by an overview of the observations and data reduction techniques in Sect.~\ref{s: obs_details}. We outline the observational results in Sect.~\ref{s: obs_results} with a brief description of the large-scale core fragmentation and kinematics, and then turn our focus onto the dense gas distribution and kinematics. In Sect.~\ref{s: analysis} we begin with an analysis of the temperature distribution and provide estimates for gas masses. We continue with a study of the candidacy of the sources as disks by studying their rotation profiles, timescales, and specific angular momenta, and finally investigate the stability of the disk candidates. The paper concludes with a summary of our findings in Sect.~\ref{s: sample_summary}.

\begin{table*}[htb]
\str{1.2}
\caption{Positions and properties of the CORE sample, grouped in track-sharing pairs, adapted from \citetads{2018A&A...617A.100B}.}
\label{t: sample_info}
\centering
\begin{tabular}{lllccccll}
\hline\hline
  Source & R.A. & Dec.  & Distance & Luminosity & $\sigma_\mathrm{cont}$\tablefootmark{a}& IR-bright & a.f.\tablefootmark{b} & Ref.\tablefootmark{c} \\
  & (J2000.0) & (J2000.0) & (kpc) & ($10^4$~\lo) & (m\jpb) & & & \\
\hline
  IRAS23151+5912  & 23:17:21.01 & +59:28:47.49 & 3.3 & 2.4 & 0.19 & + & w   & d1$^\ast$, l2 \\
  IRAS23033+5951  & 23:05:25.00 & +60:08:15.49 & 4.3 & 1.7 & 0.46 & --&cm, w  & d2, l1 \\
  \hline
  AFGL2591        & 20:29:24.86 & +40:11:19.40 & 3.3 & 20.0 & 0.60 & + & cm, w & d3$^{\ast, \dagger}$, l1 \\
  G75.78+0.34     & 20:21:44.03 & +37:26:37.70 & 3.8 & 11.0 & 0.60 & --& cm, m & d4$^{\ast, \dagger}$, l1 \\
  \hline
  S87\,IRS1        & 19:46:20.14 & +24:35:29.00 & 2.2 & 2.5 & 0.23 & + & cm   & d5$^\dagger$, l1 \\
  S106            & 20:27:26.77 & +37:22:47.70 & 1.3  & 3.4 & 1.25 & + & cm, w & d6$^\ast$, l2 \\
  \hline
  IRAS21078+5211  & 21:09:21.64 & +52:22:37.50  & 1.5 & 1.3 & 0.60 & --& cm, w &  dl1$^\ast$ \\ 
  G100.3779-03.578& 22:16:10.35 & +52:21:34.70  &3.5 & 1.5 & 0.08 & + & w    & d1$^\ast$, l2\\
  \hline
  G084.9505-00.691& 20:55:32.47 & +44:06:10.10 & 5.5 & 1.3 & 0.10 & + & w    & d2, l2\\
  G094.6028-01.797& 21:39:58.25 & +50:14:20.90  & 4.0 & 2.8 & 0.14 & + & w, m  & d7$^\ast$, l2 \\
  \hline
  Cep\,A\,HW2         & 22:56:17.98 & +62:01:49.50 & 0.8 & 1.5 & 4.00 & --& cm, w, m& d8$^\dagger$, l1 \\
  NGC7538\,IRS9     & 23:14:01.68 & +61:27:19.10  & 2.7 & 2.3 & 0.30 & + & w    & d9$^{\ast, \dagger}$, l1 \\
  \hline
  W3(H$_2$O)      & 02:27:04.60 & +61:52:24.73 &  2.0 & 8.3 & 4.50 & --& cm, w, m& d10$^\ast$, l2 \\
  W3\,IRS4          & 02:25:31.22 & +62:06:21.00 &  2.0 & 4.5 & 0.60 & + & cm, w  & d10, l1\\
  \hline
  G108.7575-00.986& 22:58:47.25 & +58:45:01.60 & 4.3 & 1.4  & 0.25 & + & w, m  & d2, l3\\
  IRAS23385+6053  & 23:40:54.40 & +61:10:28.02 & 4.9 & 1.6 & 0.25 & --& w    & dl2 \\
  \hline
  G138.2957+01.555& 03:01:31.32 & +60:29:13.20 & 2.9 & 1.4 & 0.16 & + & cm, w  & d2$^\dagger$, l1 \\
  G139.9091+00.197& 03:07:24.52 & +58:30:48.30 & 3.2$^\ddagger$ & 1.1  & 0.17 & + & cm, w  & d2, l1 \\
  \hline
  \underline{Pilot study}\\
  NGC7538\,IRS1     & 23:13:45.36 & +61:28:10.55 & 2.7 & 21.0 & 10.0 & + &cm, w, m& d9$^{\ast, \dagger}$, l1 \\
  NGC7538\,S        & 23:13:44.86 & +61:26:48.10 & 2.7 & 1.5  & 0.60 & --& w, m  & d9$^{\ast, \dagger}$, l1 \\ 
\hline
\end{tabular}
\tablefoot{
\tablefoottext{a}{rms noise in the continuum maps imaged with a robust 0.1 (uniform) weighting from \citetads{2018A&A...617A.100B}}
\tablefoottext{b}{Associated features (a.f.): cm: cm continuum; w: H$_2$O maser; m: CH$_3$OH maser}
\tablefoottext{c}{References for distances and luminosities: \\ d1: \citealt{2014ApJ...790...99C}, d2: \citeads{2011MNRAS.418.1689U}, d3: \citeads{2012A&A...539A..79R}, d4: \citeads{2011PASJ...63...45A}, d5: \citeads{2009ApJ...693..413X}, d6: \citeads{2013ApJ...769...15X}, d7: \citealt{2014ApJ...790...99C}, \citeads{2019ApJ...876...30S}, d8: \citeads{2016SciA....2E0878X}, d9: \citeads{2009ApJ...693..406M}, d10: \citeads{2006ApJ...645..337H}, \citeads{2006Sci...311...54X}, dl1: \citeads{1996A&A...308..573M}, dl2: \citeads{1998ApJ...505L..39M};\\
  l1: RMS survey database (\url{http://rms.leeds.ac.uk/cgi-bin/public/RMS\_DATABASE.cgi}) using SED fitting from \citetads{2011A&A...525A.149M} including Herschel fluxes and the latest distance determination;\\
  l2: RMS survey database, using SED fitting from \citetads{2011A&A...525A.149M} updated to the latest distance
  determination;\\
  l3: RMS survey database, calculated from the MSX\,21\,$\mu$m flux using the scaling relation derived by \citetads{2011A&A...525A.149M} and updated to the latest distance determination.}\\
  \tablefoottext{$\ast$}{Distances in agreement with \citeads{2019ApJ...885..131R}.}
  \tablefoottext{$\dagger$}{Distances in agreement with \citeads{2021A&A...646A..74M}.}
  \tablefoottext{$\ddagger$}{\citeads{2021A&A...646A..74M} report a kinematic distance of $1.6 \pm0.7$~kpc.}
  }
\end{table*}

\section{Source selection}\label{s: source_selection}

As outlined in \citetads{2018A&A...617A.100B} in more detail, the CORE sample of 20 sources includes all targets within the Red MSX (RMS) survey\footnote{\url{http://rms.leeds.ac.uk/cgi-bin/public/RMS_DATABASE.cgi}} \citepads{2013ApJS..208...11L} for which high-resolution millimetre data did not previously exist and that conform to the following criteria. To ensure that the targets are high-mass star-forming regions, we made a luminosity cut at $L>10^4$~\lo, which corresponds to a star with a mass of at least 8~\mo. To be able to observe the sample at high resolution with good image fidelity, a distance cut at $d<6$~kpc was enforced and only sources with declination~$>$24\degr\ were included to ensure good $uv$ coverage. To target the early phase of formation with active accretion, we only selected sources whose fluxes at ${21\,\mu m}$ are at least two times greater than their ${8\,\mu m}$ flux, using the RMS survey, since it is expected that the IR colour evolves from red\footnote{When the flux at ${21\,\mu m}$ is greater than at ${8\,\mu m}$.} to blue. In this way, the final sample size was narrowed down to 18 regions along with two additional sources (NGC7538\,IRS1 and NGC7538\,S) which were observed previously with almost the same setup; we will refer to these two sources as the CORE pilot regions (\citeads{2012A&A...543A..88B}; \citeyearads{2013A&A...558A..81B}; \citeads{2016A&A...593A..46F}). Table~\ref{t: sample_info} lists the names and positions of the sources with their properties, including features associated with the regions such as the presence of various masers and/or centimetre continuum emission as proxy for the star formation activity. The threshold between IR-bright and -dark is defined according to \citetads{2007A&A...476.1243M} based on cores in the Cygnus~X star-forming region at a distance $d=1.7$~kpc, such that if a core with a luminosity of 1000~\lo\ has $21\,\mu m$ flux ($S_{21\mu m}$) below 10~Jy it would be categorised as IR quiet. Therefore, regions with $S_{21\mu m} < 10\,\mathrm{Jy}\,\left(\frac{1.7~\mathrm{kpc}}{d}\right)^2\,\left(\frac{L}{1000\,L_\sun}\right)$ are classified as IR-dark. 

\begin{table*}[htb]
\str{1.2}
\caption{Frequency setup of the narrow-band correlator and important lines covered.}
\centering
\label{t: spectral_setup}
\begin{tabular}{lccc}
\hline\hline
Spectral unit & Frequency range & Important lines \\
 & (MHz) & \\
\hline 
L01 & 220\,690.6--220\,769.7 & \mckr{0}{3} \\
L02 & 220\,630.6--220\,709.7 & \mckr{4}{5}, $\mathrm{C_2H_5CN\,(25_{22,4}-24_{22,3})}$ \\
L03 & 220\,570.6--220\,649.7 & \mckr{5}{6}, \mcisokr{0}{3}, HNCO\,(10$_{1,9}$--9$_{1,8}$) \\
L04 & 220\,130.6--220\,209.7 & $\mathrm{HCOOCH_3}$\,(17$_{4,13}$--16$_{4,12}$)\,A/E, $\mathrm{H_2CCO}$\,(11$_{1,11}$--10$_{1,10}$)\\
L05 & 218\,860.6--218\,939.7 & $\mathrm{OCS}\,(18-17)$ \\
L06 & 218\,415.6--218\,494.7 & $\mathrm{H_2CO}$\,(3$_{2,2}$--2$_{2,1}$), $\mathrm{CH_3OH}$\,($4_2$--$3_1$) \\
L07 & 218\,280.6--218\,359.7 & $\mathrm{HC_3N}$\,(24--23), $\mathrm{HCOOCH_3}$\,(17$_{3,14}$--16$_{3,13}$)\,A \\
L08 & 218\,180.6--218\,259.7 & $\mathrm{H_2CO}$\,(3$_{0,3}$--2$_{0,2}$), $\mathrm{O^{13}CS}$\,(18--17)\\
\hline
\end{tabular}
\end{table*}

\section{Observations and data reduction}\label{s: obs_details}

Observations in the 1.3~mm wavelength regime of the CORE project began in June 2014 and finished in January 2017, consisting of a total of more than 400 hours of observations with originally the Plateau de Bure Interferometer (PdBI) that evolved into NOEMA over the time of the project. We observed in three different configurations of the interferometer: A-configuration being the most extended, D-configuration being the most compact array, and B-configuration covering baselines between the two. Baselines roughly in the range of $15-765$~m were covered, therefore the NOEMA observations are not sensitive to structures larger than $\sim$15\arcsec\ (0.1~pc at 2~kpc) at 220~GHz. The exact baseline ranges for all tracks are listed in Table~3 of \citetads{2018A&A...617A.100B}. On-source observations were taken in roughly 20-minute increments distributed over an observing run (\ie~a track) and interleaved with observations of various calibration sources. We observed in track-sharing mode meaning two science targets were observed in one track, with one track corresponding to a total of roughly 8 hours of observations. We paired sources with similar positions in the sky together in order to be able to use the same calibrators. With the sources being observed in three different configurations, in total at least three half tracks were observed per source. In many cases more tracks were needed to achieve the required sensitivity, better \uv-coverage, and depending on the weather conditions. The observations of the two pilot sources were carried out in early 2011 in the A- and B-array configurations of NOEMA only \citepads{2012A&A...543A..88B}. 

The spectral setup of the CORE survey is outlined in \citetads{2018A&A...618A..46A}. To summarise, we observed the frequency range of 217\,078.6--220\,859.5~MHz with the wide-band correlator WideX, in both horizontal and vertical polarisations with a fixed spectral resolution of 1.95~MHz ($\sim$2.7~\kms\ at 219~GHz). For higher spectral resolution, we used 8 narrow-band correlator units with 80~MHz bandwidth and spectral resolution of 0.312~MHz ($\sim$0.43~\kms). The exact bandwidth coverage of this correlator and the important lines observed are summarised in Table~\ref{t: spectral_setup}. The main species used in this work were observed with the narrow-band correlator, smoothed to 0.5~\kms\ spectral resolution. Even though the observations were carried out over a period of a few years when the NOEMA interferometer was upgraded from 6 to 8 antennas, the narrow-band correlator was only able to accept the signal from the 6 antennas that provided the best \uv-coverage. 

Calibration of the NOEMA observations was performed with the \textit{CLIC} program of the Grenoble Image and Line Data Analysis Software (\textit{GILDAS}\footnote{\url{http://www.iram.fr/IRAMFR/GILDAS}}) package developed by IRAM and Observatoire de Grenoble. In most cases, two phase and amplitude calibrators were observed. Strong quasars (\eg~3C84, 3C273, or 3C454.3) were used for the bandpass calibration. For most of the target observations, the source MWC349 was used for absolute flux calibration, assuming a model flux of 1.86~Jy at 220~GHz. For a few tracks, the source LKH$\alpha$101 was used for the flux calibration. The absolute flux scale is correct to within 20$\%$, according to the observatory standard. 

The imaging of the NOEMA observations was performed with the \textit{MAPPING} program of \textit{GILDAS}. The continuum was extracted from line-free regions of the WideX band as described in \citetads{2018A&A...617A.100B} and was imaged with uniform weighting yielding synthesised beam sizes of $\sim$0.32\arcsec\ -- 0.5\arcsec. The rms noise of the continuum images are listed in Table~\ref{t: sample_info} and depend on the prominence of sidelobes in the images whereby the brighter sources tend to be noisier. To study the detailed kinematics of the regions on small scales, we analysed the highest spectral resolution molecular line observations from the narrow-band correlator listed in Table~\ref{t: spectral_setup}. Each spectral cube was resampled to have 60 channels centred at the frequency of the given transition with a spectral resolution of 0.5~\kms. For each molecular line transition, the continuum was subtracted in the \uv-plane by selecting the line-free channels in the spectral unit in which the transition lies. The only exceptions were molecular transitions in spectral unit L03 because too few line-free channels were present. For this spectral unit, we used line-free channels from spectral unit L02 assuming there is no significant spectral slope across these adjacent units. The resulting visibilities were then imaged with the CLARK algorithm \citepads{1980A&A....89..377C} with different weightings, discussed further in the analysis (see Sect.~\ref{s: sample_smallscale}). 

Observations with the IRAM~30-m~telescope were taken in March and May of 2015 using the EMIR multi-band mm-wave receiver \citepads{2012A&A...538A..89C}. On-the-fly maps (whereby the telescope continuously slews smoothly and rapidly across a field) covering an area of 1.5\arcmin\ by 1.5\arcmin\ were observed for all 20 sources centred at the phase center of the NOEMA observations. The half-power-beam-width of the observations is 11.3\arcsec\ at 217~GHz and 10.6\arcsec\ at 231~GHz. Two 4~GHz wide bands were positioned at 215, and 219~GHz in the lower sideband with a small overlap to cover the spectral setup of the NOEMA observations with the WIDEX correlator. Another two 4~GHz wide units on the upper sideband were centred at 231 and 234~GHz, also with a small overlap. A Fourier Transform Spectrometer (FTS) backend was employed with a channel width of 195~kHz corresponding to a resolution of 0.27~\kms\ at 217~GHz and 0.25~\kms\ at 233~GHz. The single-dish data were reduced using the \textit{CLASS} program of \textit{GILDAS} and the procedure is detailed in Appendix~A.1 of \citetads{2020A&A...636A.118M}. In the work presented here, the single-dish observations were only used to infer outflow directions to strengthen the existence of disks using typical outflow and shock tracers. For this purpose, we also merged the single-dish and NOEMA observations for $^{13}\mathrm{CO}\,(2-1)$ using the \textit{MAPPING} software and produced images with natural weighting, and the Steer-Dewdney-Ito (SDI; \citeads{1984A&A...137..159S}) algorithm which is more appropriate when the distribution of emission is extended. The spectral resolution of this merged cube is 3.0~\kms. More details on the merging process can be found in Appendix~B of \citetads{2020A&A...636A.118M} with an extensive exploration of the effects of using different imaging algorithms and weightings.


\section{Observational results}\label{s: obs_results}

\subsection{Core fragmentation, kinematics, and evolutionary stage}

We detect strong millimetre continuum emission from all sources within the CORE survey, mainly associated with cold dust. Contours in Fig.~\ref{f: mom_velo_h2co} show the continuum observations where a diversity of fragmentation levels among the sources can be seen, with some regions harbouring as many as 20 cores (\eg~IRAS21078), while others remain singular (\eg~AFGL2591). To understand this diversity, we performed a minimum-spanning-tree analysis and found the separation between the cores to be on the order of the thermal Jeans length or smaller \citepads{2018A&A...617A.100B}. This analysis indicated that the initial density structure or magnetic fields may be among the important factors that set the fragmentation diversity we see on these clump scales. Because the smallest separation between the cores were often found to be at our resolution limit, it is expected that further fragmentation can be found on even smaller scales, likely as a result of disk fragmentation which is the focus of this paper. 

We made use of $\mathrm{H_2CO}$ and $\mathrm{CH_3OH}$ transitions that are excited at lower gas temperatures and densities than \mc\ to probe the distribution of gas on large scales. Figure~\ref{f: mom_velo_h2co} shows the intensity-weighted mean velocity (first moment) maps of $\mathrm{H_2CO\,(3_{0, 3}-2_{0, 2})}$ for the entire sample. We detect gas with velocities varying on the order of a few \kms\ in each field. For W3IRS4 that has a filamentary structure, we find a velocity gradient along the filament, consistent with flows of material along the filament onto the hot core that resides at the end of the filament \citepads{2020A&A...636A.118M}. There exist a separate velocity gradient across the hot core with a different position angle and amplitude than the larger scale flow along the filament. This finding is consistent with the filamentary paradigm of star formation (see reviews by \citeads{2014prpl.conf...27A} and \citeads{2022arXiv220309562H}). 

We further find a variety of stages in the evolutionary sequence of high-mass star formation within some of the targeted fields, hinting at the importance of the environment on triggering star formation. For example, in the case of W3IRS4 mentioned above, we find two cold cores along a filamentary structure and a hot core situated at the junction of the filament and an expanding UC\ion{H}{ii} region known as W3C. The UC\ion{H}{ii} region is seen as a circular structure to the southwest with almost no $\mathrm{H_2CO}$ emission in the interferometric observations. We can compare this source to another UC\ion{H}{ii} region in the CORE sample, W3(OH), which is at the same distance and has a similar mass sensitivity to W3IRS4. W3(OH) is the western-most feature seen in the map labeled as \W, around which we detect extended $\mathrm{H_2CO}$ emission, unlike for W3C. This difference can in part be attributed to the fact that W3C is probably at a later evolutionary stage than W3(OH) as suggested by its larger extent, hence photoionisation may have already destroyed many of the species that would often be found during an earlier evolutionary phase.

Although we attempted to observe regions at the same evolutionary stage, the CORE sample does inevitably span a range in evolutionary stage, as high-mass star formation occurs on much shorter timescales compared to the formation of low-mass stars. This is evident because some YSOs show a rich hot core chemistry (\eg~AFGL2591, G75.78, Cep\,A\,HW2, and \W), while others are weaker in line emission (\eg~S87\,IRS1, S106, G108, G138, G139). For two of these weak sources, namely S106 and G139, we barely detect any $\mathrm{H_2CO\,(3_{0, 3}-2_{0, 2})}$ emission at the position of the cores (see Fig.~\ref{f: mom_velo_h2co}). Furthermore, there is no correlation between the richness in lines and the infrared (IR) brightness of the sources (see Table~\ref{t: sample_info}). The fact that some of the most chemically-rich sources in our survey are IR-dark points to their deeply embedded nature, during which they are already capable of producing a rich chemistry and driving powerful outflows. We refer to \citetads{2021A&A...648A..66G} for a more comprehensive study of the observed chemistry in the CORE sample.


\subsection{Dense gas kinematics}\label{s: sample_smallscale}

To study candidate disks, one needs the right spectral line tracer that is chemically abundant, excited only in the innermost regions of the core, and not optically thick. \mc\ has proven to be an excellent dense gas and disk tracer, and using its higher excited transitions along with its rarer \mciso\ isotopologue, line optical thickness issues can be circumvented (\eg~\citeads{2014A&A...569A..11S}; \citeads{2017A&A...602A..59C}; \citeads{2018A&A...617A.100B}). Additionally, due to its molecular structure, \mc\ splits into a ladder of closely-spaced transitions (\ie~$K$-ladder), the level populations of which can provide a measure for core temperatures and densities. In this work, we make use of \mckr{0}{6} transitions with upper energies in the range of 69 to 325 K to study the dense gas kinematics and physical conditions of the full sample. Initially, we imaged the \mc\ cubes for the entire sample with a uniform weighting (robust 0.1 in \textit{GILDAS}) to reach the highest possible angular resolution. For sources for which \mc\ was either not detected or only one or two of the lowest $K$ transitions were detected, we re-imaged the data cubes with weightings that help in recovering more extended emission but that result in coarser spatial resolution (a higher robust parameter closer to natural weighting). The sources for which up-weighting the compact configuration helped in recovering some \mc\ emission were IRAS23033, IRAS23385, G100, and G084. Even with such a strategy, we could not detect \mc\ emission for 5 of the 20 core sources: S87\,IRS1, S106, G108, G138, and G139. Interestingly, all of these objects are IR-bright confirming our hypothesis that these sources are likely at a slightly more advanced evolutionary phase in which the warmer environment has already destroyed this species. It is still possible that these sources may harbour disks that are at a later evolutionary stage that cannot be traced by \mc. These five sources were therefore omitted from our search for circumstellar disks.

\begin{figure*}
    \centering
    \includegraphics[width=\hsize]{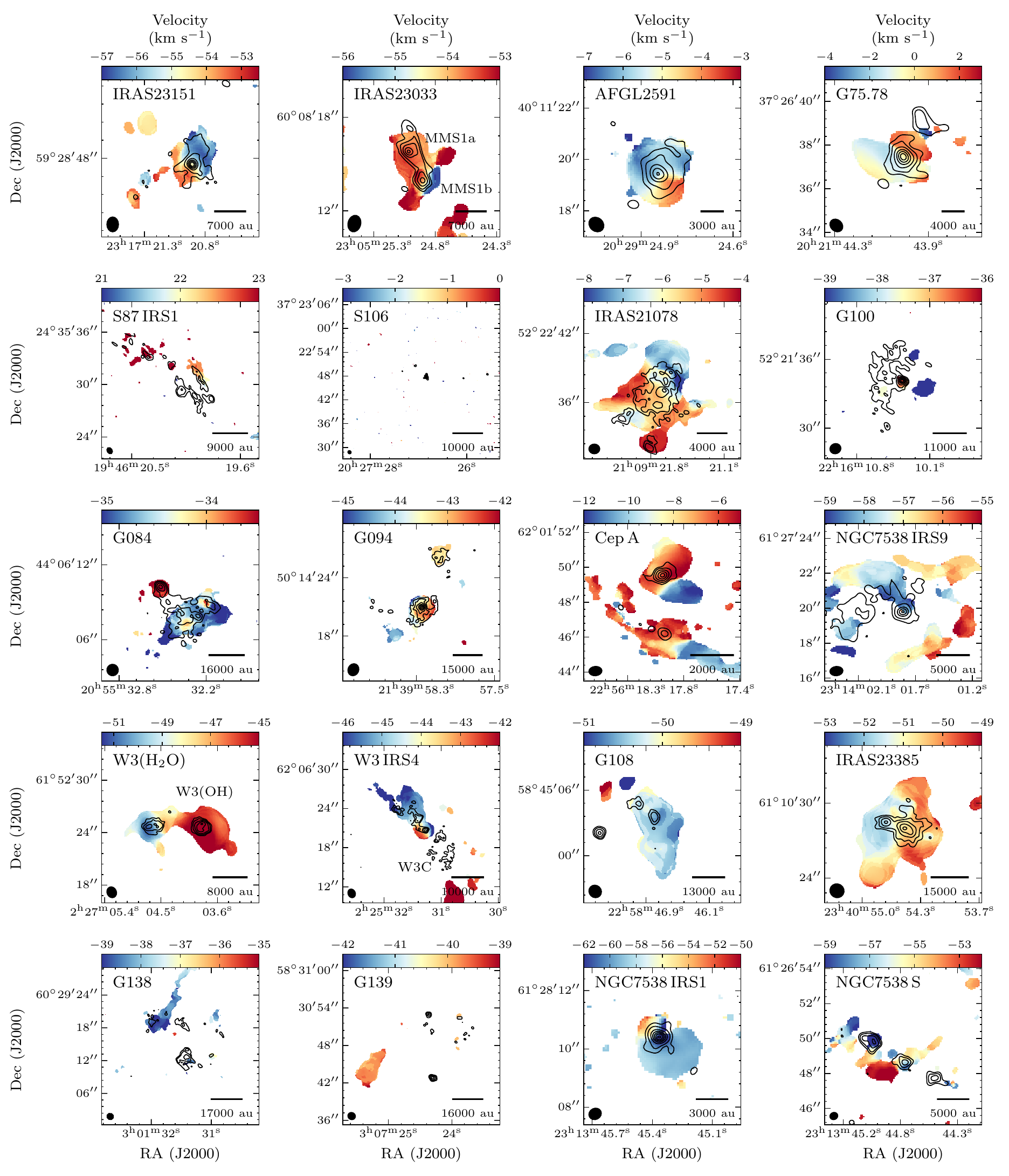}%
    \hspace{-\hsize}%
    \begin{ocg}{l:mom_vleo_ch3oh}{l:mom_vleo_ch3oh}{0}%
    \end{ocg}%
    \begin{ocg}{l:mom_velo_h2co}{l:mom_velo_h2co}{1}%
    \includegraphics[width=\hsize]{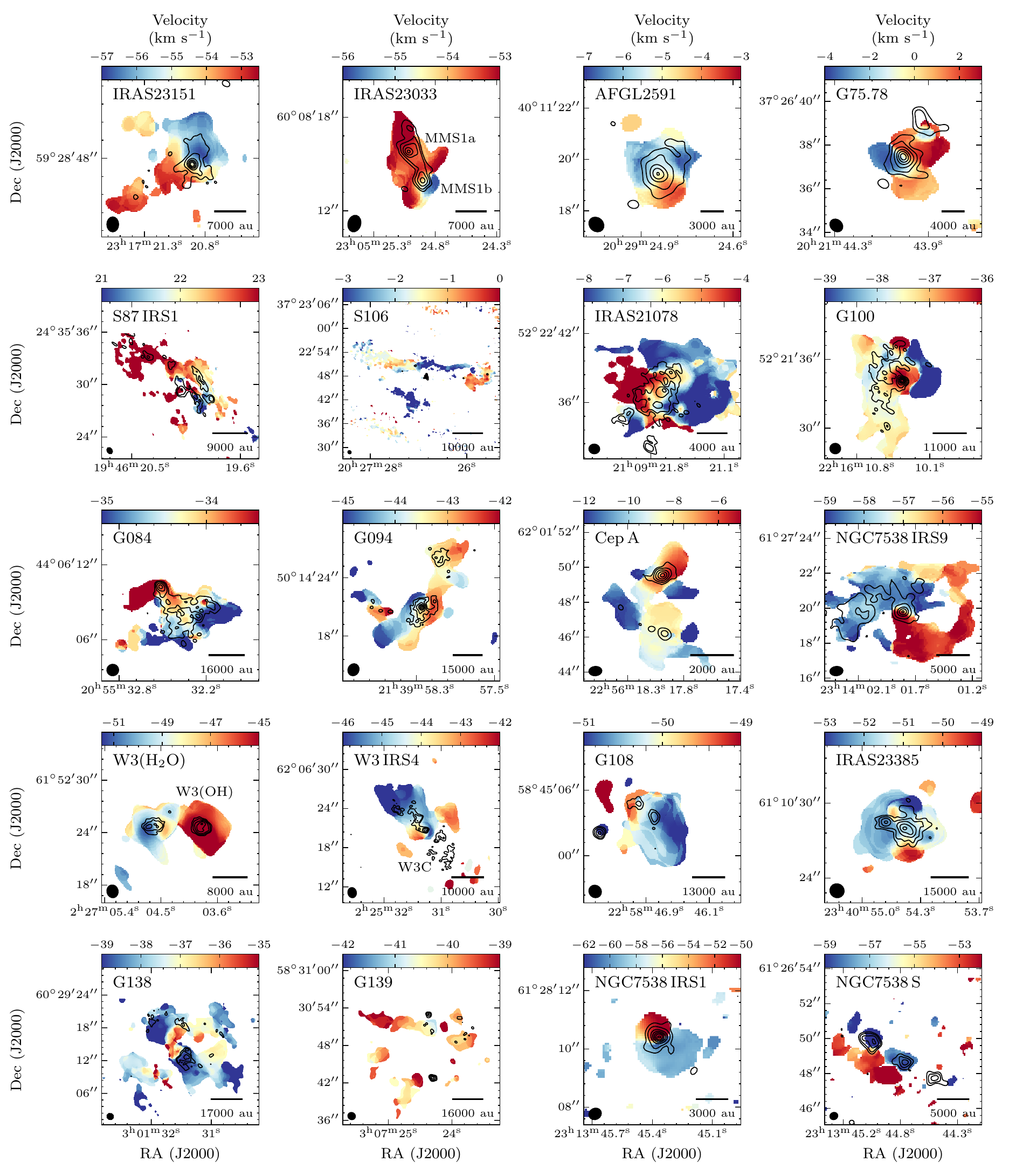}%
    \end{ocg}%
    \caption{Intensity-weighted mean velocity (first moment) maps of $\mathrm{H_2CO\,(3_{0, 3}-2_{0, 2})}$ showing the large-scale kinematics of the full CORE sample. The contours correspond to the continuum maps imaged with uniform weighting as presented in \citetads{2018A&A...617A.100B}. The outermost three contours correspond to 5, 10, and 20$\sigma$ levels, then increasing in steps of 15$\sigma$. For IRAS23151, NGC7538\,IRS1, NGC7538\,IRS9 and AFGL2591 the outermost three contours correspond to 5, 15, and 40$\sigma$ levels, then increasing in steps of 25$\sigma$ (see Table~\ref{t: CORE_disk_details} for $\sigma$ values). The synthesised beam is shown in the bottom left corner and a scale bar in the bottom right corner of each panel. When displayed in Adobe Acrobat, it is possible to switch to the \setocgs{}{l:mom_vleo_ch3oh}{l:mom_velo_h2co}{\protect\cdbox{first moment map of $\mathrm{CH_3OH (4-3)}$}} and back to the \setocgs{}{l:mom_velo_h2co}{l:mom_vleo_ch3oh}{\protect\cdbox{first moment map of $\mathrm{H_2CO\,(3_{0, 3}-2_{0, 2})}$}} for comparison.}
    \label{f: mom_velo_h2co}
\end{figure*}

\begin{figure*}
    \centering
    \includegraphics[width=0.76\hsize]{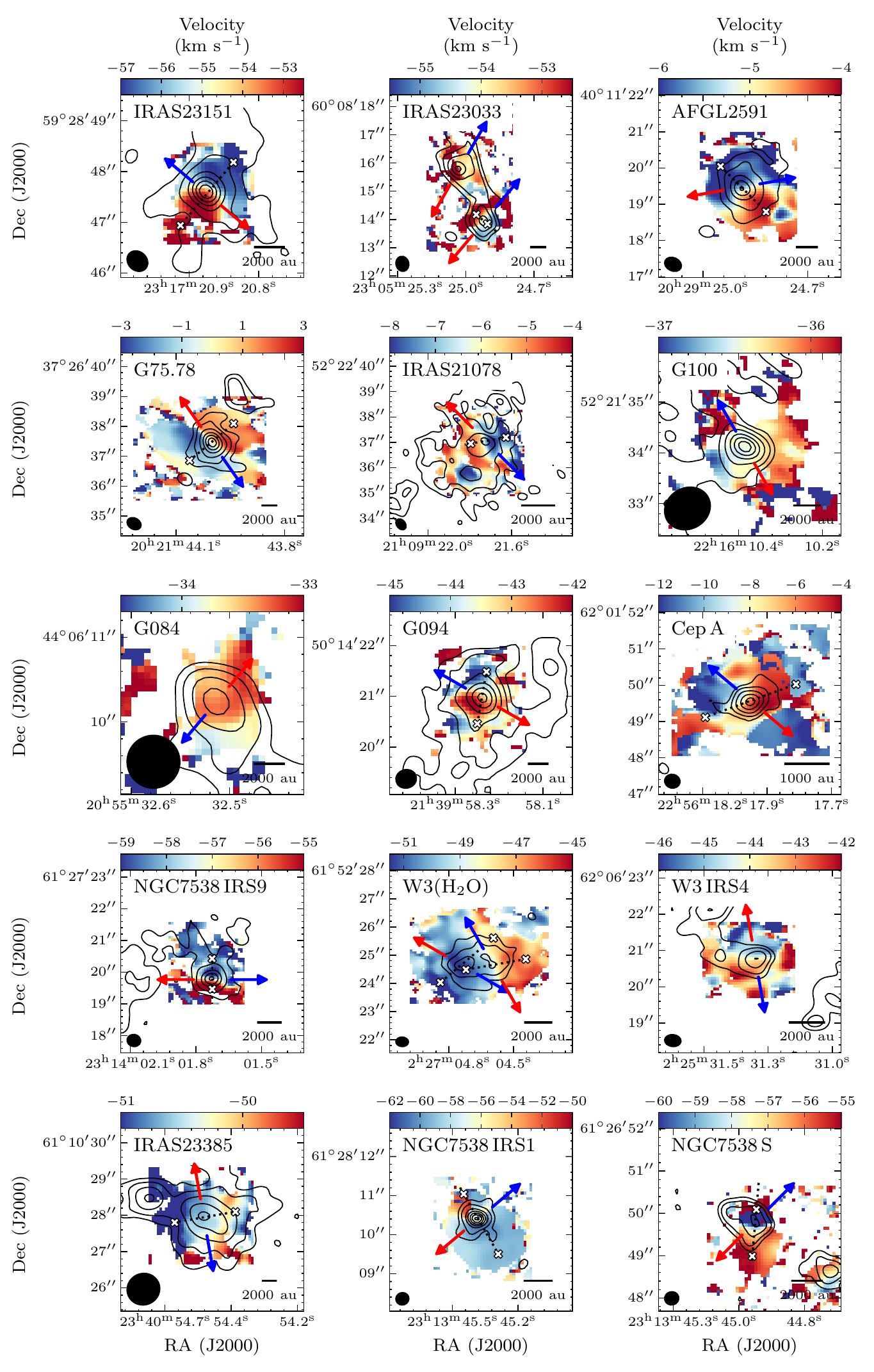}%
    \hspace{-0.76\hsize}%
    \begin{ocg}{l:decomposed_velo_ch3cn_k3}{l:decomposed_velo_ch3cn_k3}{0}%
    \end{ocg}%
    \begin{ocg}{l:mom_velo_ch3cn_k3}{l:mom_velo_ch3cn_k3}{1}%
    \includegraphics[width=0.76\hsize]{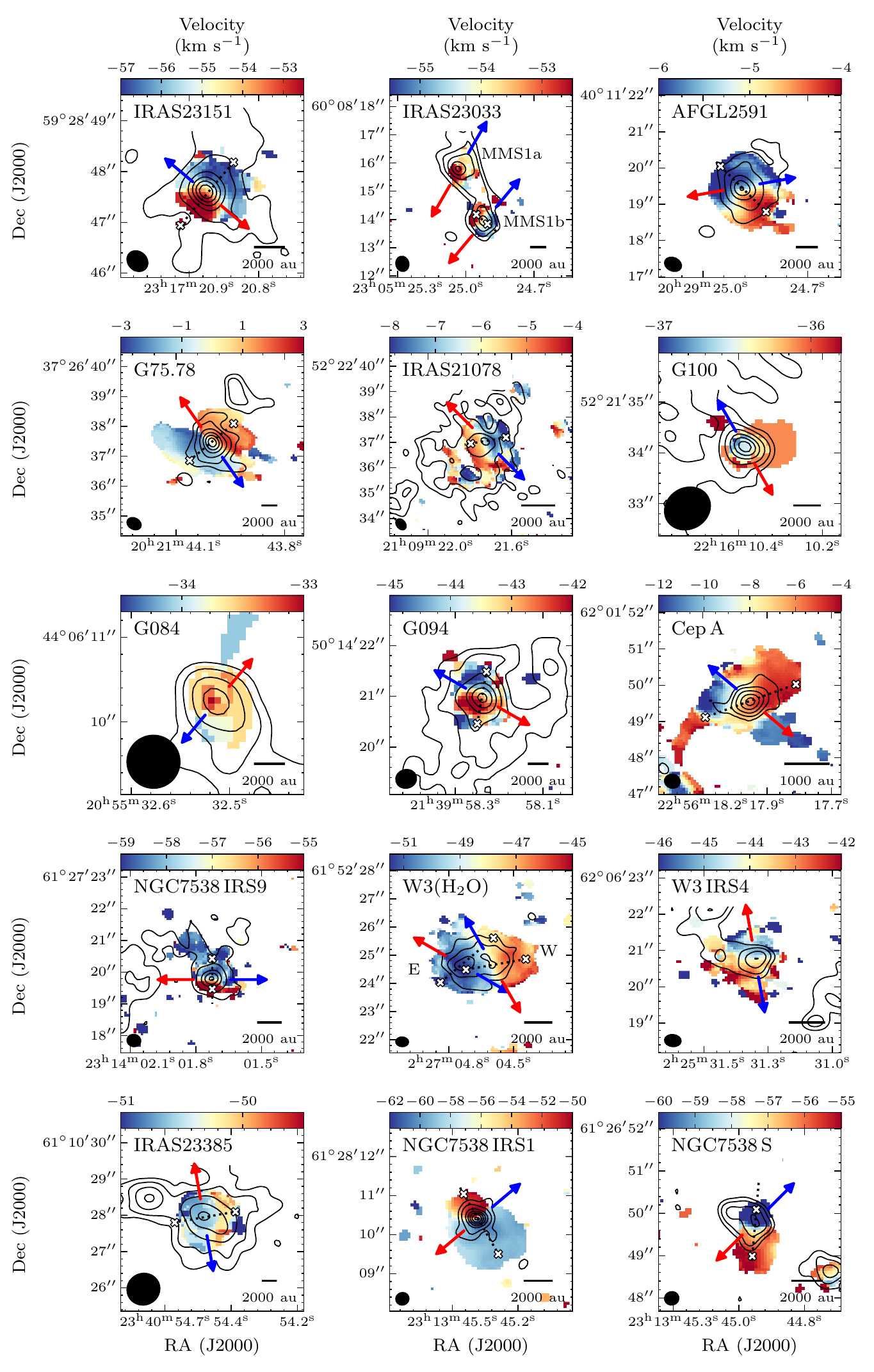}%
    \end{ocg}%
    \caption{Intensity-weighted mean velocity (first moment) maps of \mck{3} showing the dense gas kinematics for 15 of the 20 sources in the CORE survey. The contours correspond to the 1.37~mm continuum as described in Fig.~\ref{f: mom_velo_h2co}. The blue and red arrows correspond to the estimated directions of bipolar blueshifted and redshifted molecular outflows, respectively (see Sect.~\ref{s: sample_outflows}). The dotted lines indicate the position of the strongest velocity gradient tracing the disk, i.e. perpendicular to the rotation axis. The edges of the assumed disk extent are marked with an $\times$. The synthesised beam is shown in the bottom left corner and a scale bar in the bottom right corner of each panel. When displayed in Adobe Acrobat, it is possible to switch to the \setocgs{}{l:decomposed_velo_ch3cn_k3}{l:mom_velo_ch3cn_k3}{\protect\cdbox{decomposed velocity map}} and back to the \setocgs{}{l:mom_velo_ch3cn_k3}{l:decomposed_velo_ch3cn_k3}{\protect\cdbox{first moment map}} for comparison; otherwise, see Fig.~\ref{af: velo_decomp}}
    \label{f: mom_velo_ch3cn_k3}
\end{figure*}

In Fig.~\ref{f: mom_velo_ch3cn_k3} we show the intensity-weighted mean velocity (first moment) maps of \mck{3} for the remaining 15 sources in the CORE sample in search of velocity gradients that trace rotation. The details of image weighting parameters and the associated beam sizes, as well as the ranges over which the emission is integrated are listed in Table~\ref{t: CORE_disk_details}. The corresponding zeroth and second moment maps can be found in Appendix~\ref{a: mom_panels_ch3cn_k3}. The threshold used for creating the moment maps is set to 6$\sigma$, where sigma is estimated as the root mean square noise in an emission-free area in the channel corresponding to the peak of the line; this is also listed in Table~\ref{t: CORE_disk_details}. We detect velocity gradients across almost all cores with magnitudes of a few \kms, depicted by black dotted lines in the panels, spanning a physical projected distance of 2\,000--8\,000~au. The bulk of the emission for most sources is covered by many beams, allowing us to properly resolve these objects, with the exception of the sources for which we degraded the resolution in order to detect \mc. In these cases, the extent of the emission is only one or two beams across. For two of these sources in particular (G084 and G100), the velocity gradients in the first moment maps are not particularly convincing. Furthermore, IRAS23033 consists of two cores (MMS1a and MMS1b), both of which have detection of \mc, with MMS1b the brighter of the two. Similarly, \W\ consists of two cores, E and W corresponding to the eastern and western cores, respectively, approximately 2000~au apart \citepads{2018A&A...618A..46A}. Therefore, we detect \mc\ for 17 cores in total in this early stage of star formation. 

Intensity-weighted first moment maps are inherently sensitive to noise in the spectrum and any asymmetries in the line profile (see \citeads{2018RNAAS...2..173T} for a short overview). In order to overcome this issue and detect the velocity field beyond the 6$\sigma$ threshold used in the creation of moment maps, which can often mask more emission than needed, we have fitted Gaussian profiles to the spectra of all sources using the \textit{Spectral-Cube} tool within the \textit{PySpecKit} package in Python (\citeads{2011ascl.soft09001G}; \citeads{2019zndo...2573901G}, \citeyearads{2022AJ....163..291G}). Fitting one Gaussian profile assumes that there is only one velocity component at each position. In these so-called decomposed velocity maps, we detect emission that is slightly more extended in all directions for all sources (see Fig.~\ref{af: velo_decomp}). For G084 and G100, velocity gradients become slightly clearer than in the first moment maps, albeit not as strong as the other sources, with amplitudes below 1~\kms, making them less reliable disk candidates at the resolution of our observations. Furthermore, the velocity gradient across IRAS23033 MMS1a is in the same direction as the molecular outflow, depicted by blue/red arrows in Fig.~\ref{f: mom_velo_ch3cn_k3} and discussed further in the next section (see CORE case study by \citeads{2019A&A...629A..10B} for a detailed analysis of IRAS23033). Moreover, \citeads{2020A&A...636A.118M} performed an in-depth study of W3IRS4 using the CORE data and detected velocity gradients in a number of species around this core, but found that the axis of the velocity gradient changes depending on the species, hence likely tracing a different combination of infall, rotation, and outflow, depending on the tracer. They further concluded that any disk around W3IRS4 must be smaller than the resolution of our observations for this source (600~au). For these reasons, we remove G084, G100, IRAS23033 MMS1a, and W3IRS4 from our list of disk candidates. For the remaining 13 best disk candidates, the velocity gradients observed using the dense gas tracer \mc\ are consistent and coherent with the motion of more diffuse gas on larger scales, for example as traced by $\mathrm{H_2CO}$ (see Fig.~\ref{f: mom_velo_h2co}). The position angles of the \mc\ velocity gradients are listed in Table~\ref{t: CORE_disk_details}, defined with respect to the north-south axis and measured counterclockwise. 

\subsection{Molecular outflows}\label{s: sample_outflows}

We make use of outflow tracing molecules, mainly the $J=2-1$ transition of CO (from our single-dish IRAM 30-m observations) and its $^{13}\mathrm{CO}$ isotopologue (from our single-dish, NOEMA, and merged observations), to infer the directions of molecular outflows being driven by these YSOs. In Appendix~\ref{a: outflow_panels} we show the intensity maps of the blueshifted and redshifted gas with respect to the systemic velocity of the source by integrating over the blue and red wings of the emission for these transitions as well as for $\mathrm{C^{18}O}\,(2-1)$ and $\mathrm{SO}\,(6_5-5_4)$. Since the interferometer filters out most of such extended emission, we make use of the IRAM 30-m data with a field of view of 1.5\arcmin\ by 1.5\arcmin\ (and half-power beam width of $\sim$11\arcsec\ at this frequency) to infer the structure of these large-scale flows. Furthermore, the interferometer and in particular the combination of the interferometric with single-dish observations for $^{13}\mathrm{CO} (2-1)$ are useful in disentangling the ejection points of the outflows, especially in cores which have substructure as multiple outflows may exist, being driven by other objects (\eg~the case of \W\ presented in \citeads{2018A&A...618A..46A}). 

We observe extended high-velocity CO $(2-1)$ and $^{13}\mathrm{CO} (2-1)$ emission in the vicinity of the most luminous source in the field for all 17 cores where we detect \mc\  emission (see Appendix~\ref{a: outflow_panels}). In many of the sources, we detect clear bipolar molecular outflows (\eg~IRAS23151, AFGL2591, Cep\,A, \W) while the case is more complex for others. As an example, for NGC7538\,IRS1, our data suggest a bipolar outflow in the NW-SE direction confirmed by other works (\eg~\citeads{1991ApJ...371..163K}; \citeads{1998AJ....115.1118D}; \citeads{2011A&A...530A..53K}; \citeads{2011ApJ...728....6Q}; \citeads{2013A&A...558A..81B}), while an ionised jet in the N-S direction has been detected by others (\eg~\citeads{1995ApJ...438..776G}; \citeads{2009ApJ...699L..31S}; \citeads{2014A&A...566A.150M}; \citeads{2015A&A...573A.108G}). This ambiguity can be attributed to the fact that further fragmentation has been reported within this source by higher resolution observations in the centimetre regime (\citeads{2015A&A...573A.108G}; \citeads{2017A&A...605A..61B}), and made more complex because the central source is already ionising a hypercompact \ion{H}{ii} region. In fact, NGC7538\,IRS1 is the only source within the CORE survey where we detect the molecular line transitions in absorption against the strong continuum background at the peak position of the continuum emission spanning an area roughly the size of a synthesised beam \citepads{2012A&A...543A..88B}. Furthermore, for sources that are in complex environments such as IRAS23033 and W3IRS4, establishing an exact direction for the molecular outflow(s) is difficult. For IRAS23033, we adopt the directions of outflows determined by \citetads{2019A&A...629A..10B} who performed a detailed analysis of this region using the same CORE data. Similarly, for W3IRS4, \citetads{2020A&A...636A.118M} provided an in-depth study of this region using the CORE data and adopted an outflow direction roughly in the NNE-SSW direction, which is what we adopt here as well. 

The estimated directions of blueshifted and redshifted emission of molecular outflows are shown on Fig.~\ref{f: mom_velo_ch3cn_k3} as blue and red arrows, respectively. The outflow position angles are also summarised in Table~\ref{t: CORE_disk_details}. It is important to note that due to the high level of multiplicity in high-mass star formation, it is often difficult to disentangle different outflow contributions as described above for a few sources. We therefore used our outflow observations in context with other more detailed case studies for the sample to estimate the outflow directions. In almost all cases the velocity gradients seen in \mck{3} are roughly perpendicular to the direction of the molecular outflows, supporting the explanation that the velocity gradients probed by dense gas tracers are associated with rotational motions, therefore making these objects excellent disk candidates. The exceptions to this finding are W3IRS4 and IRAS23033\,MMS1a where the \mc\ velocity fields seem to trace the outflows, and G084 and G100 that have very weak velocity gradients. As described in the previous section, these four cores have been removed from our list as potential disk candidates. For the remainder of this work, we study the properties of the remaining 13 best disk candidates listed in Table~\ref{t: CORE_disk_details}. A plot showing the absolute difference between the assumed disk and outflow position angles is shown in Fig.~\ref{af: outflows_disk_pa}.

\onecolumn
\begin{landscape}
\begin{table*}
\centering
\str{1.2}
\caption{Observational parameters for the disk candidates within the CORE survey.}
\label{t: CORE_disk_details}
\begin{tabular}{lllcccccccd{3.2}Cc}
\hline\hline
  Source & Weight\tablefootmark{a} & Beam & Res.\tablefootmark{b} & $\mathrm{\sigma_{line}}$\tablefootmark{c} & Outflow & Disk & $R_\mathrm{blue}\tablefootmark{e}$ & $R_\mathrm{red}\tablefootmark{f}$ & Extent\tablefootmark{g} & \multicolumn{1}{c}{$v_\mathrm{LSR}\tablefootmark{h}$} & \multicolumn{1}{c}{Integration range\tablefootmark{i}} & $\mathrm{^{12}C/^{13}C}$\tablefootmark{j}\\
  & & (\arcsec$\times$\arcsec, PA) & (au)& (m\jpb) & PA\tablefootmark{d}  ($\degr$)  &  PA\tablefootmark{d}  ($\degr$) &(au) & (au) & (au) & \multicolumn{1}{c}{$(\mathrm{km~s^{-1}})$} & \multicolumn{1}{c}{$(\mathrm{km~s^{-1}})$} & \\
    \hline
  IRAS23151 & robust 0.1 & $0.42\times0.35$, 47\degr & 1280 & 3.60 & 230 & 140 & 2680 & 2680 & 5360 & -53.1 & -64>-48 & 93 \\
  IRAS23033\,MMS1b & robust 1 & $0.49\times0.43$, 9\degr & 1980  &  3.86 & 140 & 50 & 520 & 1300 & 1820 & -54.4 & -59>-48 & 81 \\
  AFGL2591        & robust 0.1 & $0.45\times0.34$, -115\degr & 1310 &  5.91 & 100 & 45 & 2680 & 3400 & 6080  & -5.6 & -11>0 & 68 \\
  G75.78       & robust 0.1 & $0.45\times0.34$, 59\degr & 1510 &   5.19 & 35 & 130 & 3920 & 3920 & 7840 & 0.2 &  -7>5 & 67 \\
  IRAS21078   & robust 0.1 & $0.45\times0.31$, 39\degr & 570 &  4.04  & 45 & 100 & 1220 & 990 & 2210 & -7.0 & -15 > 2 & 69 \\ 
  G094 & robust 0.1 & $0.38\times0.36$, -48\degr & 1480  & 3.92 & 60 & 170 & 2900 & 2540 & 5440 & -42.9 & -52 > -35 & 80 \\
  Cep\,A          & robust 0.1 & $0.40\times0.36$, 77\degr & 300 &  6.94 & 50 & 110 & 930 & 930 & 1860 & -5.6 & -17 >  2 & 70 \\
  NGC7538\,IRS9      & robust 0.1 & $0.40\times0.35$, 78\degr & 1020 &    4.27 & 90 & 0 & 1725 & 770 & 2495 & -56.9 & -68 > -47 & 77\\
  W3(H$_2$O)\,E       & robust 0.1 & $0.42\times0.30$, 87\degr & 720  &       6.81 & 60 & 130 & 1930 & 2970 & 4900 & -51.6 & -57 > -39 & 79 \\
    W3(H$_2$O)\,W       & robust 0.1 & $0.42\times0.30$, 87\degr & 720 &      6.81 & 30 & 100 & 1930 & 2520 & 4450 & -48.7 & -57 > -39 & 79 \\
  IRAS23385   & robust 5 & $0.93\times0.88$, 113\degr & 4440 &   5.02  & 10 & 100 & 3980 & 4700 & 8680 & -50.6& -57 > -43 & 90 \\
\hline
  \underline{Pilot study}\\
  NGC7538\,IRS1      & robust 0.1 & $0.41\times0.36$, 118\degr & 1030  & 3.88 & 130 & 30 & 2980 & 1990 & 4970 & -56.3 & -66 > -45 & 78 \\
  NGC7538\,S         & robust 0.1 & $0.31\times0.29$, 104\degr & 800  &     3.12 & 135 & 175 & 990 & 2190 & 3180 & -61.4 & -66 > -50 & 78 \\ 
\hline
\end{tabular}
\tablefoot{Images are from a combination of configurations A, B, and D arrays of NOEMA with the exception of pilot sources that lack D-array observations. \\
\tablefoottext{a}{Robust weighting parameter using \textit{GILDAS} definition whereby robust 0.1 corresponds to uniform weighting and robust 10 is closest to natural weighting.}\\
\tablefoottext{b}{Average linear resolution, using distances from Table~\ref{t: sample_info}.}\\
\tablefoottext{c}{rms noise in the emission-free region in the channel that has the peak of emission for \mck{4}, therefore the unit is per 0.5~\kms\ channel.}\\
\tablefoottext{d}{Measured counterclockwise from the north-south axis.}\\
\tablefoottext{e}{Extent of the blueshifted emission for \mck{3}.}\\
\tablefoottext{f}{Extent of the redshifted emission for \mck{3}.}\\
\tablefoottext{g}{Sum of $R_\mathrm{blue}$ and $R_\mathrm{red}$.}\\
\tablefoottext{h}{Peak velocity of \mck{3}\ at the position of the continuum peak.}\\
\tablefoottext{i}{Integration ranges for the moment maps shown in Fig.~\ref{f: mom_velo_ch3cn_k3}.}\\
\tablefoottext{j}{Isotopic ratios calculated according to \citetads{1994ARA&A..32..191W} (see text for details).}
}
\end{table*}
\vfill
\end{landscape}
\twocolumn


\section{Analysis and discussion}\label{s: analysis}
\subsection{Temperature distribution}\label{s: sample_temp}

We make use of the \textit{XCLASS} software \citepads{2017A&A...598A...7M} within \textit{CASA} to fit pixel-by-pixel the spectra of \mckr{0}{6}, including \mcisokr{0}{3}, under the assumption of LTE, which is typically the case in such dense environments. In summary, the user provides initial guesses and sets allowed ranges for a set of parameters (column density, rotational temperature, line velocity, linewidth, and source size); the software solves the radiation transfer equation and through a minimisation routine modifies these parameters until the best fit is found. The software then creates maps for each of the fitted parameters. The details of the \textit{XCLASS} fitting routine are outlined in Appendix B of \citetads{2018A&A...618A..46A} where we had fixed the source size to yield a beam filling factor of 1 and only fitted for the optically thin \mckr{4}{6} lines, including \mcisokr{0}{3} isotopologues. While the observations presented here are resolved, our extensive tests with \textit{XCLASS} have shown that having the source size parameter (and hence the filling factor) as a fitting parameter allows for much better simultaneous fits to the spectra when \mckr{0}{3} lines are optically thick and the optically thinner \mckr{4}{6} transitions\footnote{See Fig. 5 of \citetads{2015PASP..127..266M} to see the effect of beam filling factor on excitation temperatures.}. Note that the $K=3n$ transitions (\ie~3 and 6) have double the statistical weight of the other transitions. For each source, we set the expected observed isotopic ratio according to its galactocentric distance, $D_\mathrm{GC}$, based on estimates of \citetads{1994ARA&A..32..191W}: $^{12}\mathrm{C}/^{13}\mathrm{C} = (7.5\pm1.9)D_\mathrm{GC}+(7.6\pm12.9)$. Furthermore, we only fitted pixels above 6$\sigma$ detection limit ($\sigma$ values and isotopic ratios are listed in Table~\ref{t: CORE_disk_details}). The exceptional modelling approach for NGC7538\,IRS1 is summarised in Appendix~\ref{a: ngc7538_irs1_temp}. 

In Fig.~\ref{f: CORE_Trot}, we show the rotational temperature maps. Assuming the lines are thermalised, rotational temperature can be directly taken as the kinetic temperature of the gas in these dense regions. Median temperature values are listed in  Table~\ref{t: gas_masses} and are between 70--210~K with an uncertainty of $10-20\%$ (see Appendix B of \citeads{2019A&A...631A.142G}). It is useful to examine the maps of all fitted parameters, namely \mc\ column density, line velocity, linewidth, and source size, when analysing the spatial distribution of gas temperatures (see Appendix~\ref{a: xclass_panels}). The \textit{XCLASS} maps for the three cores that show \mc\ emission but that are not good disk candidates (G100, G084, W3IRS4) are shown in Figure~\ref{af: XCLASS_rest}. In the following, we summarise some of the trends we find for the physical conditions probed by \mc.

In the case of an idealised disk, the temperature and column density are expected to decrease with increasing radius. While for some of the sources in our sample we do see such behaviour (\eg~AFGL 2591, \W\,E,  \W\,W, Cep\,A), the distribution of temperature is often not as smooth and symmetric as expected. Associating substructure in the temperature maps with underlying substructure in the disk is not straight forward. In fact, we have shown through simulated observations that the temperature is overestimated in regions of the disk with non-Gaussian line shapes, as well as multi-component line profiles to which we only fit one component \citepads{2019A&A...632A..50A}. Such multi-component line profiles are in fact expected as we start to resolve the infalling material from the envelope onto the disk, which results in gas motions at various velocities along the line of sight depending on the inclination of the object. Furthermore, we do not account for density and temperature gradients along the line of sight. 

For some of the sources we find the outskirts to be warmer than the inner regions (\eg~IRAS23151, NGC7538\,IRS9). These regions have been properly modelled by \textit{XCLASS} and are not the result of fitting noisy spectra as we have masked out the regions below 6$\sigma$ (although the spectra are noisier in the outskirts than inside the cores). For these sources, it is helpful to simultaneously look at the maps of other parameters (see Figs.~\ref{af: XCLASS_Ntot}--\ref{af: XCLASS_source_size} in the Appendix). These outer hot components are often associated with large linewidths, consistent with our decomposed maps of linewidth (see Fig. ~\ref{af: width_decomp}) that are able to cover more extended emission than the second moment maps. In many cases, the hot component associated with broad linewidths are found in the ejection direction of the outflows. This hints at the possibility that these regions  are regions that may have been heated by the outflows and associated shocks and/or regions that have been carved out by the molecular outflows allowing a deeper look into the cores. This touches on an important point about high-mass star formation, that such bipolar outflows that form early may be critical in creating cavities through which intense radiation pressure is able to escape (\eg~\citeads{2002ApJ...569..846Y}; \citeads{2005ApJ...618L..33K}; \citeads{2015ApJ...800...86K}, \citeyearads{2016ApJ...832...40K}; review by \citeads{2019FrASS...6...54P}), allowing the accretion of mass onto their disks. The case of \W\ is perhaps the most impressive (see the large linewidths and positions of outflows in Fig.~\ref{af: width_decomp}).

\begin{figure*}
    \centering
    \includegraphics[width=0.97\hsize]{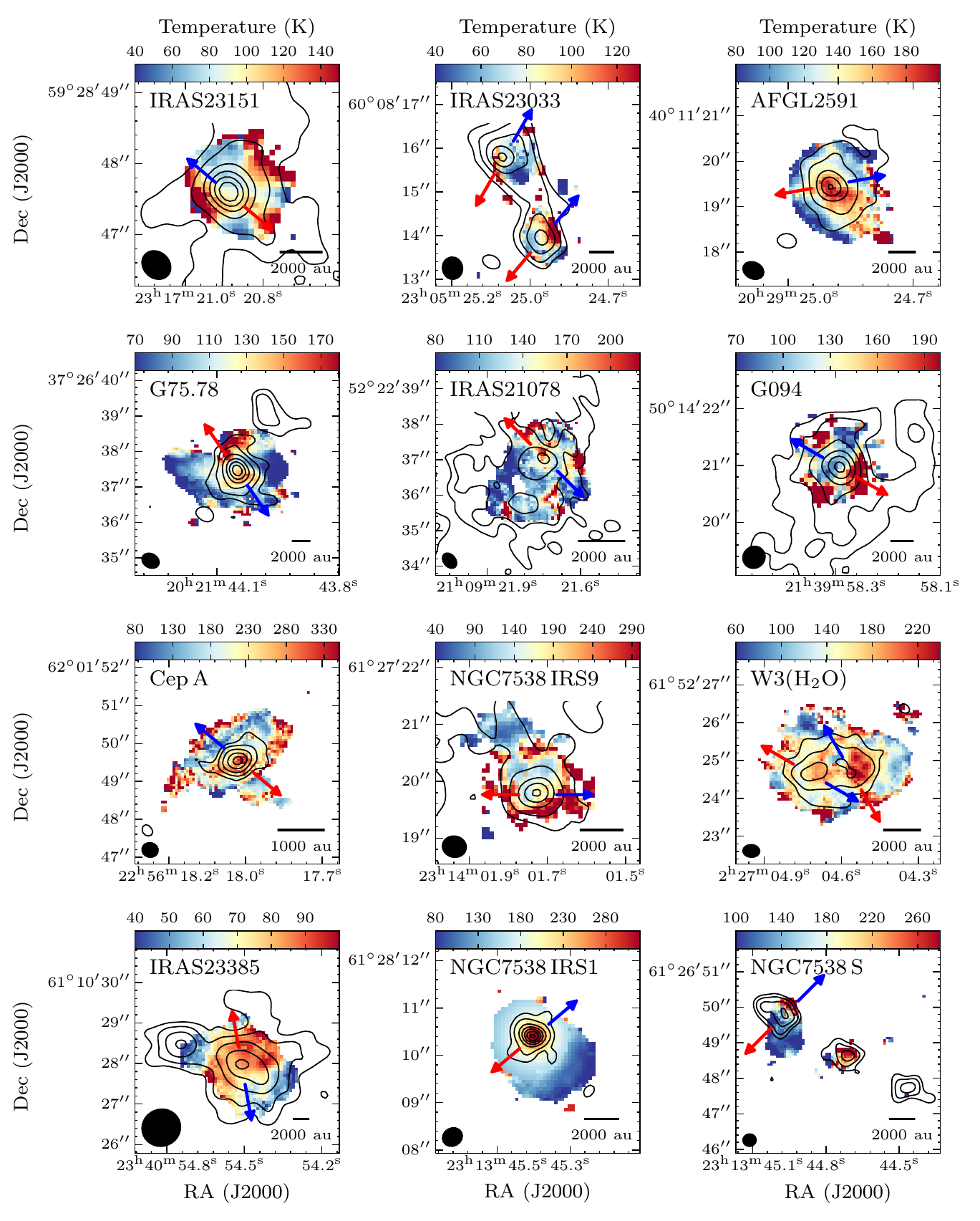}
    \caption{Rotational temperature maps obtained by fitting \mckr{0}{6} and \mcisokr{0}{3} lines with \emph{XCLASS}. The contours correspond to the 1.37~mm continuum as described in Fig.~\ref{f: mom_velo_h2co}. For NGC7538\,IRS1, only the region outside the continuum to the south-west is modelled by \emph{XCLASS} and was scaled towards the continuum peak position following a temperature power-law distribution $T\propto r^{-0.4}$. The blue and red arrows correspond to the estimated directions of bipolar blueshifted and redshifted molecular outflows, respectively. The synthesised beam is shown in the bottom left corner and a scale bar in the bottom right corner of each panel. Maps of column density, velocity offset, linewidth, and source size are shown in Appendix~\ref{a: xclass_panels}.}\vspace{1cm}
    \label{f: CORE_Trot}
\end{figure*}

\begin{table}
\str{1.2}
\centering
\caption{Gas mass estimates.}
\label{t: gas_masses}
\begin{tabular}{lcccc}
\hline\hline
  Source & $F_\nu$\tablefootmark{a} & $T_\mathrm{med}$ & $M_\mathrm{gas}(T_\mathrm{med})$ & $M_\mathrm{gas}$ \\
  & (mJy)  & (K) & (\mo) & (\mo) \\
\hline
  IRAS23151                   & 54 & 99 & 3.4 & 3.6 \\
  IRAS23033\,MMS\,1b  & 47 & 84 & 3.9 &3.9  \\ 
  AFGL2591                   & 156 & 132 & 7.1 & 6.2 \\
  G75.78                        & 123 & 102 & 9.7 & 8.1 \\
  IRAS21078                  & 301 & 120 & 3.2 & 3.3 \\ 
  G094                           & 27 & 131 & 2.8 & 3.0 \\
  Cep\,A                         & 457 & 212 & 0.8 & 0.6 \\
  NGC7538\,IRS9           & 55 & 212 & 0.9 & 1.2 \\
  W3(H$_2$O)\,E           & 669 & 149 & 9.9 & 9.4 \\
  W3(H$_2$O)\,W          & 544 & 148 & 8.1 & 7.0 \\
  IRAS23385                 & 96 & 72 & 18.3 & 19.1  \\
\hline
  \underline{Pilot study}\\
  NGC7538\,IRS1   & 2341\tablefootmark{$\ast$} & 155 & 53.9 & 31.7 \\
  NGC7538\,S         & 67 & 132 & 2.0 & 1.8 \\ 
\hline
\end{tabular}
\tablefoot{
\tablefoottext{a}{Continuum flux density at 1.37~mm within $6\sigma$ threshold (see text for details).}
\tablefoottext{$\ast$}{Subtracting the free-free contribution corresponds to a flux density of 2078~mJy at 1.37~mm (see text for details). Gas masses are calculated using the continuum emission maps with the free-free contribution subtracted.}
}
\end{table}
 
\begin{table*}
\str{1.2}
\centering
\caption{\textit{KeplerFit} parameters and dynamical mass estimates.}
\label{t: dynamical_masses}
\begin{tabular}{lcccccc}
\hline\hline
  Source & $M_\mathrm{PV}$ & $v_\mathrm{cont}$\tablefootmark{a} & $v_\mathrm{fitted}$\tablefootmark{b} & $p_\mathrm{fitted}$\tablefootmark{c} &  $p_\mathrm{masked}$\tablefootmark{d} & Best fitted transition \\
  &(\mo) & (\kms) & (\kms) & (au) & (au) &  \\
\hline
  IRAS23151                   & $23\pm2$ &  $-53.9$ & $-54.8$  & $120$ & $-400\rightarrow500$ & \mck{{3,4,5}} stacked \\
  IRAS23033\,MMS\,1b  &  $14\pm1$ & $-54.4$ & $-53.6$ & $88$ & $-400\rightarrow400$ &  \mck{3} \\ 
  AFGL2591                    &  $23\pm1$ & $-5.6$ & $-5.8$ & $-623$ & $-1000\rightarrow1000$ & \mck{3}\\
  G75.78                          &  $19\pm3$ & $0.7$ & $0.0$  & $-1078$ & $-1800\rightarrow2300$ & \mck{5}\\
  IRAS21078                    & $13\pm1$ & $-6.6$ & $-6.5$  &  $100$  & $-250\rightarrow250$ & \mck{5} \\ 
  G094                           & $20\pm3$ & $-42.9$ & $-43.9$  & $0$  & $-500\rightarrow1800$ &  \mck{3}\\
  Cep\,A                         & $24\pm1$ &  $-7.3$ & $-8.3$  &  $10$  & $-500\rightarrow500$ & $\mathrm{OCS}\,(18-17)$ \\
  NGC7538\,IRS9           & $25\pm3$ & $-56.7$ &  $-57.1$  & $370$  & $-200\rightarrow700$ & $\mathrm{HC_3N}\,(24-23)$ \\
  W3(H$_2$O)\,E\tablefootmark{$\ast$}  & $9\pm1$ & $-51.0$ & $-51.3$ & $0$ & $-500\rightarrow500$ & \mck{4} \\
  W3(H$_2$O)\,W\tablefootmark{$\ast$} & $10\pm1$ & $-47.0$ & $-48.0$ & $99$ & $-\infty\rightarrow200$ &  \mck{3} \\
  IRAS23385                 & $17\pm2$ & $-50.7$ & $-51.0$  &  $1215$ & $-1200\rightarrow2000$ & $\mathrm{HC_3N}\,(24-23)$ \\
\hline
  \underline{Pilot study}\\
  NGC7538\,IRS1   &  $22\pm2$ & $-56.0$ & $-55.1$ &  $310$ & $0\rightarrow900$ & $\mathrm{C_2H_5CN}\,(25_{22, 4}-24_{22, 3})$ \\
  NGC7538\,S         & $12\pm1$ & $-60.8$ & $-60.7$ &  $0$  & $-\infty\rightarrow950$ & \mck{5} \\ 
\hline
\end{tabular}
\tablefoot{
\tablefoottext{a}{The velocity of the line at the position of the continuum peak.}
\tablefoottext{b}{The fitted central velocity in the PV diagram. The central velocity of the PV diagram was initiated at the velocity of the line at the position of the continuum peak ($p_\mathrm{cont}$) and allowed to shift $\pm1$~\kms\ in the fitting routine (see dashed horizontal lines in Fig.~\ref{f: pvpanels}).}
\tablefoottext{c}{The fitted central position in the PV diagram. The central position of the PV diagram was initiated at the continuum peak position and allowed to shift $\pm$ a synthesised beam in the fitting routine (see dashed vertical lines in Fig.~\ref{f: pvpanels}).}
\tablefoottext{d}{Masked out (\ie~excluded) positions with respect to the fitted position, $p_\mathrm{fitted}$ (see dotted vertical lines in Fig.~\ref{f: pvpanels}).}
\tablefoottext{$\ast$}{Based on A-array observations presented in \citetads{2018A&A...618A..46A}.}
}
\end{table*}
 
\subsection{Gas Masses}\label{s: sample_gas_mass}

Assuming optically thin dust emission at 1.3~mm and a gas-to-dust mass ratio $R$ of 150 \citepads{2011piim.book.....D}, we can convert the flux density $F_\nu$ of the continuum observations to a mass via
\begin{equation}
  \label{e: sample_mass}
  M=\frac{d^2\,F_\nu\,R}{B_\nu (T_D)\,\kappa_{\nu}},
\end{equation}
where $d$ is the distance to the source and $\kappa_\nu=0.9\,\mathrm{cm^2\,g^{-1}}$ the dust absorption coefficient of thin ice mantles after $10^5$ years of coagulation at a density of $10^6$~cm$^{-3}$  \citepads{1994A&A...291..943O}. $B_\nu (T_D)$ is the Planck function for blackbody radiation, which at the wavelength under investigation nearly follows the Rayleigh-Jeans law and is linearly dependent on the dust temperature $T_D$. We assume that gas and dust are in equilibrium and use the temperature maps presented in the previous section (Fig.~\ref{f: CORE_Trot}) and the continuum maps (shown as contours in e.g. Fig.~\ref{f: mom_velo_h2co}) to create maps of gas mass for each core. The only source with significant free-free contribution is NGC7538\,IRS1 \citepads{2017A&A...605A..61B} that we subtract from our 1.37~mm continuum map (see Appendix~\ref{a: ngc7538_irs1_freefree}). Summing over all pixels with continuum emission above 6$\sigma$ gives a final mass value for each core, listed in Table~\ref{t: gas_masses}. We also list gas mass estimates calculated using the median temperature values, as it is common practice to use one temperature value when good thermometers are not observed. For this calculation, the flux density is calculated by converting the units of each map from \jpb\ to \jpp, masking out the continuum emission below 6$\sigma$ as well as regions that do not have any detection for \mc, and summing over the remaining pixels. For IRAS23033\,MMS1b the emission is integrated over a circular region encompassing this core only. For \W\ E and W, we simply define a vertical boundary between the two cores as presented in \citetads{2018A&A...618A..46A}. It is reassuring that gas mass estimates calculated using median temperatures are, in almost all cases, very close to the estimates that use the full temperature maps. The only exception is NGC7538\,IRS1, which is the brightest continuum source in the sample. This large difference is due to the fact that gas temperatures are warm ($>200$~K) where the bulk of continuum emission is found, while regions outside of the continuum emission are colder, lowering the value of the median temperature (155~K), which results in a higher mass estimate. 


\subsection{Rotation curves and dynamical mass estimates} \label{ss: sample_pv_masses}

\begin{figure*}
    \centering
    \includegraphics[width=0.83\hsize]{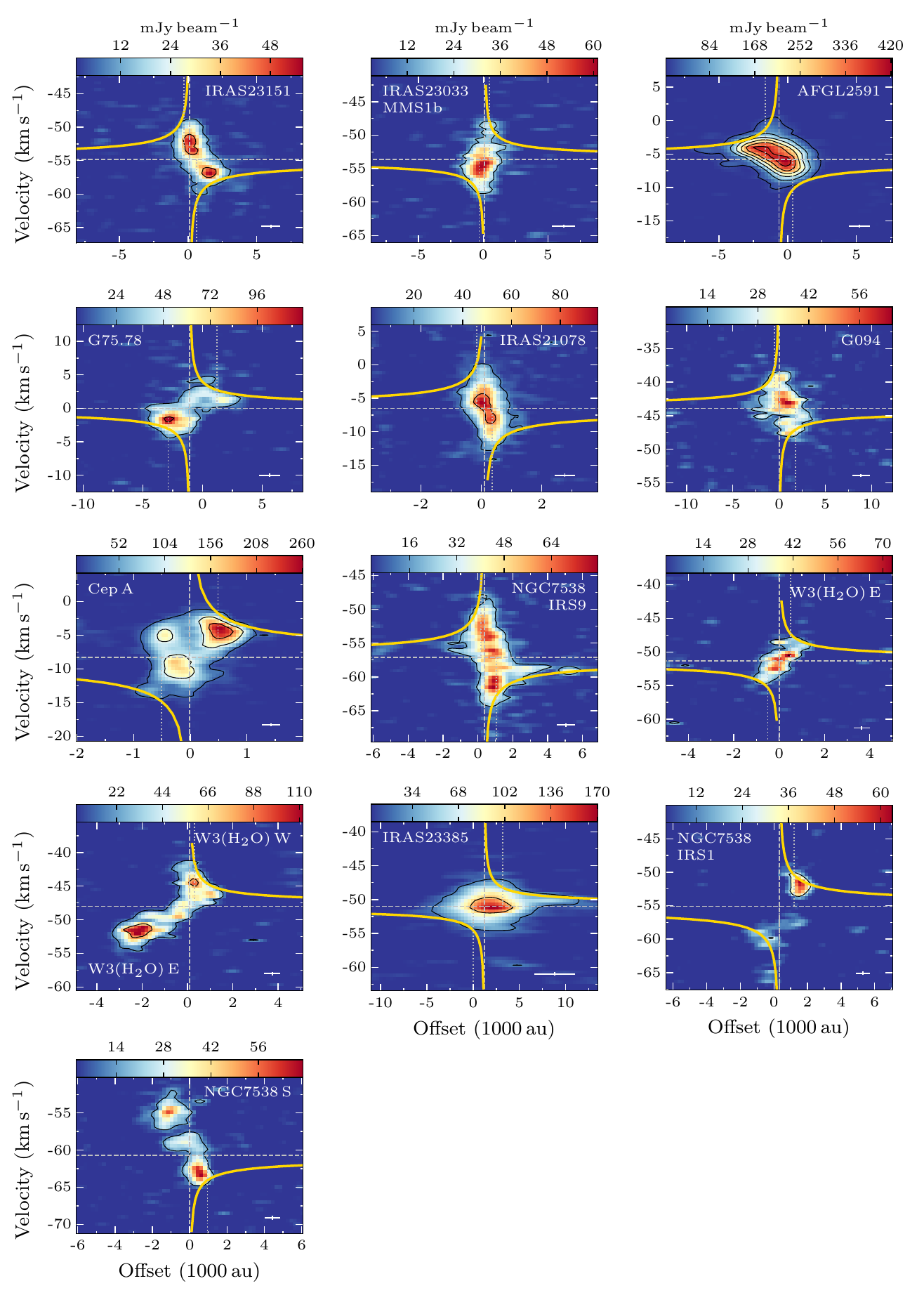}
    \caption{Position--velocity (PV) plots for best fitted transitions listed in Table~\ref{t: dynamical_masses} along cuts in the direction of rotation as depicted by dotted lines in Fig.~\ref{f: mom_velo_ch3cn_k3}. The width of the cut is the size of a synthesised beam to increase the signal-to-noise ratio. The PV plots of \W~E and \W~W make use of the A-array only observations in order to gain more resolution and disentangle the two cores from the circumbinary material. PV plots for all other sources make use of the full dataset observed in the CORE survey. The contours correspond to the 6$\sigma$ level increasing in steps of 24$\sigma$. Yellow lines show the Keplerian rotation curves for enclosed masses, $M_\mathrm{PV}$, listed in Table~\ref{t: dynamical_masses} for each source. The inner radii that were masked out (\ie~excluded) in the fitting routine are shown as vertical dotted lines. The fitted central velocities and positions are shown as dashed lines. The zero offset corresponds to the continuum peak position. The crosses in the bottom right corners correspond to the spatial and spectral resolutions.}
    \label{f: pvpanels}
\end{figure*}

In Fig.~\ref{f: pvpanels} we show the position-velocity (PV) maps of best fitted transitions listed in Table~\ref{t: dynamical_masses} along cuts across the strongest velocity gradients as depicted by dotted lines in Fig.~\ref{f: mom_velo_ch3cn_k3}. The yellow curves correspond to Keplerian rotation profiles with $v(r)=\sqrt{GM/r}$ for enclosed masses listed in Table~\ref{t: dynamical_masses} (described further below). It is clear that in many of the sources the PV diagrams do not resemble differential rotational motion of the gas well, but rather mimic rigid-body-like rotation (\ie\ $v(r) \propto r$). Furthermore, the existence of excess emission in the two quadrants opposite to where the rotational motions of the disk are expected hints to the presence of infalling material from the envelope (\eg~see models of \citeads{1997ApJ...475..211O} and \citeads{2012ApJ...748...16T}), outflowing gas,  and/or coarse spatial resolution. While weak (or lack of) differential rotation could be an indicator that the \mc\ emission may not be tracing disks in these observations, we have shown through synthetic observations in \citetads{2019A&A...632A..50A} that in fact PV diagrams of poorly resolved disks do often mimic rigid-body-like rotation. In the PV plots of poorly resolved disks, the molecular emission appears stretched in the position direction with excess infall contributions appearing in opposite quadrants as the disk and envelope material become blended. Therefore, using PV diagrams to rule out the existence of disks is a challenging task. 

From the PV plots, we see that the rotation curves of some sources like AFGL2591, G75.78, IRAS21078, and NGC7538\,IRS1 better resemble differential rotation of gas in a disk as some gas is found at high-velocities close to the central object at zero offset, with lower velocity gas further out. The PV plot of Cep\,A\,HW2 is particularly interesting, as it has a very extended nature, with a significant amount of emission present in the top-left quadrant associated with infall motions. In fact, using the 3D velocity field of 6.7~GHz $\mathrm{CH_3OH}$ masers, \citetads{2017A&A...603A..94S} show that the velocity field in Cep\,A\,HW2 is dominated by an infall component from a radius of 900~au ($\sim$1.3\arcsec) down to 300~au ($\sim$0.4\arcsec), where a rotational component associated with an accretion disk becomes dominant. 

To determine protostellar masses, we fit Keplerian curves to the $6\sigma$ edges of emission in the PV diagrams shown in Fig.~\ref{f: pvpanels} (the outermost contours) using the \textit{KeplerFit}\footnote{Developed by Felix Bosco, available at \url{https://github.com/felixbosco/KeplerFit}.} package \citepads{2019A&A...629A..10B} in \textit{Python}. This method was first introduced by \citetads{2016MNRAS.459.1892S} who showed that fitting the edge of emission in PV diagrams provides a more accurate estimate for the enclosed mass than other methods. We tested this method with simulated observations for our purposes in \citetads{2019A&A...632A..50A} where we showed that depending on how resolved disk/envelope systems are, fits to either the inner or outer parts of the PV diagrams may provide more accurate protostellar mass estimates. We further showed that fitting the PV plots of poorly resolved disk/envelope systems result in masses that are overestimated because emission is stretched over a larger area and high-velocity components of a disk that typically span small radii close to the central object are beam-smeared to lower velocities. For this reason, often velocities have a constant value at regions closest to the central object and have to be masked out in order to obtain better fits to the PV curves. 

Using the \textit{KeplerFit} code, we fit Keplerian rotation profiles ($v(r)=\sqrt{GM/r}$) to the $6\sigma$ edges of emission in the PV diagrams of \mckr{3}{5} transitions as they are not heavily blended with other (detected) lines, as well as $\mathrm{HC_3N}\,(24-23)$ and $\mathrm{OCS}\,(18-17)$. We also create and fit PV plots of \mckr{3}{5} transitions stacked together, which is particularly useful for sources where the emission of individual \mc\ transitions is weak. The central positions of the PV plots are set to the continuum peak positions and allowed to be fitted $\pm$ a synthesised beam in the fitting routine. The central velocities are set to the line velocities at the continuum peak positions and allowed to be fitted $\pm1$~\kms\ in the fitting routine. To obtain estimates on the fitted masses due to the choice of the edge of emission, we fix the best fitted central positions and central velocities and fit Keplerian rotation profiles to the $5\sigma$ and $7\sigma$ edges of emission in the PV diagrams.  In Table~\ref{t: dynamical_masses} we present the fit parameters corresponding to the PV diagrams of the best fitted transitions chosen based on the goodness of the fit determined according to the lowest combined $\chi^2$ value and confirmed by eye. The reported masses are the error-weighted averages for the $5-7\sigma$ solutions. The estimated dynamical masses, which for a Keplerian disk is the mass of the central object, are in the range $\sim10-25$~\mo\ for our sample of disk candidates. The Keplerian curves corresponding to the best mass estimates are shown as solid yellow lines on the PV diagrams shown in Fig.~\ref{f: pvpanels}. The inner radii that were excluded in the fitting routine are shown as dotted vertical lines. The fitted central velocities and positions are also listed in Table~\ref{t: dynamical_masses} and are shown as dashed lines in Fig.~\ref{f: pvpanels}.

The PV plot of NGC7538\,IRS1 shows the emission of \mc\ to be very extended in position, such that the best fitted PV curves yield masses that are extremely large and unrealistic considering the luminosity of the region. Observations of thermal $\mathrm{CH_3OH}$ in the centimetre regime have shown that this source harbours two disk-like structures embedded within the same rotating circumbinary envelope \citepads{2017A&A...605A..61B}, consistent with earlier observations of multiple clusters of methanol masers at 6.7~GHz \citepads{2014A&A...566A.150M}. Therefore, fitting Keplerian curves to the PV diagrams of \mc, which likely traces this circumbinary envelope, is not valid. For this reason, the PV diagrams of the rarer molecule $\mathrm{C_2H_5CN}$ that better trace regions closer to the protostars resulted in better fits. For the circumbinary envelope encompassing \W~E and W, we were able to disentangle the gas contributions from each of the two cores by using only the observations from the most extended configuration (A-array) of NOEMA \citepads{2018A&A...618A..46A}. In fact, the emission in the bottom-left quadrant of the PV plot for \W~W corresponds to \W~E, therefore the expected Keplerian curve corresponding to a mass of 10~\mo, shown as a yellow line, properly fits the contribution from \W~W in the top-right quadrant. Similarly, NGC7538\,S likely harbours multiple protostars as the continuum emission has an elongated shape and the \mck{3} emission has multiple peaks along this elongation (see Fig.~\ref{af: mom0_panels_ch3cn_k3}). Therefore, for this source, only one side of the PV plot was fitted. Furthermore, the emission of \mc\ lines in IRAS23385 lacks the high velocity components expected in Keplerian rotation, and can be better traced by $\mathrm{HC_3N} (24-23)$ and OCS\,(18-17). In fact, thorough analysis of the CORE data by \citetads{2019A&A...627A..68C}, revealed that the velocity field of the main core in this source can be described by a self-gravitating disk rotating about a $\sim9$\mo\ star.

We compare the mass estimates from fitting the PV diagrams, $M_\mathrm{PV}$, against masses derived from the source luminosities $M_\mathrm{lum}$ using the spectral type-temperature calibration of \citetads{2005IAUS..227..389C} (see Table~\ref{t: sample_info}). Masses derived from source luminosities are in the range $19-41$~\mo . Since the reported luminosities are from multi-wavelength SED fitting of lower resolution observations (see \citeads{2011A&A...525A.149M}), estimates for $M_\mathrm{lum}$ are upper limits, especially in sources with a high degree of fragmentation and multiple cores detected (\eg~IRAS23033 or \W). Our mass estimates from fitting the PV diagrams are either lower than or consistent with these upper mass limits, within the estimated errors. The reported dynamical masses and their uncertainties should be used cautiously as they rely on the choice of fitted molecular transition and whether the modelled transition traces the disk well. We select the upper limit of the dynamical mass estimates for the protostellar masses in the remainder of this work. When the dynamical mass upper limits are higher than the masses derived from the source luminosities, we select the luminosity derived masses. For sources that were studied in more depth using the same CORE data, we adopt the mass estimates reported in these works (see Table~\ref{t: masses_Q_CORE} for a summary).

\subsection{Timescales}
Following \citetads{2006Natur.444..703C}, \citetads{2011A&A...525A.151B} defined the term `toroids' for rotating structures surrounding the most massive stars that span thousands of au with masses of several 10~\mo, comparable to or greater than the mass of the central protostar(s). These objects are thought to be transient, feeding a central cluster of forming stars, each surrounded by their own centrifugally-supported disk on smaller scales (\eg~see Fig.~2 of \citeads{2018A&A...616A.101K}). They are analogous to rotating infalling envelopes around forming low-mass stars. \citetads{2011A&A...525A.151B} studied the stability of rotating structures around O- and B-type stars by plotting the ratio of free-fall timescale to the rotational period at the outer radius, $t_\mathrm{ff}/t_\mathrm{rot}$, versus $M_\mathrm{gas}$. The free-fall timescale represents the time it takes the gas to collapse under its own gravity. The rotational period at the outer radius is the time it takes the rotating structure to re-adjust its internal structure and stabilise after accreting new material. Simplifying the equations given in \citetads{2014A&A...569A..11S}, we calculate the rotational period via
\begin{equation}
  \begin{split}
  t_\mathrm{rot} &= \frac{2\,\pi\,R}{v_\mathrm{rot}} \\
  & = 29.8\,\left(\frac{R}{\mathrm{au}}\right)\left(\frac{v_\mathrm{rot}}{\mathrm{km\,s\textsuperscript{-1}}}\right)^{-1}\,\mathrm{yr},
  \end{split}
\end{equation}
where $v_\mathrm{rot}$ is the rotational velocity at the outer radius, $R$. The free-fall timescale is calculated according to 
\begin{equation}
  \begin{split}
   t_\mathrm{ff} &= \sqrt{\frac{3\,\pi}{32\,G\,\rho}} \\
  & = 0.195\,\left(\frac{M_\mathrm{gas}}{M_\sun}\right)^{-0.5}\left(\frac{R}{\mathrm{au}}\right)^{1.5}\,\mathrm{yr}.
  \end{split}
\end{equation}
While this expression is valid if there is no central protostar (\ie\ for a starless core), we use this expression to be consistent with the analysis presented in \citetads{2016A&ARv..24....6B}.

Because the rotating structures traced by \mc\ are not always symmetrically distributed on the position of the continuum peak, we calculate the rotational period and free-fall timescale for the redshifted and blueshifted sides of the rotating structures separately. We estimate the radii of the blueshifted and redshifted sides ($R_\mathrm{blue}$ and $R_\mathrm{red}$) as the distance from the continuum peak to the edge of the \mc\ emission along the strongest velocity gradient (dotted lines on Fig.~\ref{f: mom_velo_ch3cn_k3}). We manually set the positions of $R_\mathrm{blue}$ and $R_\mathrm{red}$ if the velocity gradient appears to have a break close to the edge of the emitting region (\eg~blueshifted side of Cep~A) or if the cores are blended (\eg~blueshifted side of \W~W). The positions of $R_\mathrm{blue}$ and $R_\mathrm{red}$ are marked by an $\times$ in Fig.~\ref{f: mom_velo_ch3cn_k3}. $v_\mathrm{rot}$ is estimated by subtracting the peak velocity at these positions from the LSR velocity measured at the peak of the continuum emission (listed in Table~\ref{t: CORE_disk_details}). 

In Fig.~\ref{f: tff_trot_Mgas}, we show the plot of $t_\mathrm{ff}/t_\mathrm{rot}$ versus $M_\mathrm{gas}$ for the redshifted and blueshifted sides of the 13 cores under investigation. The black dashed line corresponds to the theoretical curve of $t_\mathrm{ff}/t_\mathrm{rot}$ for a disk of mass $M_\mathrm{gas}$ rotating about a star with mass $M_\ast=10~M_\sun$. This is obtained by substituting $v_\mathrm{rot} = \sqrt{G(M_\mathrm{gas}+M_\ast)/R}$, \ie\ the expected velocity at radius $R$ if the rotation follows a Keplerian profile, in the above equations such that  
\begin{equation}\label{e: tff_trot}
  \frac{t_\mathrm{ff}}{t_\mathrm{rot}}=\sqrt{\frac{M_\mathrm{gas}+M_\mathrm{\ast}}{32\,M_\mathrm{gas}}}.
\end{equation}
In a review by \citetads{2016A&ARv..24....6B}, the authors created the $t_\mathrm{ff}/t_\mathrm{rot}$ versus $M_\mathrm{gas}$ plot for the largest sample of intermediate and high-mass protostars from the literature and confirmed the findings of \citetads{2011A&A...525A.151B} that disks and toroids are kinematically and dynamically different structures. Toroids contain more gas mass and are less rotationally supported with lower $t_\mathrm{ff}/t_\mathrm{rot}$, while for centrifugally supported disks in Keplerian rotation the infalling material has enough time to settle onto the disk, therefore the rotational timescale is shorter and the $t_\mathrm{ff}/t_\mathrm{rot}$ is higher. Comparing Fig.~\ref{f: tff_trot_Mgas} to Fig.~14 of \citetads{2016A&ARv..24....6B}, all of our sources lie in the regime corresponding to centrifugally supported disks, while toroids reside in a regime below the theoretical curve where gas masses are larger than $40~M_\sun$, beyond the limits plotted in Fig.~\ref{f: tff_trot_Mgas}.

\begin{figure}
    \centering
    \includegraphics[width=\hsize]{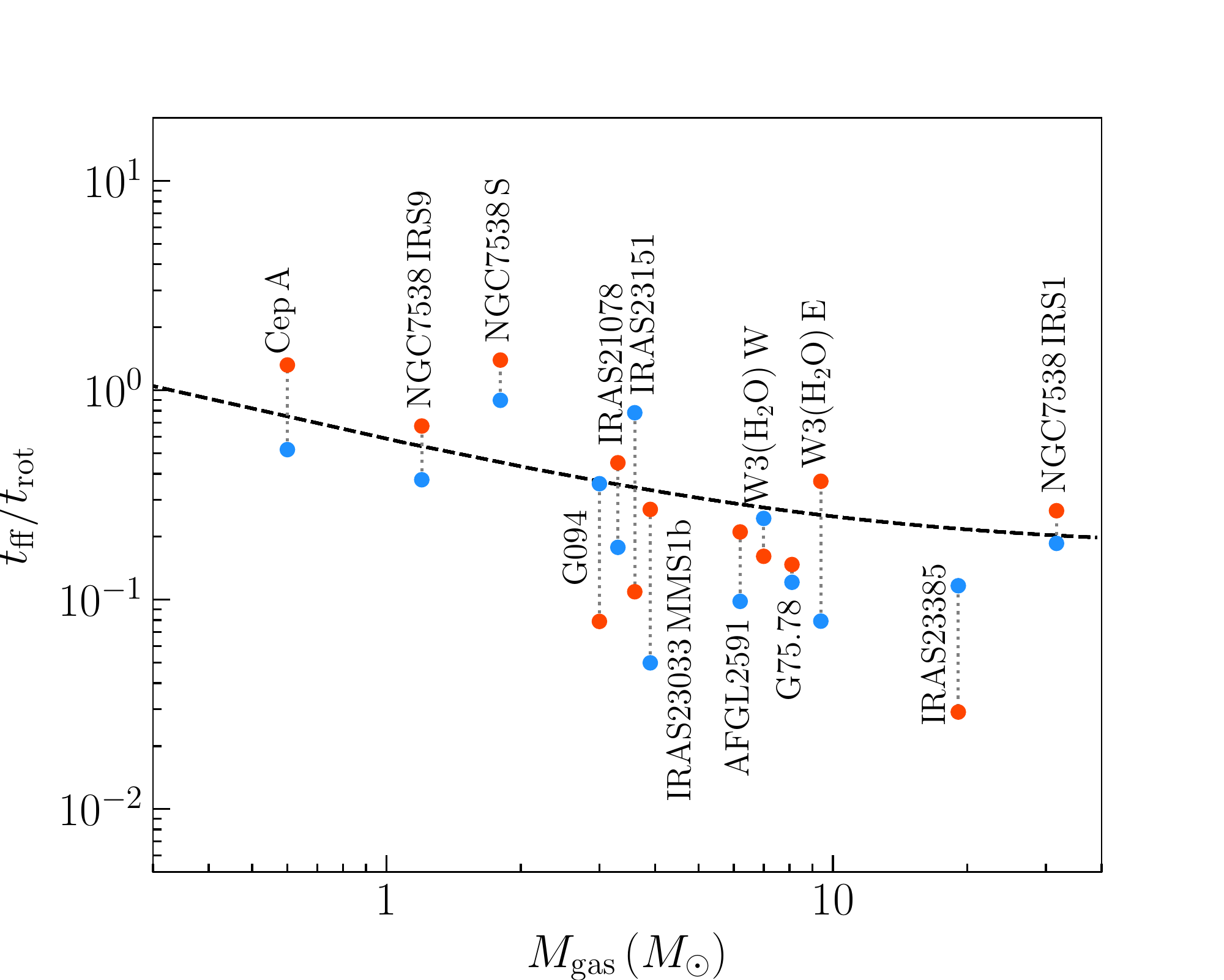}
    \caption{Ratio of free-fall to rotational timescale as a function of disk gas mass, showing most of the sources with \mc\ velocity gradients are rotationally supported. The blue and red dots correspond to the values calculated based on the rotational velocities at the edges of the disks on the blueshifted and redshifted sides, as depicted by crosses in Fig.~\ref{f: mom_velo_ch3cn_k3} and listed in Table~\ref{t: CORE_disk_details}. The black dashed line corresponds to theoretical curve of $t_\mathrm{ff}/t_\mathrm{rot}$ for a disk of mass $M_\mathrm{gas}$ in Keplerian rotation about a star with mass $M_\ast=10~M_\sun$ (see Eq.~\ref{e: tff_trot}).}
    \label{f: tff_trot_Mgas}
\end{figure}

\subsection{Specific angular momentum}

For a molecular cloud to form a star, it must lose many orders of magnitude in specific angular momentum ($j$). Various mechanisms at a multitude of scales can contribute to this loss (and transport) of angular momentum (\eg~disks, jets, outflows, winds, magnetic fields). Specific angular momenta have predominantly been measured in regions of low-mass star formation. In particular, it has been shown that on dense core scales (0.1~pc), the local specific angular momentum  (\ie\ the product of $v_\mathrm{rot}$ and $r$) scales as $r^{1.6}$ \citepads{1993ApJ...406..528G}, down to a scale of about $\sim$6000~au where the specific angular momentum is constant down to disk scales ($\sim$100~au). \citetads{1997ApJ...488..317O} was first to measure a constant $j$ value of $\sim10^{-3}$~\kms\,pc for this intermediate regime and attribute it to protostellar envelopes that are dynamically collapsing. Below $\sim$100~au, Keplerian rotation dominates and once the disk becomes unstable, material begins to move inward and angular momentum outward. A review by \citetads{2013EAS....62...25B} compiles most of the relevant observations from literature on this topic and presents the correlation of the specific angular momentum with radius from disk to core scales in their Fig.~8. In a recent work, \citetads{2019ApJ...882..103P} calculated the specific angular momentum radial profiles toward two Class 0 objects and a hydrostatic core candidate in Perseus and found a power-law relation with $j \propto r^{1.8}$ from 800--10000~au. The authors concluded that because the power-law index is between that of solid-body rotation ($j\propto r^2$) and pure turbulence ($j\propto r^{1.5}$), the influence of the dense core's initial level of turbulence is still present even down to 1000~au, inherited from the turbulent parental molecular cloud. Furthermore, \citetads{2020A&A...637A..92G} calculated the specific angular momentum profiles of 11 Class 0 protostellar envelopes from the CALYPSO survey from $\sim50-5000$~au scales and found a decreasing power-law relation with $j \propto r^{1.6\pm0.2}$ down to $\sim$1600~au and relatively constant down to $\sim50$~au. Since these findings are based on observational evidence from low-mass star-forming regions, it is important to calculate the specific angular momenta for our more massive sources for comparison. Additionally, it is important to note that the same scaling relations may not apply in high-mass star forming regions as it remains unclear whether high-mass stars form via ordered collapse of pre-existing cores or in a more dynamic environment.

In an effort to determine whether the rotational signatures that we have observed are more in agreement with core rotation, conservation of angular momentum, or differential rotation of a disk-like object, we calculate the specific angular momenta as a function of radius following the prescription in \citetads{2013EAS....62...25B}. The rotational kinetic energy of an object in solid body rotation with mass $M$ and radius $r$ is $E_\mathrm{rot}=\frac{1}{2}I\Omega^2$, where $\Omega$ is the angular velocity and $I=\frac{2}{3}M\,r^2\left(\frac{3-\alpha}{5-\alpha}\right)$ is the moment of inertia \citepads{2013EAS....62...25B}. The value of $\alpha$ is determined based on the power law density profile of the object such that $\rho \propto r^{-\alpha}$. \citetads{2021A&A...648A..66G} fitted the continuum visibility profiles of the sources in the CORE sample and found a mean density power law index of $2\pm0.2$. Therefore, setting $\alpha=2$ and knowing the angular momentum $J=I\Omega$, the specific angular momentum is
\begin{equation}\label{eq: j_definition}
  j=\frac{J}{M}=\frac{2}{9}\Omega r^2.
\end{equation} 
The angular velocity is calculated by subtracting the LSR velocity, defined to be the velocity at the position of the continuum peak (see Table~\ref{t: CORE_disk_details}), from the centroid velocity $v_\mathrm{cen}$, at each position and dividing by the radius at that position with a sin$(i)$ correction factor included for the unknown inclination $i$:
\begin{equation}\label{eq: angular_velo}
  \Omega=\frac{|v_\mathrm{cen}-v_\mathrm{LSR}|}{r\,\mathrm{sin}(i)}.
\end{equation}
Putting together Eq.~\ref{eq: j_definition} and Eq.~\ref{eq: angular_velo}, we can calculate the specific angular momentum as a function of radius times the unknown $\mathrm{sin}(i)$ correction for each of the sources in the sample, such that 
\begin{equation}\label{eq: j_obs}
  j_\mathrm{obs} (r) \,\mathrm{sin}(i) = \frac{2}{9}\,|v_\mathrm{cen} - v_\mathrm{LSR}|\,r.
\end{equation}

The observed specific angular momentum for the CORE sample were calculated using the decomposed velocity maps of \mck{3} for each source (see Fig. ~\ref{af: velo_decomp}) along the cut with the strongest velocity gradient (dotted lines) for the redshifted and blueshifted sides separately. The resulting radial profiles are presented in Fig.~\ref{f: CORE_j_jmax} as circular data points with their errors, propagated from the error in velocity from the Gaussian decomposition, shown as regions with filled colours. The upward and downward pointing triangles correspond to the same procedure but performed for a cut with a position angle that is respectively 10\degr\ larger and smaller than the position angle of the cut with the strongest velocity gradient. The distance between two data points corresponds to half-beam spacing to ensure the data points are independent of each other, and the specific angular momenta plotted are averaged over a region of $3\times3$ pixels. Only the sources for which we had at least 4 data points (\ie\ more than 2 beams across) for either the redshifted or blueshifted sides are shown. Therefore, this criterion excludes the core IRAS23385.

\begin{figure*}
    \centering
    \includegraphics[width=0.495\hsize]{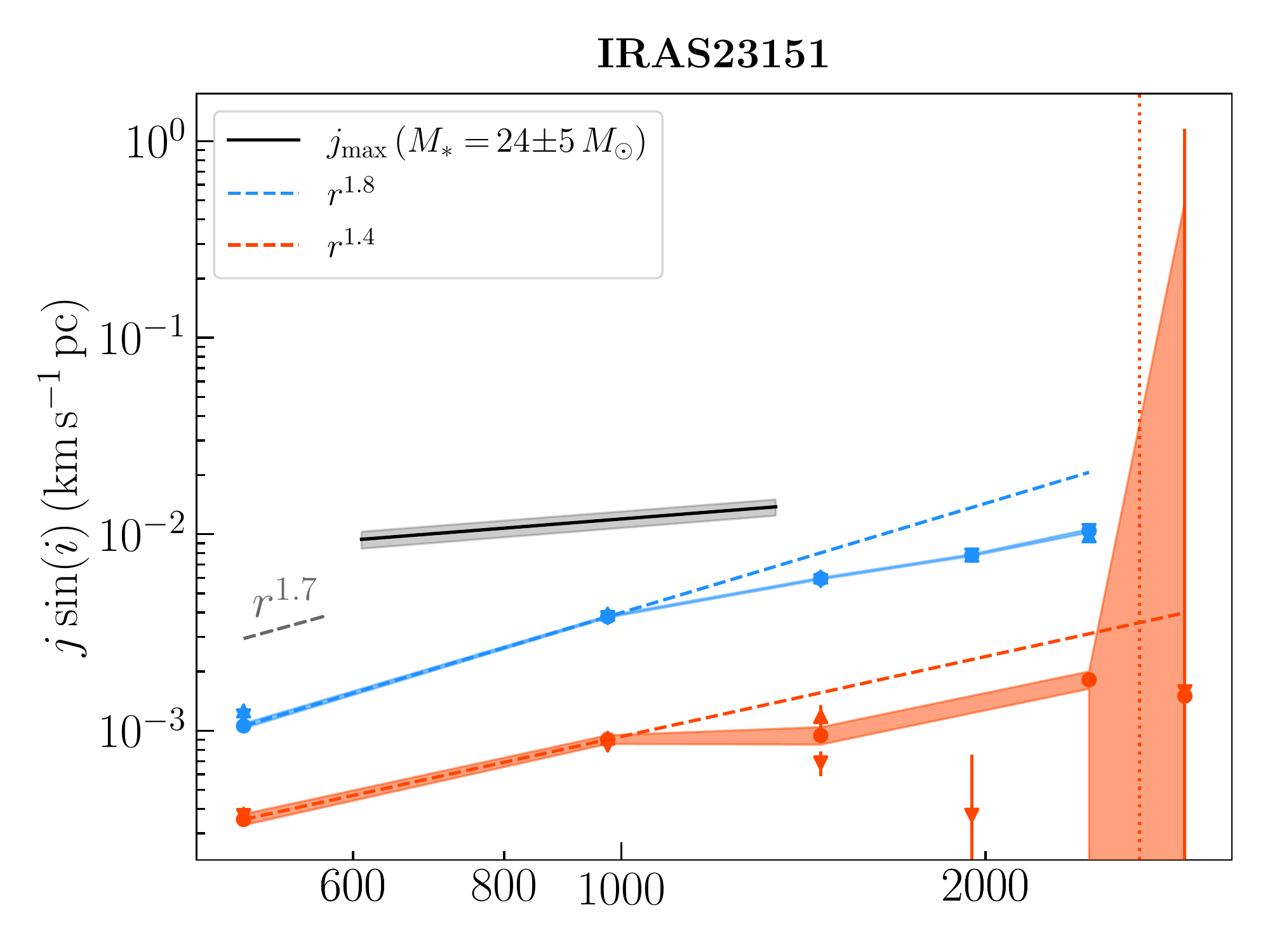} 
    \includegraphics[width=0.495\hsize]{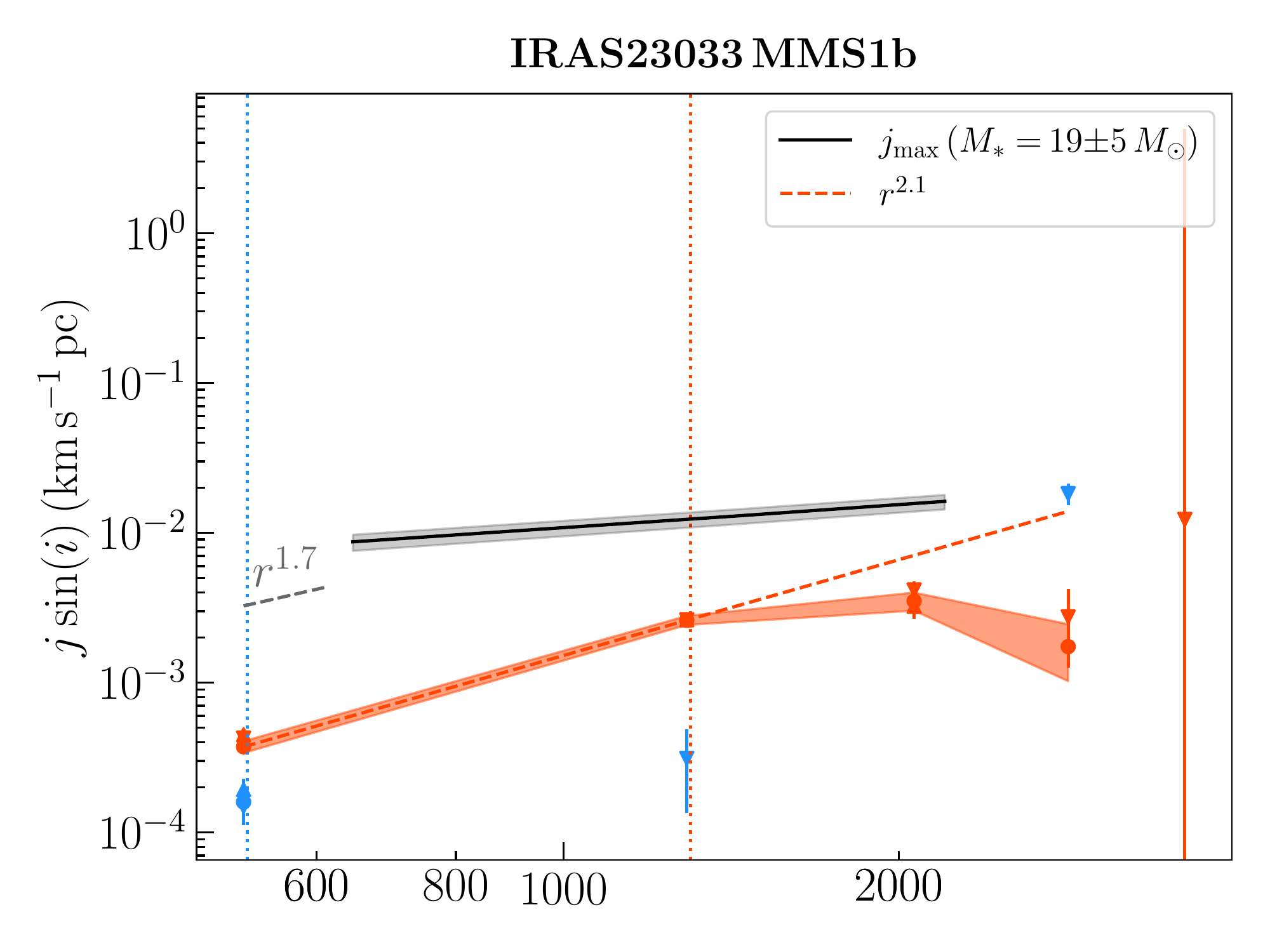} \\
    \includegraphics[width=0.495\hsize]{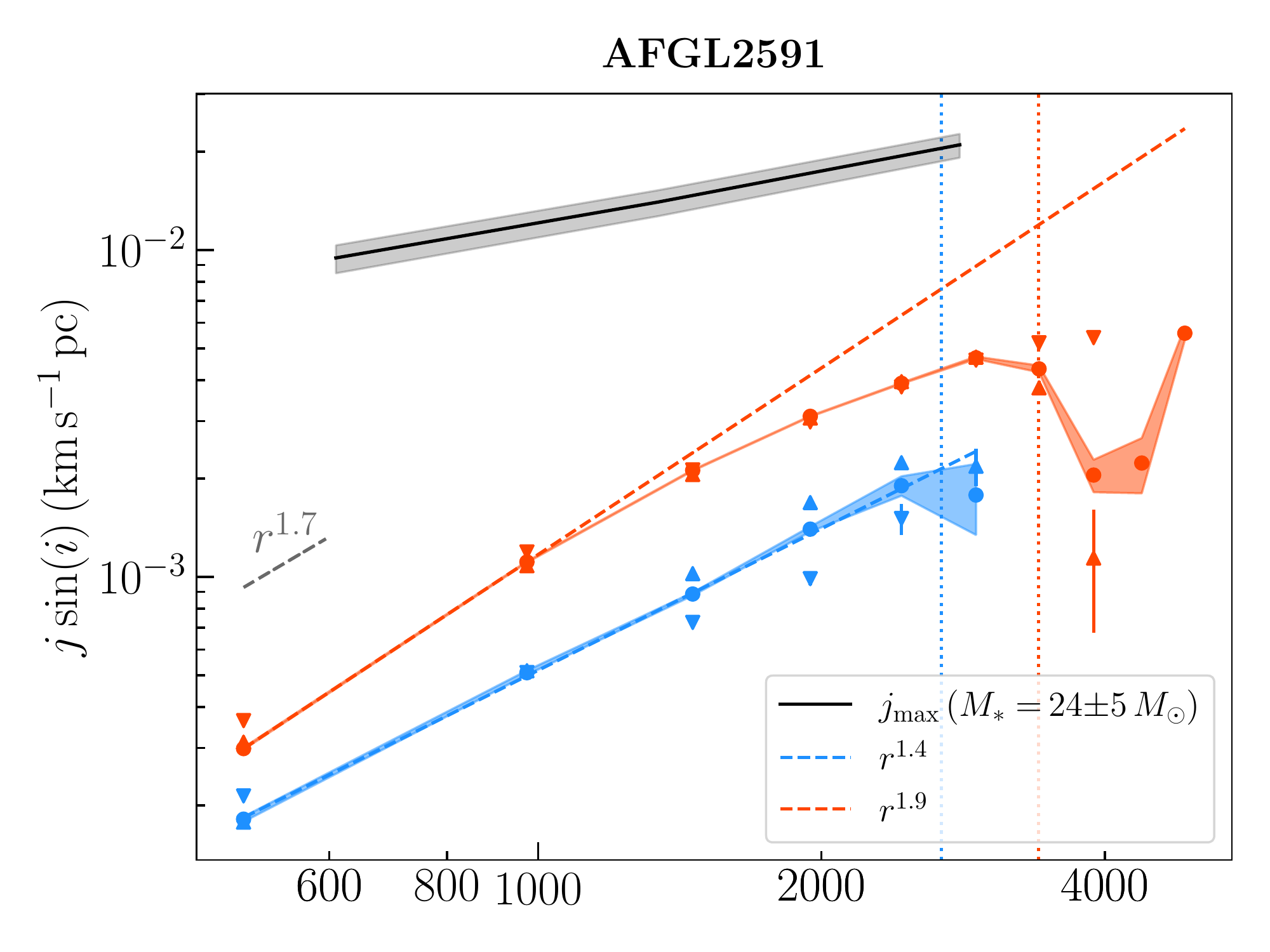} 
    \includegraphics[width=0.495\hsize]{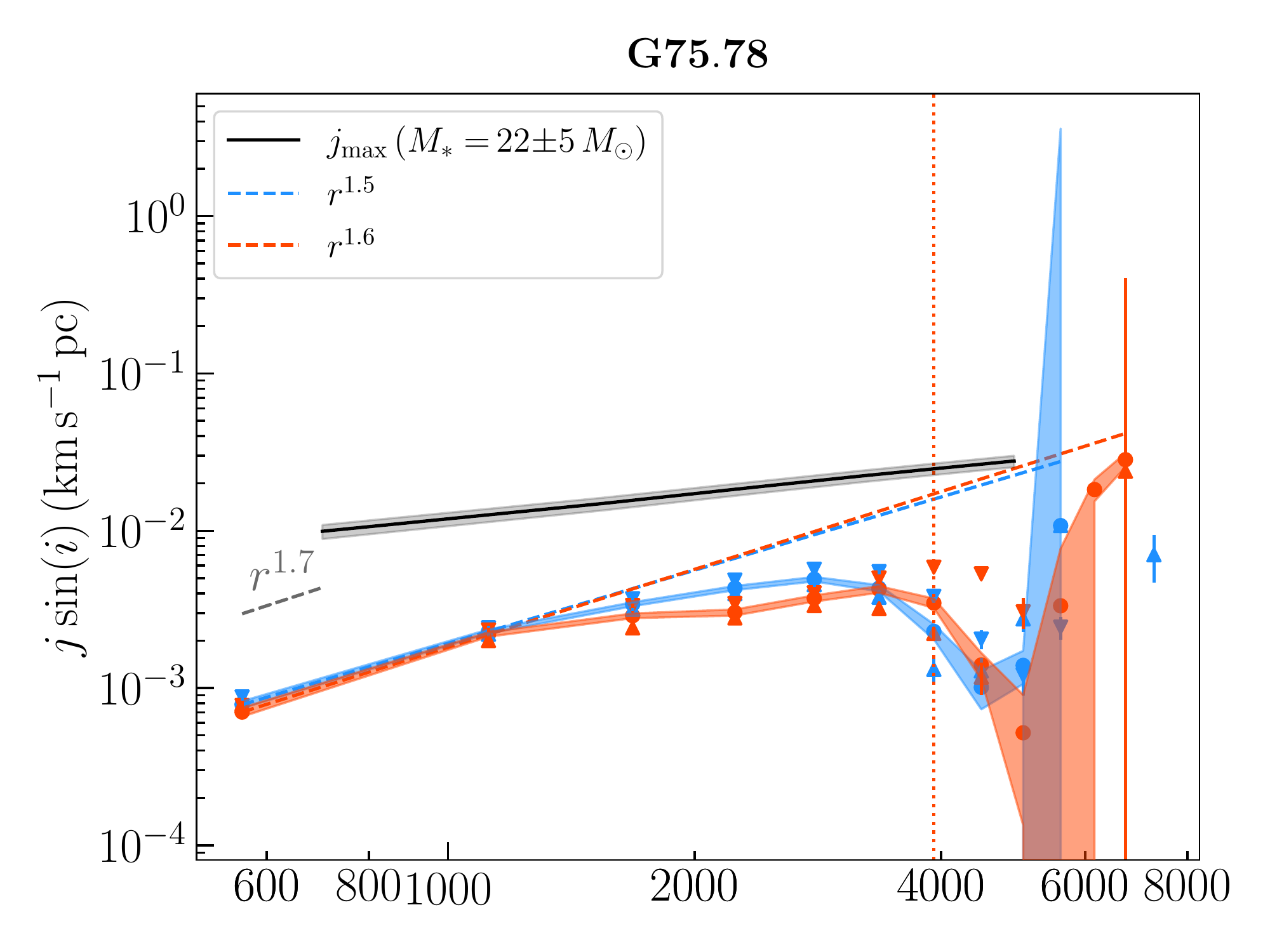} \\
    \includegraphics[width=0.495\hsize]{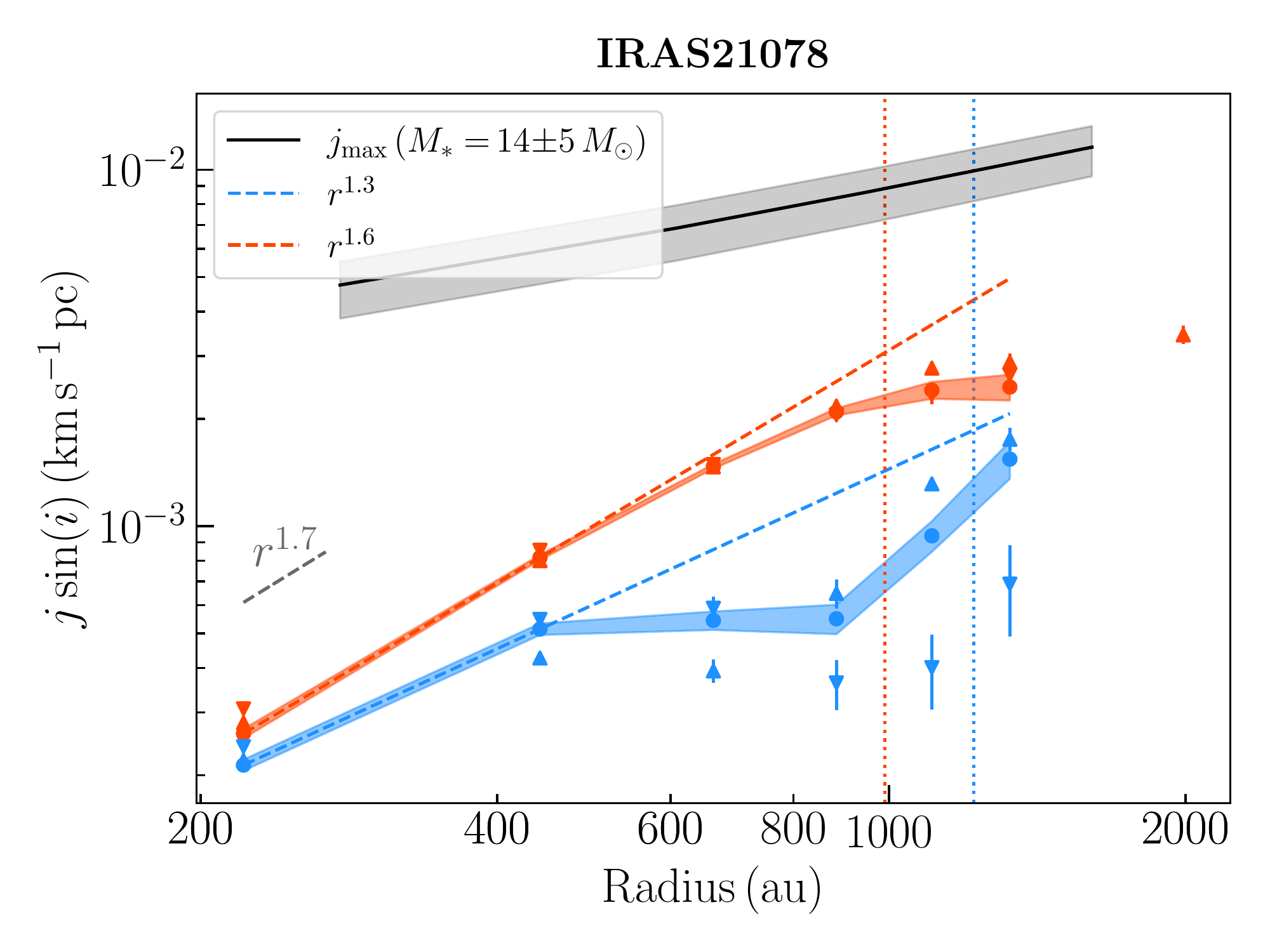}
    \includegraphics[width=0.495\hsize]{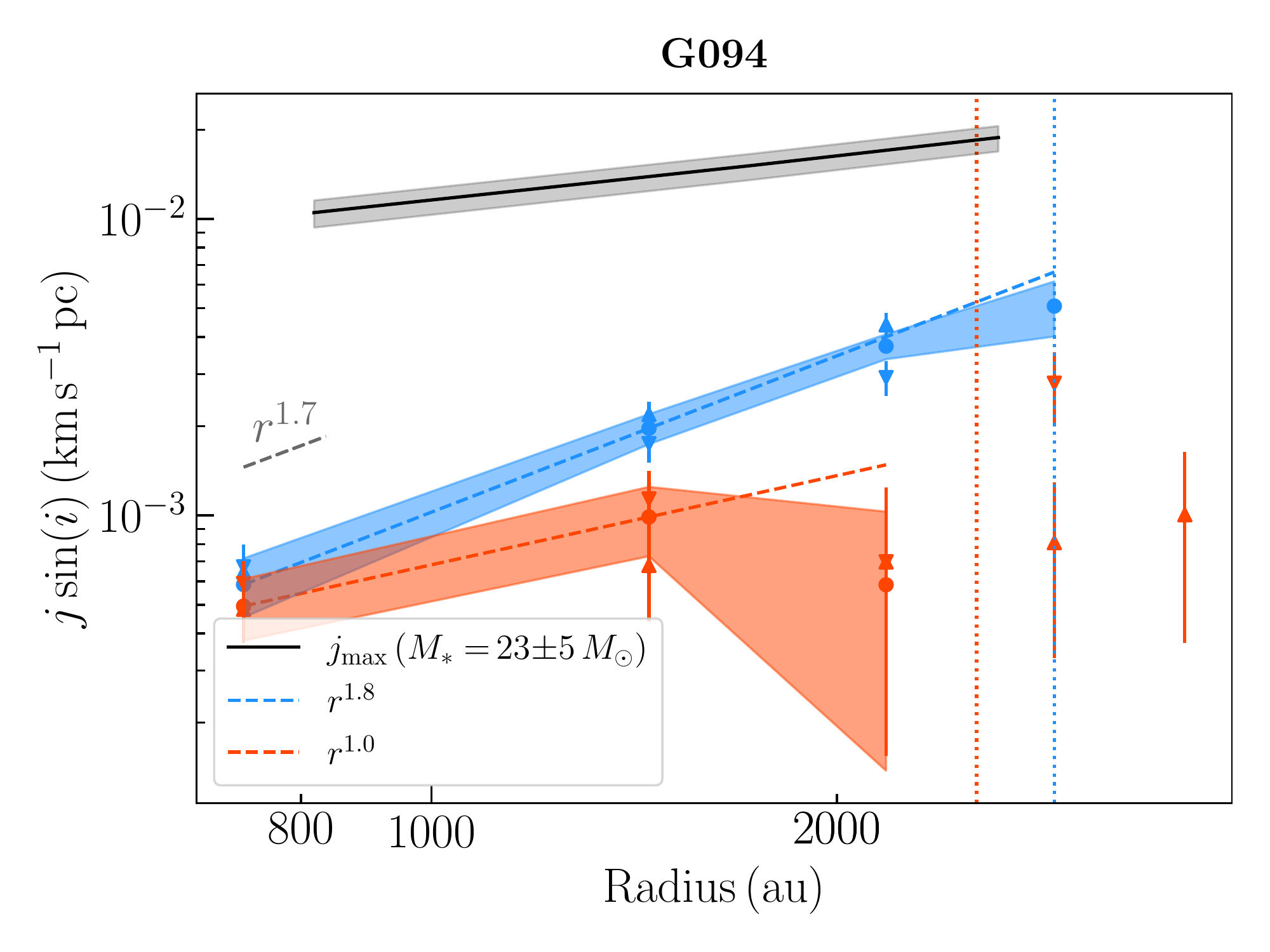}    
    \caption{Specific angular momentum radial profiles calculated using Eq.~\ref{eq: j_obs} along the cut with the strongest velocity gradient (dotted lines in Fig.~\ref{f: pvpanels}) \emph{(circles)}, a cut with position angle $+10$\degr\ \emph{(triangles pointing up)}, and a cut with position angle $-10$\degr\ \emph{(triangles pointing down)} with respect to the strongest velocity gradient, for the redshifted \emph{(red)} and blueshifted \emph{(blue)} sides of the emission. The blue and red filled regions show the error in $j~\mathrm{sin}(i)$ for the circles. Note that the inclinations are not corrected for as they are not known. The blue and red dashed lines show least-squared power-law fits to the first two data points and extrapolated to larger radii. The values of the fitted slopes are noted in the legends and listed in Table~\ref{t: CORE_j_slopes} along with the intercepts. The blue and red dotted vertical lines correspond to radii associated with the edges of the disks depicted with  $\times$ markers in Fig.~\ref{f: mom_velo_ch3cn_k3}. The black curve shows the maximum specific angular momentum calculated using Eq.~\ref{eq: jmax} for $M_\ast$ listed in the sub-panels with the grey region corresponding to $M_\ast\pm5$~\mo. The grey dashed line shows the $j\,\mathrm{sin}(i)\propto r^{1.7}$ slope.  \vspace{1cm}}
    \ContinuedFloat
    \label{f: CORE_j_jmax}
\end{figure*}
\begin{figure*}
   \centering
   \includegraphics[width=0.495\hsize]{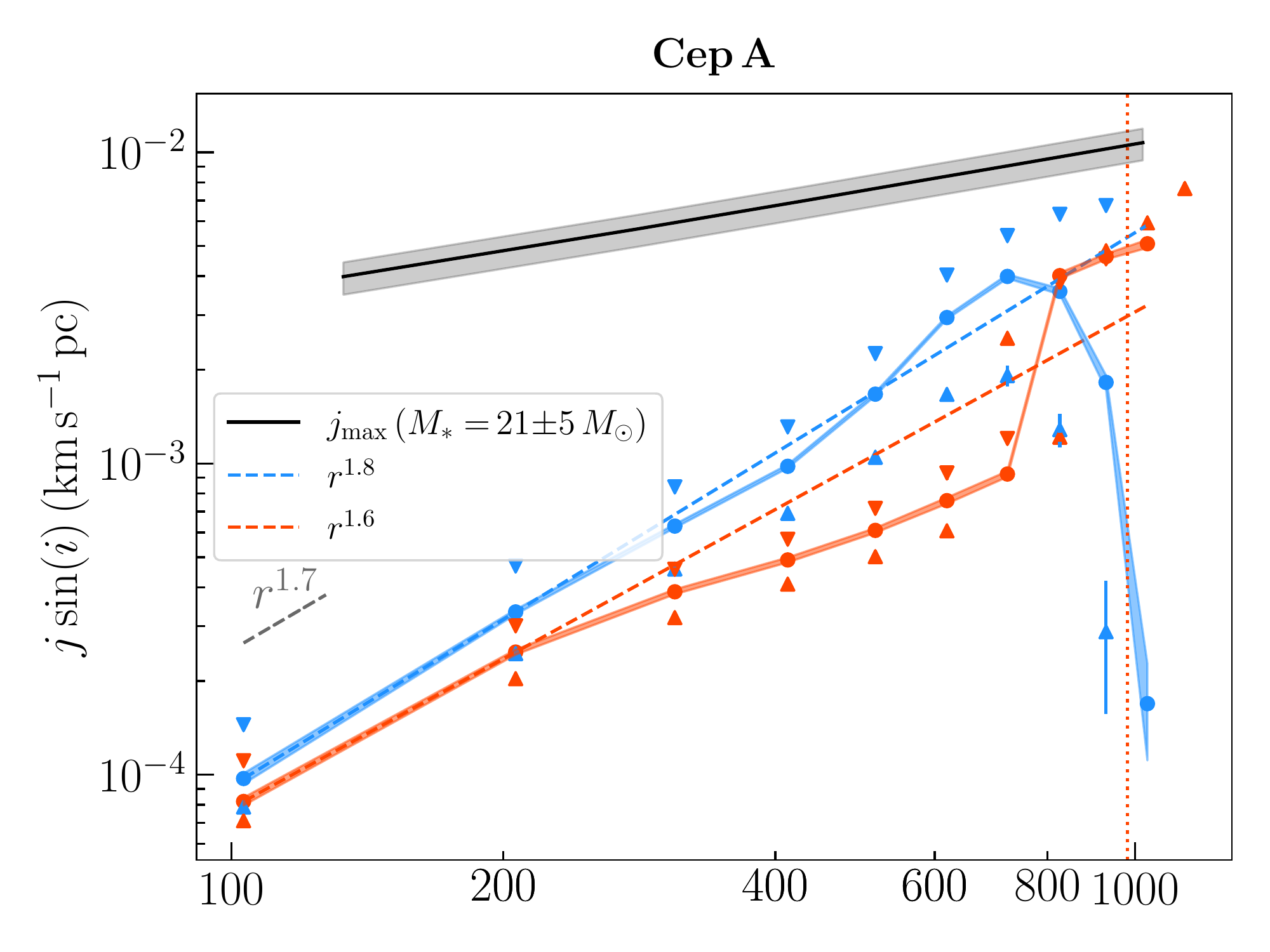} 
   \includegraphics[width=0.495\hsize]{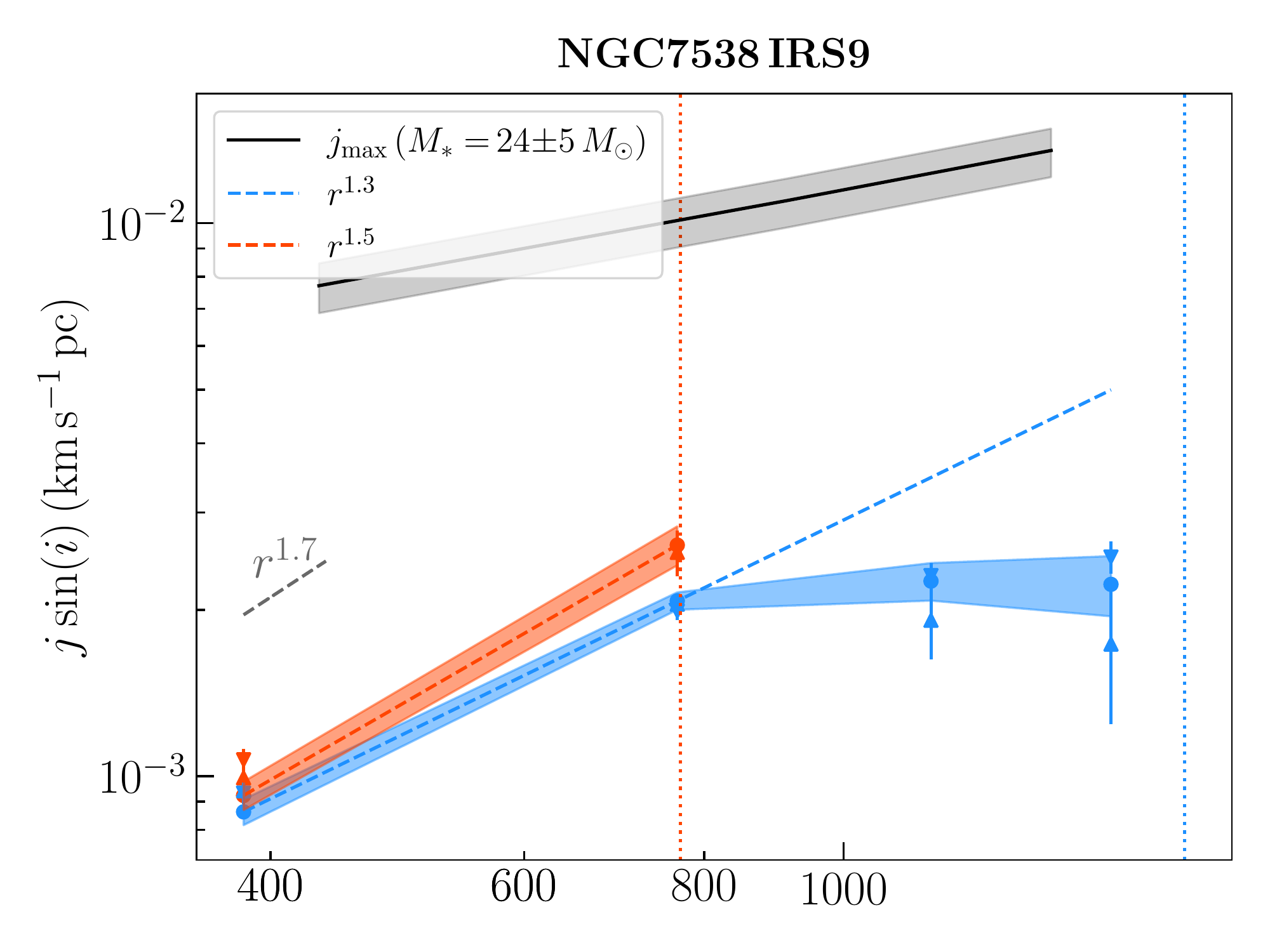} \\
   \includegraphics[width=0.495\hsize]{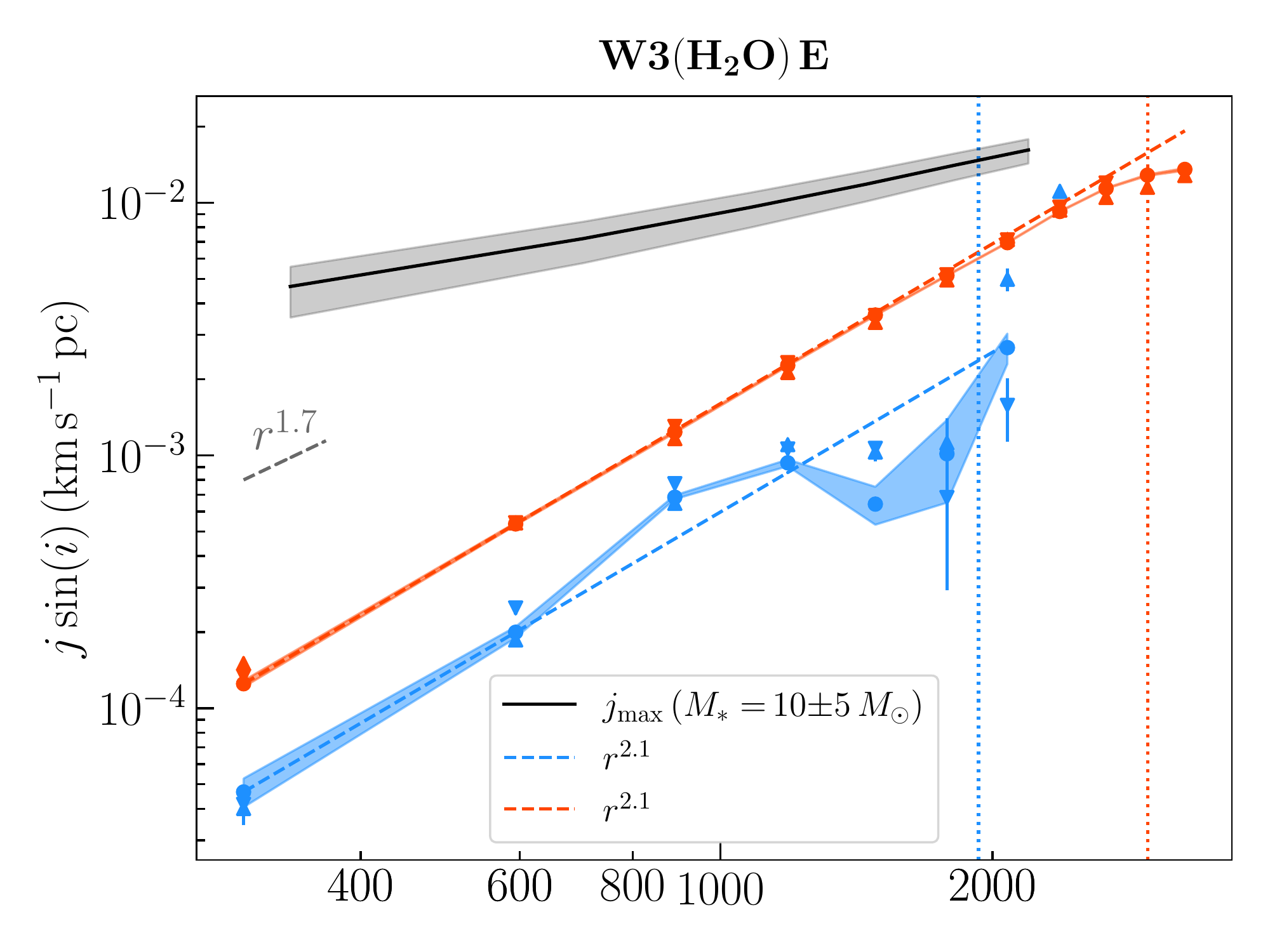} 
   \includegraphics[width=0.495\hsize]{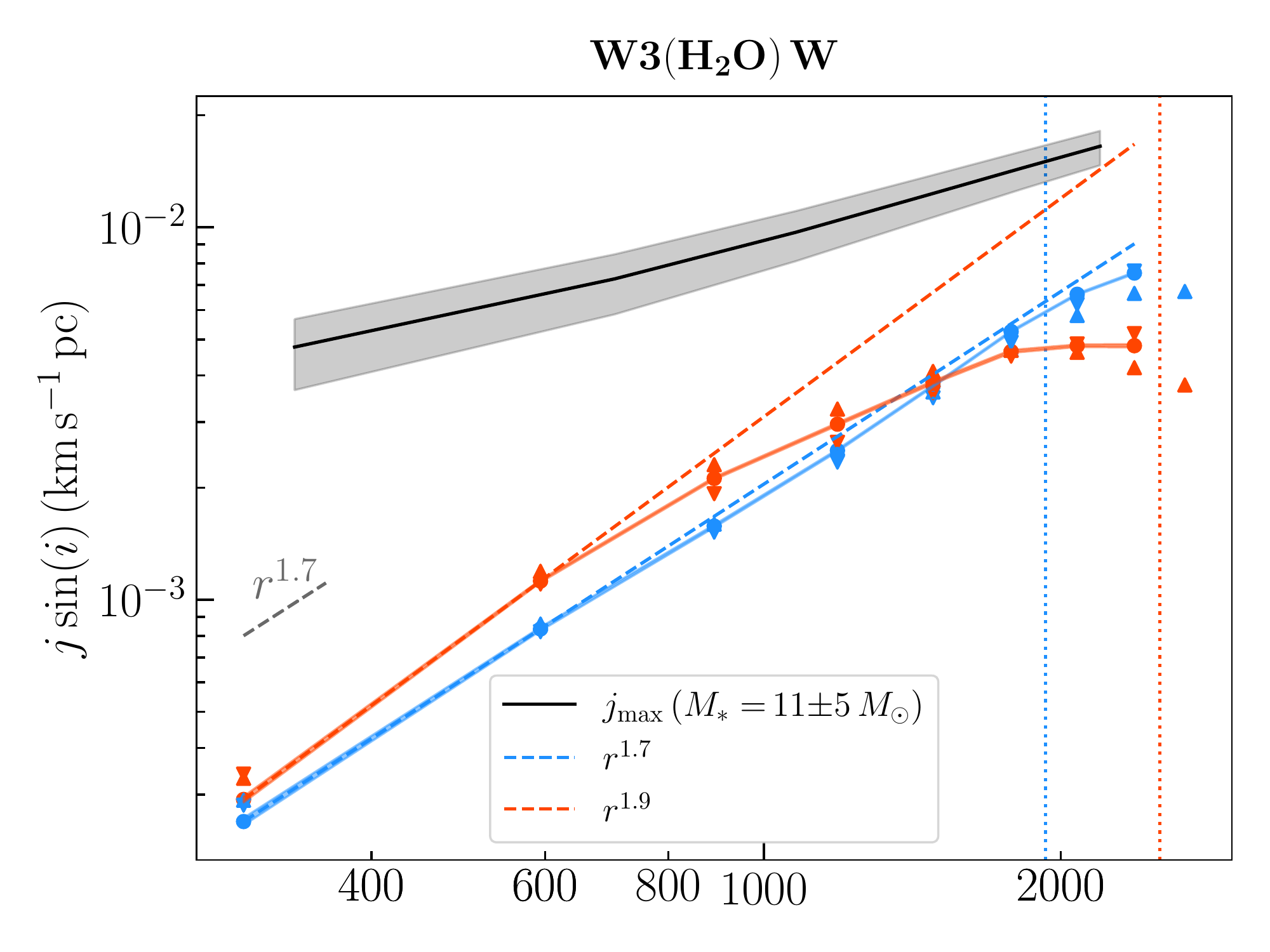} \\
   \includegraphics[width=0.495\hsize]{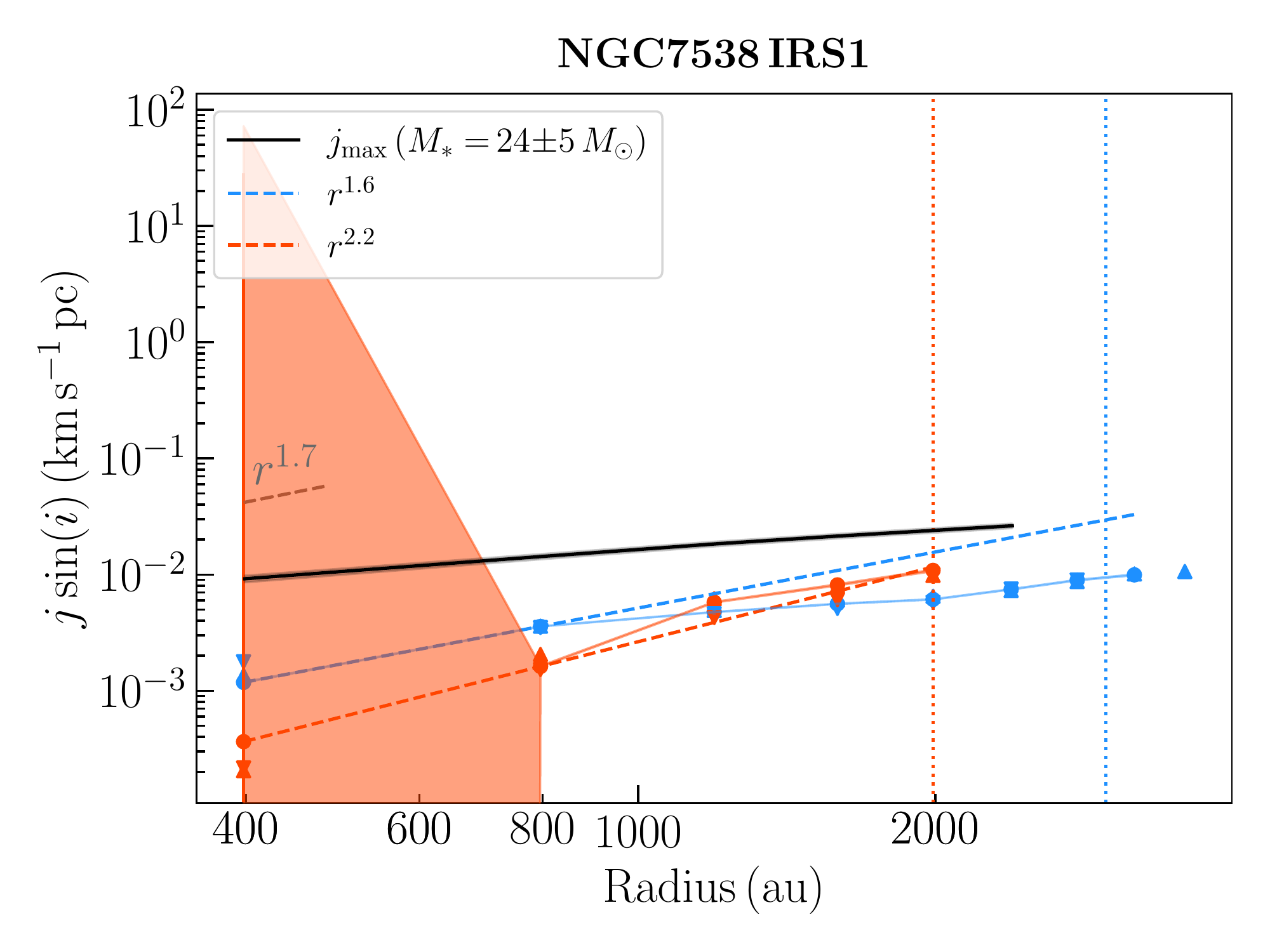} 
   \includegraphics[width=0.495\hsize]{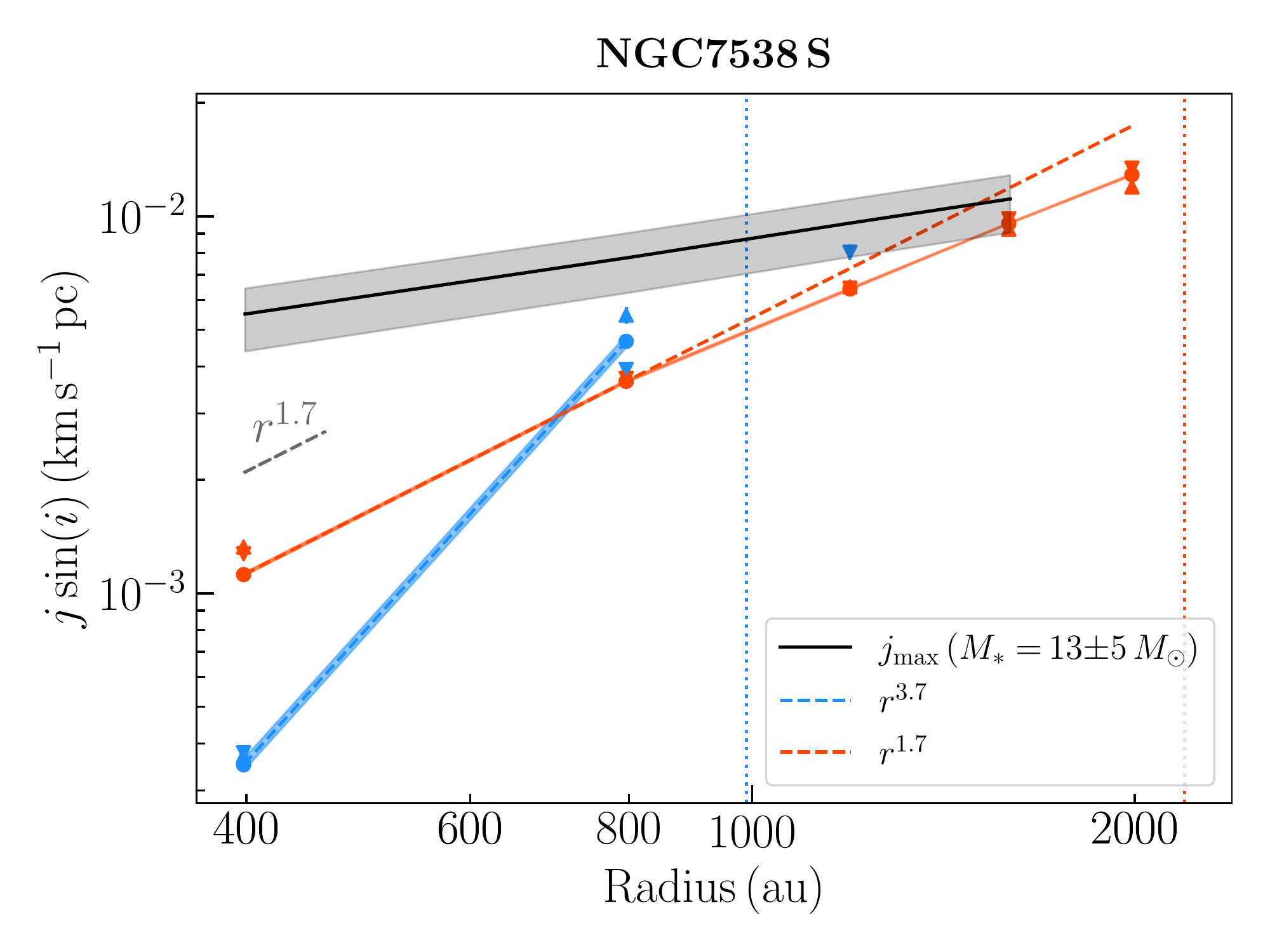}
   \caption[]{Continued.}
\end{figure*}

We can compare the angular momentum of each source with some estimate for the maximum value above which the core becomes gravitationally unbound. Because neither the rotation nor the density profile is expected to be uniform, we follow the approach of \citetads{2018MNRAS.480.5495Z}, assuming the structure is flattened with its mass growing linearly with radius (see Mestel disk: \citeads{1963MNRAS.126..553M}). The maximum specific angular momentum can be written as
\begin{equation}\label{eq: jmax}
  j_\mathrm{max}(r) = \frac{1}{2}\sqrt{G\,M(r)\,r},
\end{equation}
where $M(r)$ is the mass interior to $r$ and is equal to $M_\ast+M_\mathrm{gas}(r)$. For the protostellar masses, we used the estimates from fitting the PV diagrams of dense gas tracers (see Sect.~\ref{ss: sample_pv_masses}) unless other detailed studies of the sources using our data provided better estimates (see Table~\ref{t: masses_Q_CORE}). We used the gas mass maps created using Eq.~\ref{e: sample_mass} and the temperature maps obtained from our radiative transfer modelling with \textit{XCLASS} to calculate $M(r)$. The maximum specific angular momenta are over-plotted on Fig.~\ref{f: CORE_j_jmax} with black lines, and the grey shaded areas surrounding the black curves correspond to the estimate for $j_\mathrm{max}$ with $M_\ast\pm$5~\mo. The maximum radius for which we calculate $j$ shown in blue and red data points corresponds to the extent of the blueshifted and redshifted gas emission, while the extent of the black curves corresponding to $j_\mathrm{max}$ mark the edge of the continuum emission at 6$\sigma$. Comparing this curve with the observed $j$ radial profiles, we see that almost all cores are gravitationally bound as they lie below $j_\mathrm{max}$ with the exception of NGC7538\,S which has an elongated shape in both the continuum and \mc\ towards the redshifted side -- a sign that further fragmentation has taken place.

From the specific angular momentum radial profiles (see Fig.~\ref{f: CORE_j_jmax}) we see that all sources span a range in $j$ of 1--2 orders of magnitude, on average around $\sim10^{-3}$~\kms\,pc, and exhibit a shallowing of their power-law slopes with increasing radius. The dashed lines correspond to least-squared power-law fits to the first two datapoints and extrapolated to larger radii to clearly show this behaviour. The power-law slopes and intercepts corresponding to these dashed lines are listed in Table~\ref{t: CORE_j_slopes}, with a median slope of $j \propto r^{1.7}$. For some sources (\eg~IRAS23151, IRAS23033\,MMS1b, G094, NGC7538\,S), the flattening of $j$ on larger scales is more subtle than other sources (\eg~G75.78 and Cep~A). For AFGL2591 and G75.78 the decrease in $j$ is followed by a sharp increase as a result of the complex velocity structures in the outskirt of these objects (see Fig.~\ref{f: mom_velo_ch3cn_k3}), likely as a result of material flowing from larger scales. The redshifted side of \W~E and the blueshifted side of \W~W are contaminated by the emission from each other and have a similar shape, while the blueshifted side of \W~E and the redshifted side of \W~W which actually correspond to the edges of these rotating structures show the decrease in $j$ more clearly. 

In order to better understand the observational findings, we analyse the expected specific angular momentum profiles for the numerical simulations presented in \citetads{2020A&A...644A..41O} as well as synthetic observations of the system at varying inclinations and angular resolutions to study their effects on $j$ (see Appendix~\ref{a: j_sims}). We have purposefully not corrected for the inclinations of the simulated observations in order to compare the findings with our CORE observations with unknown inclinations. Similar to the observational findings for low-mass star formation summarised in the first paragraph of this section, we find three distinct regimes in the $j$ radial profile of the simulations (see Fig.~\ref{f: sims_j}). The inner disk region ($r<100$~au) actively accreting material onto the central protostar with $j\propto r^{0.6}$ close to Keplerian rotation ($j\propto r^{0.5}$), an intermediate region where $j$ is quite flat but not exactly constant, and the region beyond $\sim$500~au is rather consistent with the $j \propto r^{1.6}$ relation that had been found for the rotation of low-mass dense cores and Class 0 envelopes (\eg~\citeads{2020A&A...637A..92G}). As resolution worsens, the substructure in the $j$ radial profile becomes smoother with the power-law relations for the different regimes becoming more continuous and not as clear to disentangle. At the scale of synthetic NOEMA observations at 2~kpc, the envelope and disk components become completely blended, with only one power-law relation seen across all scales, with $j \propto r^{1.6}$. This is consistent with our findings for the CORE sample with a median $j \propto r^{1.7}$ (see Fig.~\ref{f: CORE_j_jmax}). It is also evident that at the spatial resolution of our observations, the lack of information on the inclination is not an important factor and does not impact our findings.

There exists a small shallowing of the slope (a kink) in the $j$ profiles of poorly resolved simulated observations and the CORE observations. If the decrease in $j$ is in fact due to the transition between a Keplerian disk and the collapse of the envelope for which the angular momentum is constant (\ie~$j$ is constant), one could put a limit on the size of the inner disk that is actively accreting onto the protostar. We refrain from calculating exact numbers for each source due to the uncertainties in such an estimate and our limited resolution, but note that for most of the sources this transition is seen around $1000-2000$~au. The only exceptions are Cep~A and IRAS21078 that are the closest sources in the sample and hence resolved the best, for which this transition is seen around 300--400~au, and 500--700~au, respectively. 

\begin{table}
\str{1.2}
\centering
\caption{Fit parameters to the specific angular momentum radial profiles shown in Fig.~\ref{f: CORE_j_jmax} (dashed lines).}
\label{t: CORE_j_slopes}
\begin{tabular}{lcccc}
\hline \hline
  Source & $a_\mathrm{blue}$ & $a_\mathrm{red}$ & $b_\mathrm{blue}$ & $b_\mathrm{red}$ \\
\hline
  IRAS23151                   & 1.8 & 1.4 & -2.4 & -3.0 \\
  IRAS23033\,MMS\,1b  &  (...) & 2.1 & (...) & -2.8 \\ 
  AFGL2591                   & 1.4 & 1.9 & -3.3 & -2.9 \\
  G75.78                        &  1.5 & 1.6 & -2.7 & -2.7 \\
  IRAS21078                  & 1.3 & 1.6 & -2.8 & -2.5 \\ 
  G094                           & 1.8 & 1.0   & -3.0 & -3.2 \\
  Cep\,A                         & 1.8 & 1.6 & -2.3 & -2.5 \\
  NGC7538\,IRS9          &  1.3 & 1.5 & -2.5 & -2.4 \\
  W3(H$_2$O)\,E           &  2.1 & 2.1 & -3.2 & -2.8 \\
  W3(H$_2$O)\,W          &  1.7 & 1.9 & -2.7 & -2.5 \\
\hline
  \underline{Pilot study}\\
  NGC7538\,IRS1   & 1.6 & 2.2 & -2.3 & -2.6 \\
  NGC7538\,S         & 3.7 & 1.7 &  -2.0 & -2.3 \\ 
\hline
\end{tabular}
\tablefoot{The heading parameters $a$ and $b$ satisfy the following relation: $j\,\mathrm{sin}(i)=10^b\,(\frac{R}{1000~\mathrm{au}})^a~\mathrm{km\,s^{-1}\,pc}$, for the blueshifted and redshifted sides.}
\end{table}


\subsection{Toomre stability}\label{s: sample_toomre}

We determine the stability of the disk candidates against axisymmetric gravitational collapse using the \tq\ parameter \citepads{1964ApJ...139.1217T},
\begin{equation}
  Q=\frac{c_s\,\Omega}{\pi\,G\,\Sigma}.
\end{equation}
The Toomre equation determines the balance between the stabilising force of thermal pressure (through the sound speed, $c_s$) and shear forces (through the angular velocity, $\Omega$) against the local gravity of the disk (through the disk surface density, $\Sigma$). For a disk that is sufficiently massive, the self-gravity of the disk could cause it to fragment. The Toomre~Q parameter quantifies this scenario such that if $Q<1$, the disk is prone to fragmentation. While this critical threshold has been derived for axisymmetric instabilities in an infinitely thin disk, non-axisymmetric instabilities can form in thicker disks at higher values of $Q$ (discussed further below and in Sect.~\ref{ss: disk_thickness}).

\begin{table*}
\str{1.2}
\centering
\caption{Overview of mass estimates and \tq\ results.}
\label{t: masses_Q_CORE}
\begin{tabular}{lcccccccc}
\hline\hline
  Source & $M_\mathrm{gas}$ & $M_\mathrm{PV}$\tablefootmark{a} & $M_\mathrm{lum}$\tablefootmark{b} & $M_\ast$\tablefootmark{c}  & $Q_\mathrm{min}$ & $Q_\mathrm{median}$ & $\left(\frac{H}{r}\right)\tablefootmark{d}_\mathrm{median}$ & Toomre stable \\
  & ($M_\sun$) & ($M_\sun$) & ($M_\sun$) & ($M_\sun$) & & &  \\
  \hline 
  IRAS23151                   & 3.6 & $23\pm2$ & 24 &  24 & 1.2 & 4.4 & 0.16 & no \\
  IRAS23033\,MMS\,1b   & 3.9 & $14\pm1$ &  21 & 19\tablefootmark{$\ast$} &  0.5 &  2.0 & 0.15 & no \\ 
  AFGL2591                    & 6.2 & $23\pm1$ & 40 &  24 &  1.2 & 2.7 & 0.21 & no \\
  G75.78                         & 8.1 & $19\pm3$ &  34 & 22 &  0.3 & 2.1  & 0.23 & no \\
  IRAS21078                   & 3.3 & $13\pm1$ & 19 & 14 &  0.4 & 3.3  & 0.23 & no \\ 
  G094                            & 3.0 & $20\pm3$ & 25 & 23 & 0.8 & 6.4 & 0.23 & no \\
  Cep\,A                         & 0.6 & $24\pm1$ & 21 & 21 & 5.1 & 11.9 & 0.17 & yes \\
  NGC7538\,IRS9           & 1.2 & $25\pm3$ & 24 & 24 & 4.4 & 15.0 & 0.24 & yes \\
  W3(H$_2$O)\,E            & 9.4 & $9\pm1$ & 33 &  10\tablefootmark{$\dagger$} & 0.1 &  0.9 & 0.23 & no \\
  W3(H$_2$O)\,W           & 7.0 & $10\pm1$ & 33 &  11\tablefootmark{$\dagger$} & 0.1 & 1.0 & 0.23 & no \\
  IRAS23385                  & 19.1 & $17\pm2$ & 21 & 9\tablefootmark{$\ddagger$} & 0.1 & 0.9 & 0.25 & no\\
\hline
  \underline{Pilot study}\\
  NGC7538\,IRS1           & 31.7 & $22\pm2$ & 41 &  24 &  0.3 & 0.7 & 0.16 & no \\
  NGC7538\,S                & 1.8 & $12\pm1$ & 21 &  13 &  0.9 & 4.2 & 0.26 & no \\ 
\hline
\end{tabular}
\tablefoot{
\tablefoottext{a}{Dynamical mass estimates from fitting Keplerian curves to the $6\sigma$ edges of emission in the PV diagrams shown in Fig.~\ref{f: pvpanels} (See Sect.~\ref{ss: sample_pv_masses}.)}
\tablefoottext{b}{Luminosity mass extrapolated from Table~1 of \citetads{2005IAUS..227..389C} using luminosities reported in Table~\ref{t: sample_info}.}
\tablefoottext{c}{Protostellar mass used in the \tq~analysis.}
\tablefoottext{d}{Median ratio of disk scale height to radius.}
\tablefoottext{$\ast$}{Based on \citetads{2019A&A...629A..10B} using data from the CORE survey.}
\tablefoottext{$\dagger$}{Based on A-array observations presented in \citetads{2018A&A...618A..46A} using data from the CORE survey.}
\tablefoottext{$\ddagger$}{Based on \citetads{2019A&A...627A..68C} using data from the CORE survey.}}
\end{table*}

The sound speed is calculated using the temperature maps obtained from our radiative transfer modelling of \mc\ lines with \textit{XCLASS}, presented in Sect.~\ref{s: sample_temp}, such that 
\begin{equation}\label{eq: SAMPLE_cs}
  c_s=\sqrt{\frac{\gamma k_\mathrm{B} T}{\mu m_\mathrm{H}}},
\end{equation}
where $\gamma$ is the adiabatic index with a value of $7/5$, $k_\mathrm{B}$ is the Boltzmann constant, $\mu$ is the mean molecular weight with a value of 2.8 \citepads{2008A&A...487..993K}, and $m_\mathrm{H}$ is the mass of the hydrogen atom. Using the obtained temperature maps and assuming gas and dust temperatures are coupled, the surface density of the disk is calculated according to
\begin{equation}
\Sigma = \frac{S_\nu\,R}{B_\nu(T_D)\,\theta_B\,\kappa_\nu},
\end{equation}
where $S_\nu$ is the continuum intensity and $\theta_B$ is the beam solid angle. Furthermore, assuming the disk is in gravito-centrifugal equilibrium following a Keplerian rotation profile, the angular velocity at a given radius is 
\begin{equation} \label{eq: SAMPLE_omega}
\Omega(r)=\sqrt{\frac{G(M_\mathrm{\ast}+M_\mathrm{disk}(r))}{r^3}},
\end{equation}
where $M_\ast$ is the protostellar mass and $M_\mathrm{disk}(r)$ is the gas mass interior to $r$ calculated from the continuum maps (see Sect.~\ref{s: sample_gas_mass} and Table~\ref{t: masses_Q_CORE}). Protostellar masses were estimated from fitting the PV diagrams of \mc\ transitions and other dense gas tracers as discussed in Sect.~\ref{ss: sample_pv_masses}. We select the upper limit of the dynamical mass estimates for the protostellar masses to calculate upper limits for the \tq\ parameter. When the dynamical mass upper limits are higher than the masses derived from the source luminosities, we select the luminosity derived masses. For sources that were studied in more depth using the same CORE data (IRAS23033: \citeads{2019A&A...629A..10B}; IRAS23385: \citeads{2019A&A...627A..68C}), we adopt the mass estimates reported by these authors. All mass estimates and the chosen protostellar masses are summarised in Table~\ref{t: masses_Q_CORE}.

\begin{figure*}
    \centering
    \includegraphics[width=0.96\hsize]{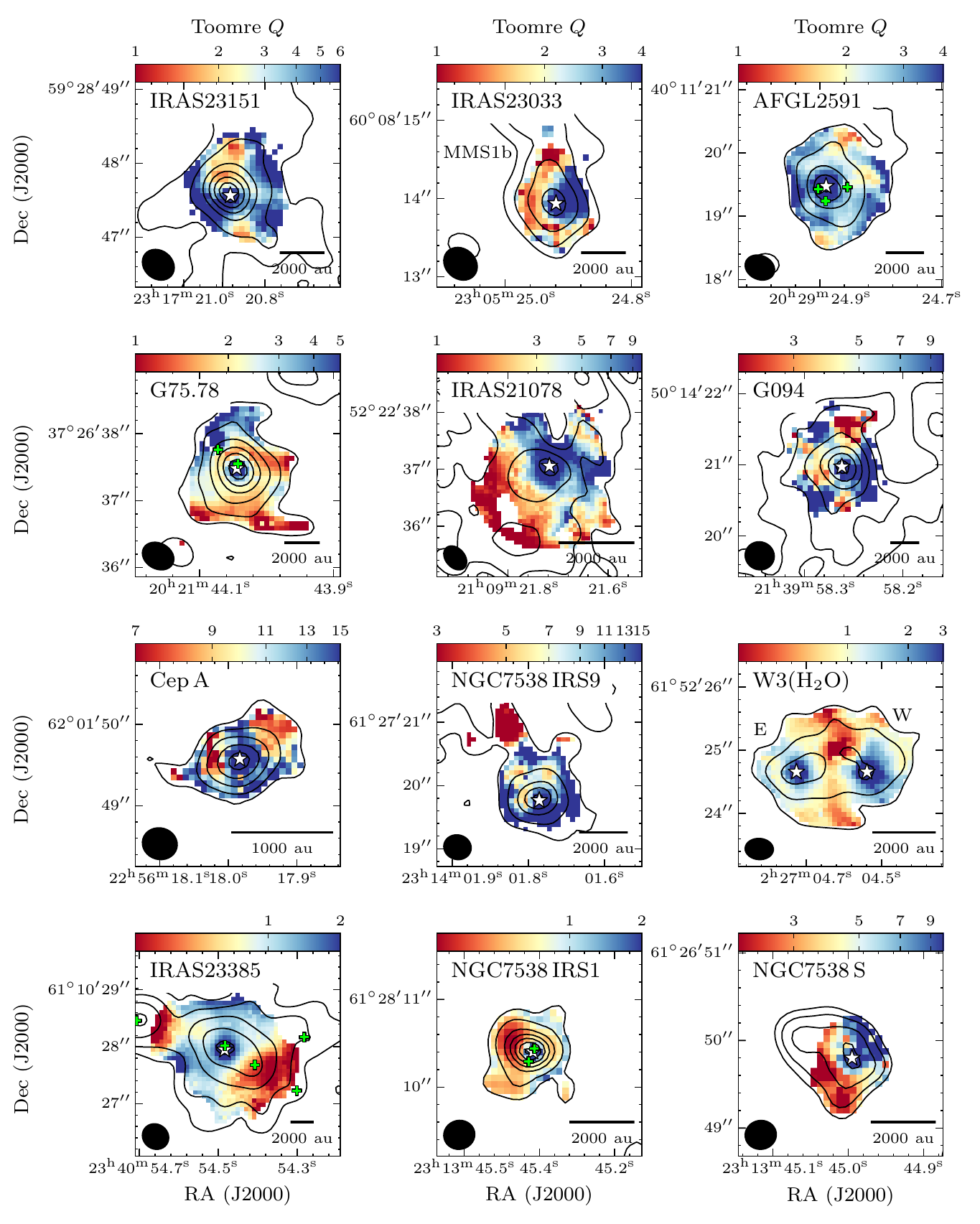}
    \caption{\tq\ maps for the best disk candidates in the CORE survey assuming a protostar is located at the position of the continuum peak as depicted by a star and accounting for the self-gravity of the disk (protostellar and gas mass values are listed in Table~\ref{t: masses_Q_CORE}). The contours correspond to the 1.37~mm continuum as described in Fig.~\ref{f: mom_velo_h2co}. Regions outside of 6$\sigma$ continuum contours are masked out. Fragments detected by ancillary observations are marked by plus symbols in green for AFGL2591 \citepads{2021A&A...655A..84S}, IRAS23385 \citepads{2019A&A...627A..68C}, and NGC7538\,IRS1 \citepads{2017A&A...605A..61B}. The synthesised beam is shown in the bottom left corner and a scale bar in the bottom right corner of each panel.}
    \label{f: toomre_CORE}
\end{figure*}

Assuming the protostar is positioned at the location of the continuum peak, we calculate the \tq\ parameter for each disk candidate, accounting for the self-gravity of the disk by including the enclosed gas mass at each radius. The resulting $Q$ maps are presented in Fig.~\ref{f: toomre_CORE} for the 13 disk candidates in the CORE survey. Regions outside of the 6$\sigma$ continuum contours are masked out since the continuum maps are used as input for the calculation of $Q$. The minimum and median $Q$ values for each source are provided in Table~\ref{t: masses_Q_CORE}.

The critical value of the Toomre parameter, $Q_\mathrm{crit}$, which determines the stability of a disk is in theory equal to 1, while effectively it lies somewhere between 1--2 (\citeads{1989ApJ...344..685K}; \citeads{1991ApJ...378..139E}, \citeyearads{1994ApJ...433...39E}). Considering the uncertainties in our analysis, especially in determining temperatures from radiative transfer modelling of \mc\ $K$-ladder (10--20\% uncertainty, see Appendix B of \citeads{2019A&A...631A.142G}) as well as the fact that the derived temperatures may be probing layers above the disk mid-plane and warmer regions shocked by outflows, we assume that regions where $Q \lesssim 2$ are unstable against gravitational collapse. Furthermore, it is important to keep in mind that this Toomre analysis depends on the spatial resolution and gives average values on size-scales of the beam. Toomre-stable disks could have smaller-scale substructures that may be gravitationally unstable which only higher spatial resolution observations can disentangle. We studied these caveats in depth by creating synthetic NOEMA observations of a numerical simulation of a gravitationally unstable and fragmented disk and confirmed the $Q_\mathrm{crit} \approx 2$ to be a reasonable threshold for our CORE observations presented here \citepads{2019A&A...632A..50A}. 

\begin{figure*}
    \centering
    \includegraphics[width=\hsize]{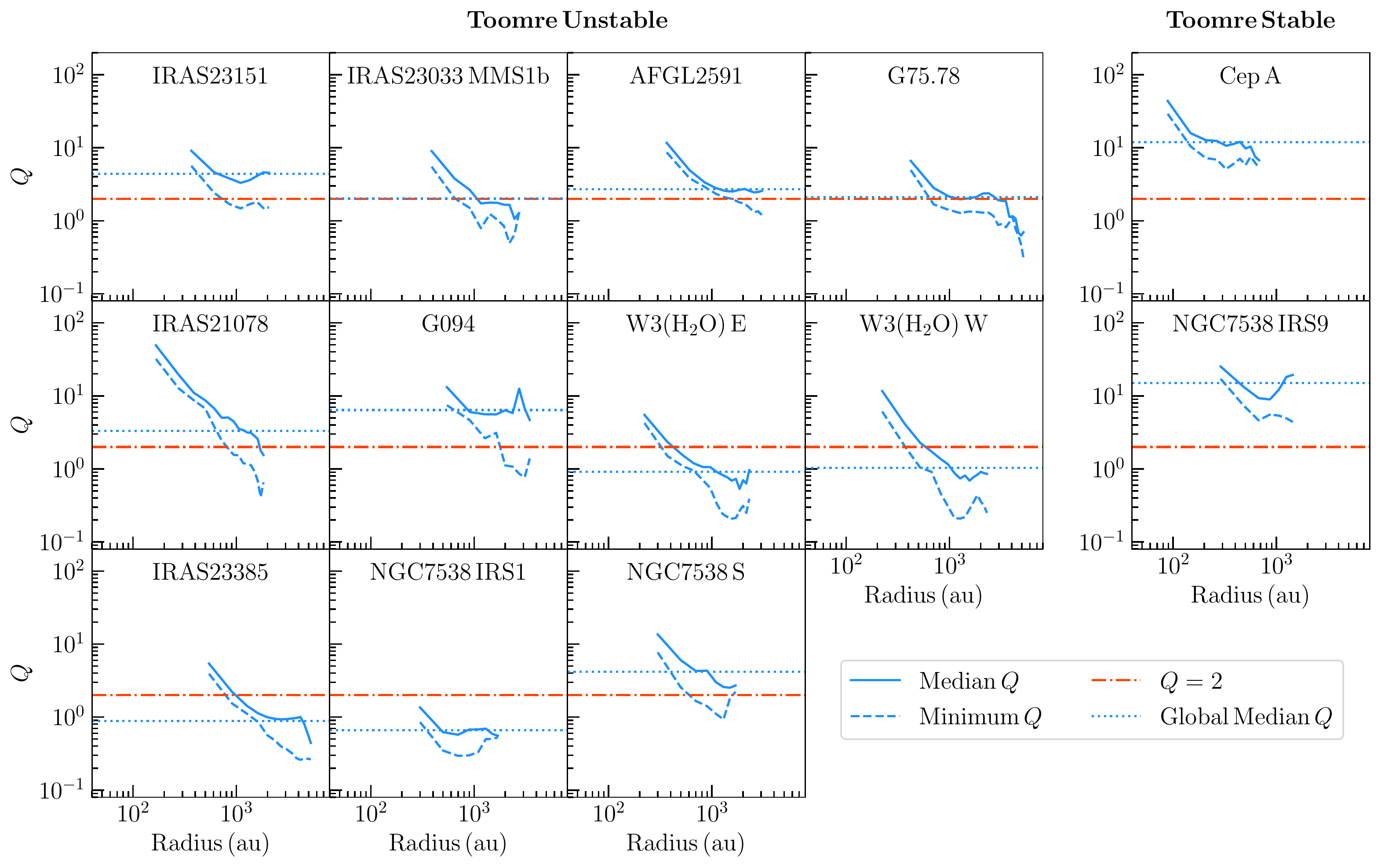}     
    \caption{Distribution of median and minimum $Q$ as a function of radius shown in solid and dashed blue lines. The dotted horizontal line corresponds to the global median Q  computed over the entire disk and listed in Table~\ref{t: masses_Q_CORE}. The dash-dotted red horizontal line shows the critical $Q=2$ threshold.}
    \label{f: Q_vs_radius}
\end{figure*}

Because the Toomre parameter quantifies \textit{local} stability in a disk, the global median values listed in Table~\ref{t: masses_Q_CORE} should be used with caution as a determining factor for whether a disk is stable against gravitational instabilities. In fact, numerical simulation of a massive fragmenting disk has shown that the disk can be globally Toomre-stable while locally Toomre-unstable and undergoing gravitational collapse \citepads{2020A&A...644A..41O}. Hence having observations with sufficient spatial resolution and analysing the spatial distribution of the \tq\ parameter rather than calculating a global value are crucial in studying the stability of such disks. In order to understand the local variations of $Q$, Fig.~\ref{f: Q_vs_radius} shows the distribution of median and minimum $Q$ values as a function of radius. While the global median $Q$ value lies above the critical $Q$ threshold for a number of sources, the minimum $Q$ dips below the threshold in the outskirts of these disk candidates for all sources but two (Cep~A and NGC7538\,IRS9). Therefore, we find 11 out of 13 disks to be Toomre unstable against gravitational collapse.

\subsubsection{Toomre unstable disks}

There exist regions with $Q \lesssim 2$ for most disk candidates (see Fig.~\ref{f: toomre_CORE}), meaning they are unstable against gravitational instabilities and prone to fragmentation. While some candidate disks appear to be Toomre unstable almost entirely (\eg~G75.78, \W~E and W, IRAS23385, NGC7838\,IRS1), others show signs of instability only in their outskirts (\eg~NGC7538\,S). For some sources, the low $Q$ values are coincident with the elongation of the continuum, hinting at the existence of underlying substructure. The best examples for such cases are IRAS23385, and NGC7538\,S. In fact, the in-depth analysis of \citetads{2019A&A...627A..68C} using the CORE survey data confirms that IRAS23385 is forming a cluster of massive stars (marked by green plus symbols in Fig.~\ref{f: toomre_CORE}). Another interesting case worth noting is IRAS21078, for which the low $Q$ values are coincident with a secondary source detected in continuum, $\sim$0.5\arcsec\ to the southeast of the continuum peak. 

Similarly, other sources for which we know further fragmentation has taken place are AFGL2591 and NGC7538\,IRS1, marked by green plus symbols in Fig.~\ref{f: toomre_CORE}. For AFGL2591, NOEMA observations at $843\,\mathrm{\mu m}$ (0.19\arcsec\, $\sim$700~au - roughly a factor of 2 better than the work presented here and providing the highest achievable spatial resolution at (sub-)mm wavelengths in the northern sky) revealed multiple fragments within the inner $\sim$1000~au, supporting the disk-fragmentation scenario \citepads{2021A&A...655A..84S}. For NGC7538\,IRS1, \citetads{2017A&A...605A..61B} observed the fragmentation in centimetre continuum emission (with 0.055\arcsec\ resolution, $\sim$150~au) and found two HC\ion{H}{ii} regions with a projected separation of $\sim$430~au. The authors resolved velocity gradients in thermal $\mathrm{CH_3OH}$ across each HC\ion{H}{ii} region, painting a picture of two disk-like structures embedded within the same rotating circumbinary envelope. Methanol masers at 6.7~GHz have also been detected, consistent with the accretion disk scenario \citepads{2014A&A...566A.150M}. The position angles of these disks are misaligned from that of the rotating structure we detect by about 30\degr. This scenario is similar to the synthetic observations of the numerical simulation we presented in \citetads{2019A&A...632A..50A} whereby the disk on large scales has a different rotation axis than the inner small-scale disks that have formed around each of the fragments. Whether the large-scale rotating structure is a circumbinary envelope rather than a flattened disk-like structure is difficult to establish, especially because our line observations at the location of the protostar(s) are seen in absorption. The large extent of \mc\ emission in the PV diagrams (see Fig.~\ref{f: pvpanels}) at first glance may hint at the possibility that the structure is not a disk, especially because the mass estimates from fitting the PV diagram are much in excess of the mass estimated from the luminosity. However, as we showed in \citetads{2019A&A...632A..50A}, PV diagrams of poorly resolved disks often have very large sizes resulting in mass estimates that are significantly higher than the true enclosed mass. Due to the small separation between the two embedded disks, it is likely that they formed via disk fragmentation. 

\subsubsection{Toomre stable disks}

The candidate disks that appear to be completely Toomre stable against gravitational instabilities include Cep\,A\,HW2 and NGC7538\,IRS9. The stability of Cep\,A is in part a result of warmer temperatures across the entire structure as compared with other sources as well as the low gas mass (0.5~\mo) to stellar mass (20~\mo) ratio. Therefore, the self-gravity of the disk is not very strong and the disk is warm prohibiting any possible fragments to effectively cool and survive. The $Q$ values for Cep\,A are so high that even if we have overestimated the mass of the protostar by a factor of 2 (\eg~due to contamination of the blueshifted emission by a hot core that is located 0.3\arcsec away from the main continuum peak \citepads{2009ApJ...703L.157J}, the disk would still appear stable at the scales probed by our observations. The low gas to stellar mass ratio of these two systems could be an indication that these systems may be somewhat closer to the end of their accretion phase, though Cep\,A\,HW2 is considered IR-dark while NGC7538\,IRS9 is IR-bright.

\subsubsection{General trends} 
\begin{figure}
    \centering
    \includegraphics[width=\hsize]{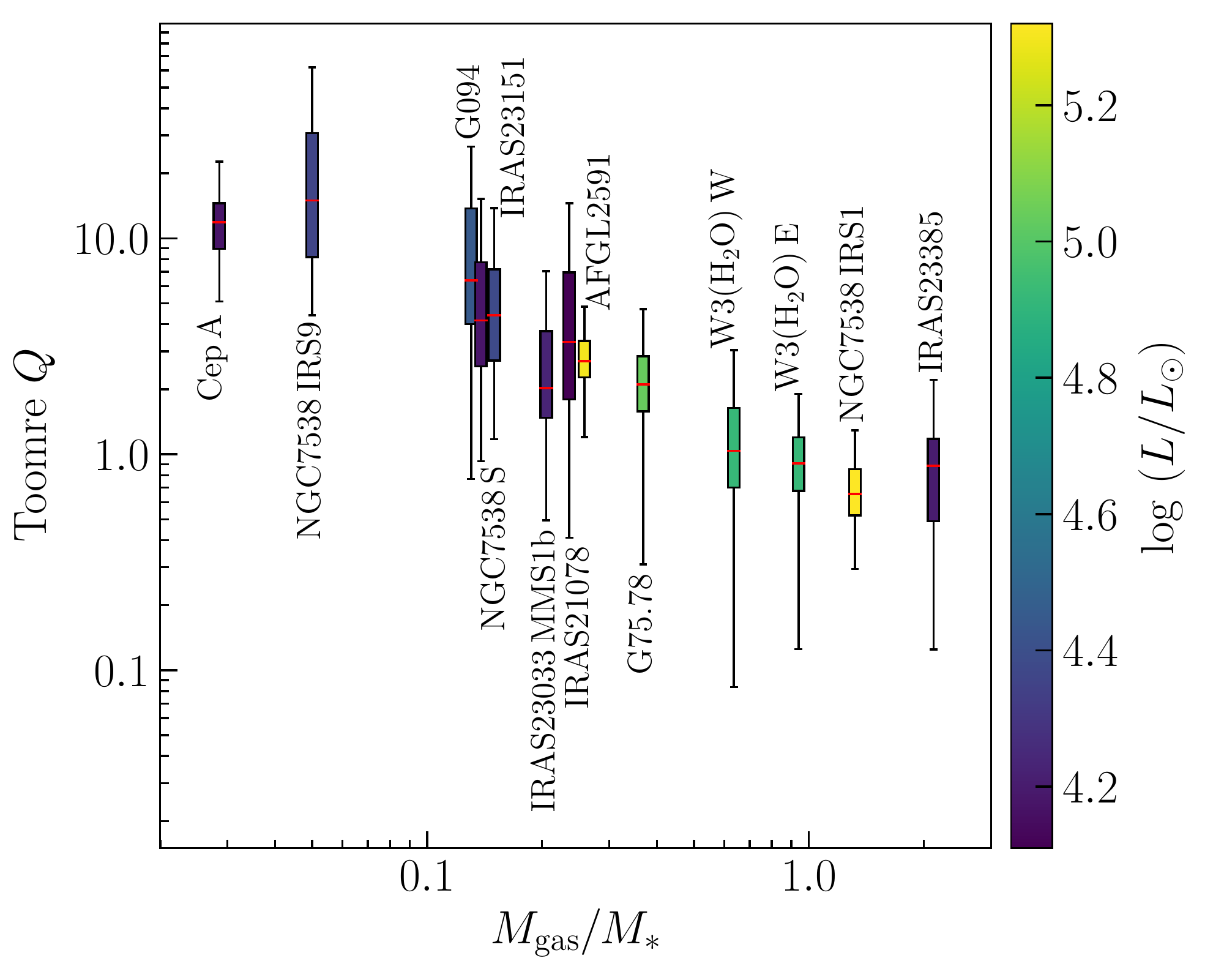}     
    \caption{Box-plot showing the $Q$ distribution versus the ratio of gas to stellar mass, coloured according to the luminosity of the regions within which they reside. The boxes extend from the first to the third quartiles, with a red line at the median~$Q$ value. The whiskers extend from the box by $\pm1.5$ times the box size. Disks with $Q\lesssim 2$ are Toomre unstable (\ie\ all but Cep~A \& NGC7538\,IRS9).}
    \label{f: Q_vs_Mratio}
\end{figure}

\begin{figure*}
    \centering
    \includegraphics[width=0.495\hsize]{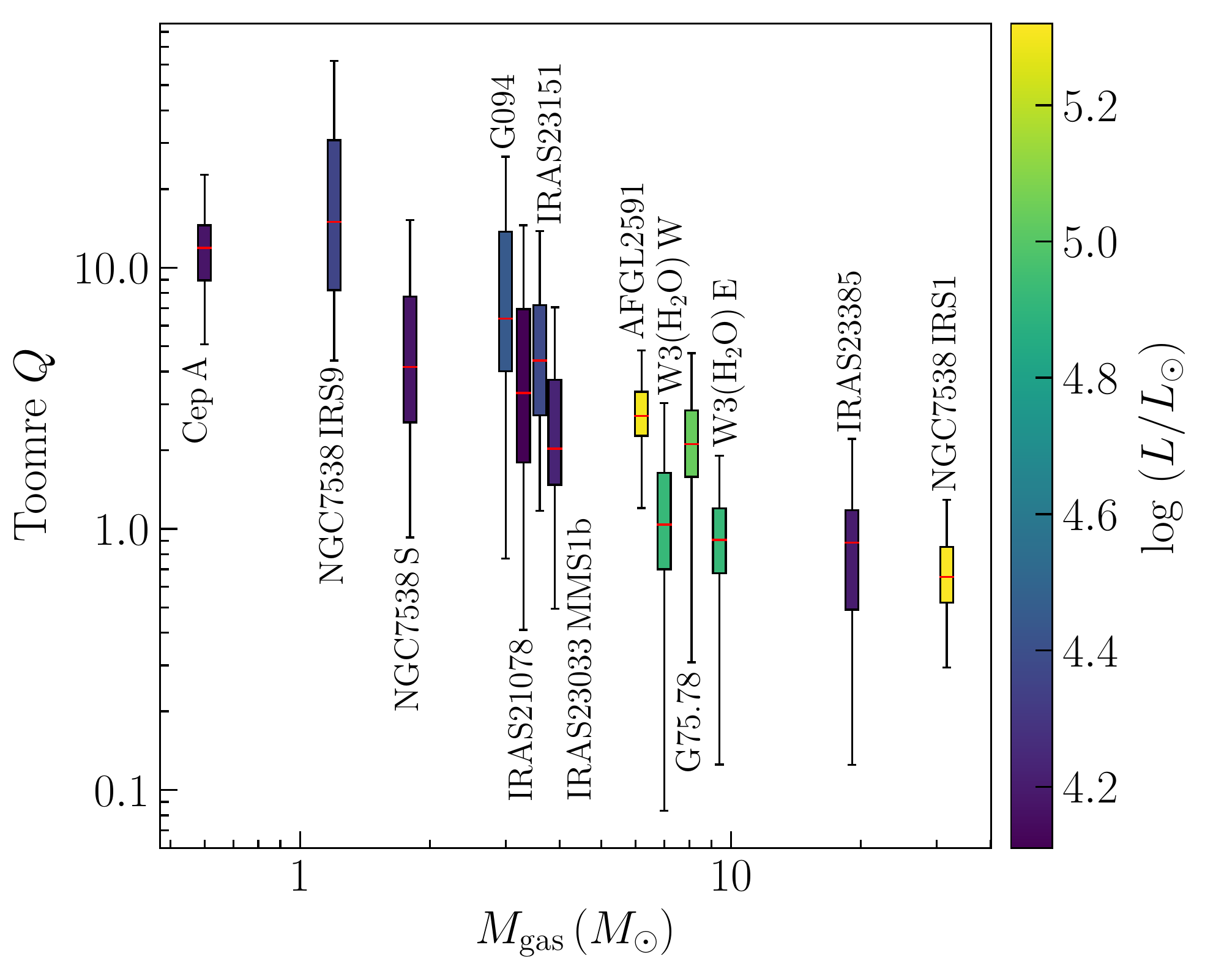}  \hspace{1mm}\includegraphics[width=0.495\hsize]{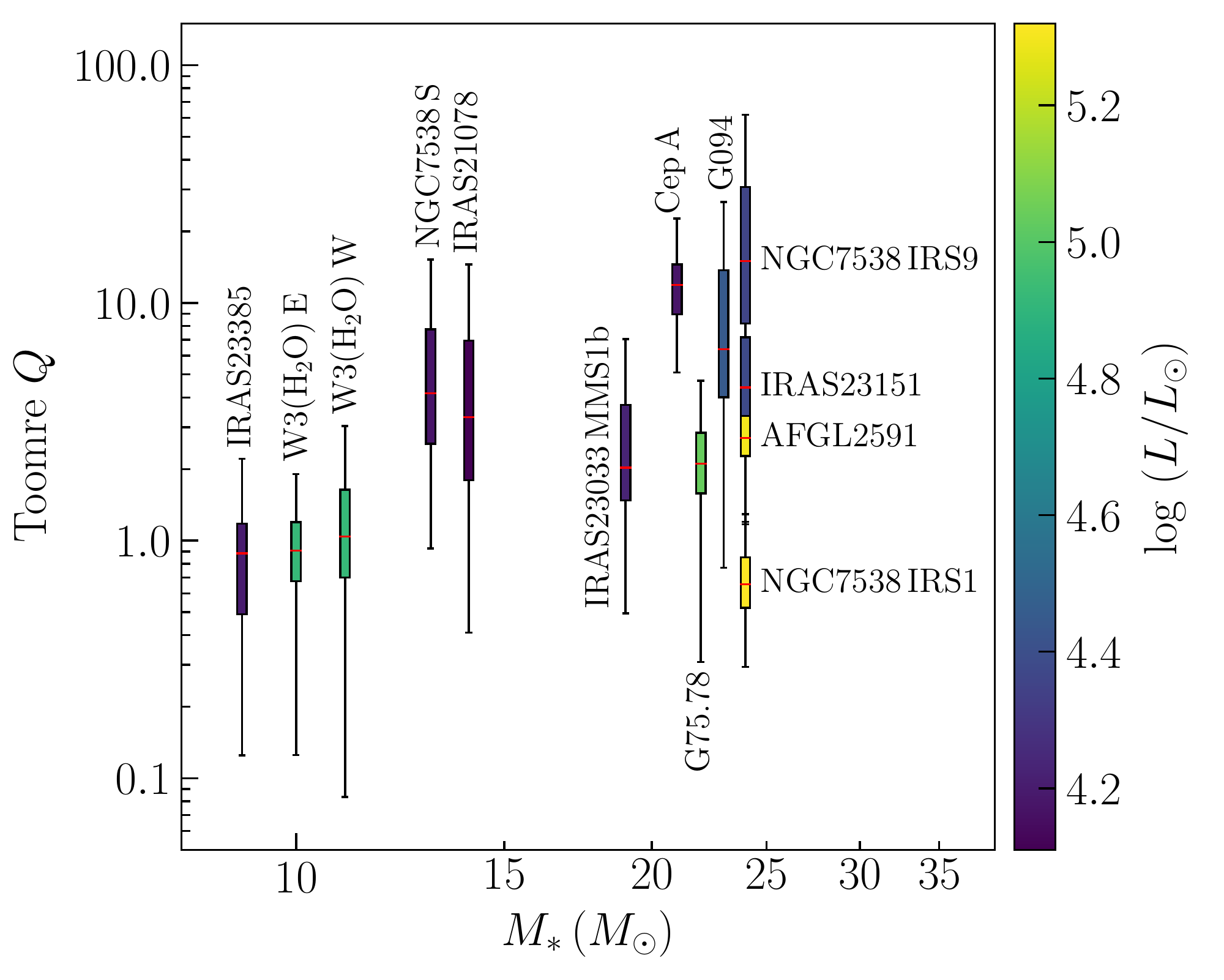}
    \caption{Median $Q$ plotted against gas mass (\emph{left}) and stellar mass (\emph{right}) for 13 candidate disks within the CORE survey, coloured according to the luminosity of the regions within which they reside. The Toomre-stable disks are marked by triangles.}
    \label{f: Q_vs_M}
\end{figure*}

In Fig.~\ref{f: Q_vs_Mratio} we present the $Q$ distribution versus the ratio of gas to stellar mass, coloured according to the YSO's luminosity. The boxes extend from the first to the third quartiles, with a red line at the median~$Q$ value (also listed in Table~\ref{t: masses_Q_CORE}). As explained in the previous section, since Toomre stability is a measure for the \textit{local} stability of the disk against axisymmetric gravitational instabilities (see also \eg\ \citeads{2020A&A...644A..41O}), simply using $Q_\mathrm{crit} \approx 2$ to determine the stability of these disks from the median~$Q$ values is incorrect. Based on the analysis presented above and by looking at the distribution of $Q$ for each individual case against $Q_\mathrm{crit}$ (see Figs.~\ref{f: toomre_CORE}, \ref{f: Q_vs_radius}, \& \ref{f: Q_vs_Mratio}), disks with masses greater than $\sim 10-20\%$ of the mass of their host (proto)stars are subject to fragmentation through gravitational instabilities. Numerical experiments of \citetads{2010ApJ...708.1585K} find that the maximum disk mass before disk fragmentation occurs is on the same order as the stellar mass, with a review on the subject by \citetads{2016ARA&A..54..271K} providing a value of $M_\mathrm{disk}/M_\ast \geq 0.1$ for gravitational instabilities to become important. Our observations confirm that disk fragmentation must be prevalent in high-mass star formation as we find 11 of 13 disks studied here to be gravitationally unstable, all of which have $M_\mathrm{gas}/M_\ast \geq 0.1$. Furthermore, we find an inverse relationship between $Q$ and $M_\mathrm{gas}/M_\ast$ a result of the expected tight anti-correlation between $Q$ and $M_\mathrm{gas}$ and less so due to a correlation between $Q$ and $M_\ast$ (see Fig.~\ref{f: Q_vs_M}). We also find indications that more luminous YSOs tend to have disks that are more massive and as a result more prone to fragmentation (see left panel of Fig.~\ref{f: Q_vs_M}). This could be a selection effect as high-mass protostars with lower mass disks must exist at a later stage of accretion. Additionally, the luminosities are derived from lower resolution observations corresponding to a different number of cores in each star forming region.

\citetads{2020A&A...644A..41O} investigated the formation and early evolution of an accretion disk surrounding a forming high-mass protostar in their 3D radiation-hydrodynamic simulation with high enough resolution to resolve the pressure scale height and Jeans length of the disk. The accretion disk with Keplerian-like rotation is globally Toomre-stable but locally Toomre-unstable and therefore fragments. The fragments are dynamic within the system, migrating inwards and can even get flung out of the system if they are on highly eccentric orbits. More massive fragments that remain on more stable orbits form their own accretion disks and become hydrostatic cores, some of which go on to form second Larson cores. The authors further study the fate of these companions and conclude that the hydrostatic core fragments migrate inwards and accrete onto the central protostar causing accretion bursts, while the second cores would likely become spectroscopic companions to the protostar. They find that companions produced through disk fragmentation can exist at a range of distances from the primary protostar, from spectroscopic multiples to companions at distances between $1000-2000$~au. Synthetic ALMA and NOEMA observations of this numerical simulation are presented in \citetads{2019A&A...632A..50A}.

Whether fragments survive as companions to the central protostar depends largely on whether they can cool fast enough. The cooling time has been confirmed to be short enough relative to the orbital period both numerically (\citeads{2016ApJ...823...28K}; \citeads{2018MNRAS.473.3615M}) and observationally \citepads{2018A&A...618A..46A} such that fragments that are beyond $\sim$50~au of the central protostar should be rapidly cooling (see review by \citeads{2016ARA&A..54..271K}). This is an important finding as nearly all high-mass stars are found in binary or multiple systems (\citeads{2012MNRAS.424.1925C}; \citeads{2012Sci...337..444S}, \citeyearads{2014ApJS..215...15S}). With evidence that  massive binaries start their lives in wide pairs and harden  into tighter massive binaries over time \citepads{2021A&A...645L..10R}, disk fragmentation may be a possible mechanism creating the initial binary and multiple systems at large separations before the inward migration occurs. Additionally, disk fragmentation can have profound implications for the formation of high-mass stars, as the mass flow is shared by multiple accretors rather than a single protostar (\eg\ \citeads{1997MNRAS.285...33B}; \citeads{2000MNRAS.314...33B}; \citeads{2010ApJ...725..134P}).

It is important to note that we have neglected the effect of magnetic fields as we do not have information about the magnetic field strengths in our survey. Theoretically, one can incorporate the Alfv\'en velocity in the Toomre equation in order to determine the magnetic $Q$ parameter, $Q_\mathrm{mag}$ \citepads{2001ApJ...559...70K}. In their numerical collapse simulations of strongly magnetised and turbulent molecular cloud cores, \citetads{2013MNRAS.432.3320S} compare $Q$ measurements with $Q_\mathrm{mag}$. They find that $Q_\mathrm{mag}$ rises above the value of $Q$, confirming that disk fragmentation is suppressed by the presence of magnetic fields as magnetic pressure provides added support against collapse. However, fragmentation in a magnetised environment can still proceed through other mechanisms (\eg~high amplitude fluctuations in the core initially, ambipolar diffusion and fragmentation during the second collapse; see \eg~\citeads{2008A&A...477...25H}). In fact, if the disk is well within the non-ideal MHD limit, the circumstellar disk structure is very similar to the non-magnetic case (\eg~\citeads{2018A&A...620A.182K}). Numerical simulations of massive protostellar disks that include non-ideal MHD and radiative transfer have shown that magnetic pressure is not as supportive as thermal pressure \citepads{2021A&A...652A..69M}. Furthermore, on slightly larger scales, \citetads{2011ApJ...729...72P} show that while magnetic fields reduce fragmentation, they do not prevent it and gravitational instability eventually sets in. Nevertheless, without information about the magnetic field strengths in the CORE sources, it is not trivial to estimate to what extent that may affect our conclusions.

\begin{figure}
    \centering
    \includegraphics[width=\hsize]{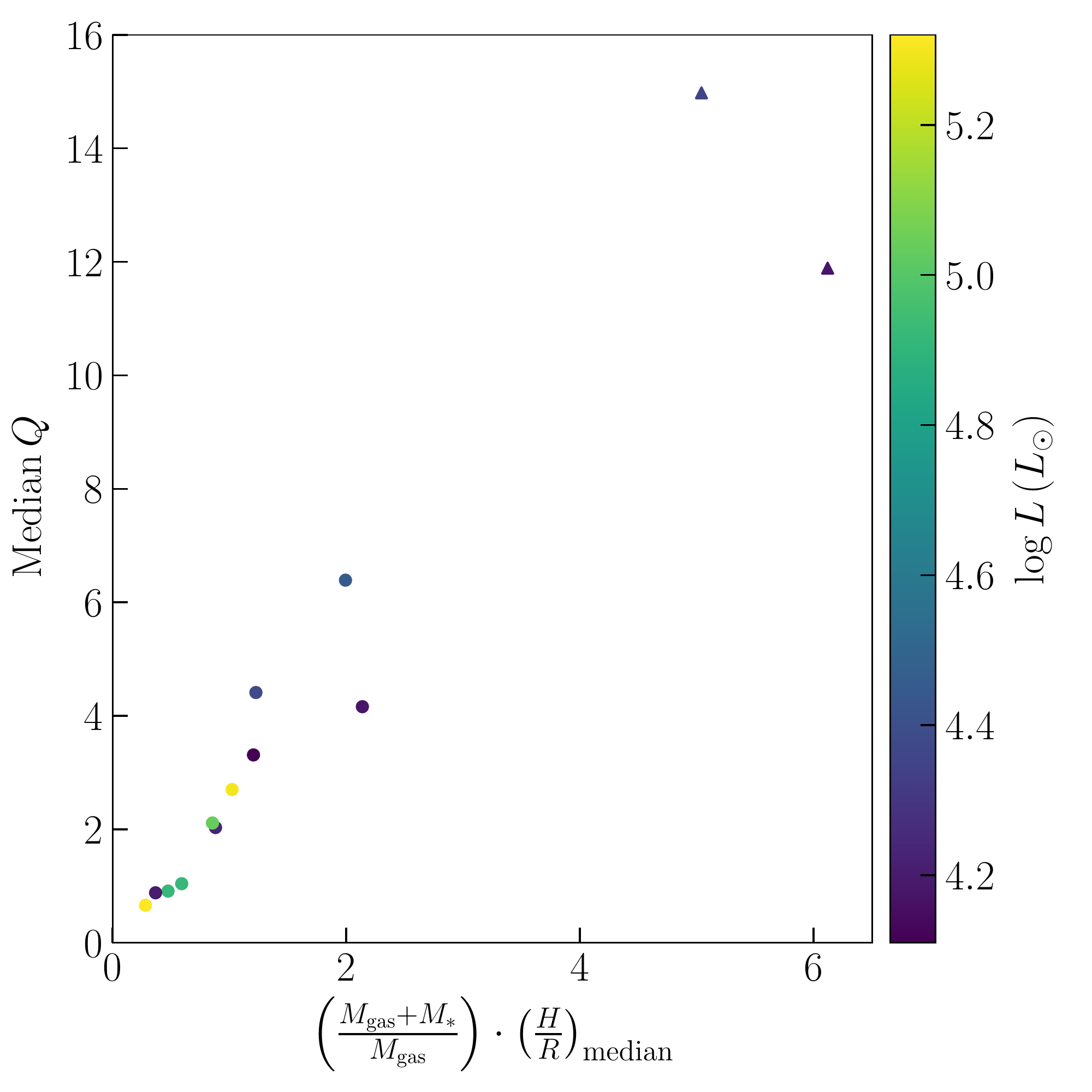}
    \caption{Median $Q$ as a function of $\left(\frac{H}{r}\right)\left(\frac{M_\ast+M_\mathrm{disk}}{M_\mathrm{disk}}\right)$. The Toomre-stable disks are marked by triangles.}
    \label{f: Q_correlation}
\end{figure}

\subsubsection{Disk thickness}\label{ss: disk_thickness}

One of the main assumptions for the Toomre analysis is the thin-disk approximation, such that $Q_\mathrm{crit}$ for an isothermal disk of finite thickness is lower $\sim$0.7 \citepads{1965MNRAS.130..125G}. This is because the thicker a disk, the more diluted its self-gravity becomes. To get an estimate for the importance of disk thickness, \citetads{2001ApJ...553..174G} derived an additional stability criterion taking into account the disk scale height, $H\simeq c_s/\Omega$, such that a disk is unstable against gravitational instabilities if 
\begin{equation}
  M_\mathrm{disk}\gtrsim \frac{H}{r}M_\ast.
\end{equation}
 
Accounting for the self-gravity of the disk (\ie~replacing $M_\ast$ with $M_\ast+M_\mathrm{disk}$) and with the surface density of the disk $\Sigma=M_\mathrm{disk}/(\pi r^2)$, the Toomre criterion can be rewritten as
\begin{equation}\label{eq: disk_thickness}
  Q=\left(\frac{H}{r}\right)\left(\frac{M_\ast+M_\mathrm{disk}}{M_\mathrm{disk}}\right)
\end{equation} 

Using the disk scale height, $H\simeq c_s/\Omega$, and Eqs.~\ref{eq: SAMPLE_cs} and \ref{eq: SAMPLE_omega}, we create maps of $H/r$ ratio and report the median $H/r$ values in Table~\ref{t: masses_Q_CORE}. The observed median disk scale heights are generally too high ($0.15-0.31$) for a thin disk-like structure, possibly because at our resolution the envelope and disk components are likely blended, and the fact that temperatures derived from \mc\ emission probe a layer higher than the disk midplane. Figure~\ref{f: Q_correlation} shows the median $Q$ as a function of $\left(\frac{H}{r}\right)\left(\frac{M_\ast+M_\mathrm{disk}}{M_\mathrm{disk}}\right)$. We find a tight relationship between the two stability criteria, validating the analytically motivated scalings and the naïve expectations on the disk physics. However, we observe a normalisation value of $\sim$2 (\ie\ $Q \approx 2 \left(\frac{H}{r}\right)\left(\frac{M_\ast+M_\mathrm{disk}}{M_\mathrm{disk}}\right)$) instead of the expected value of unity, which is a direct consequence of the high disk scale heights.

\begin{figure*}
    \centering
    \includegraphics[width=\hsize]{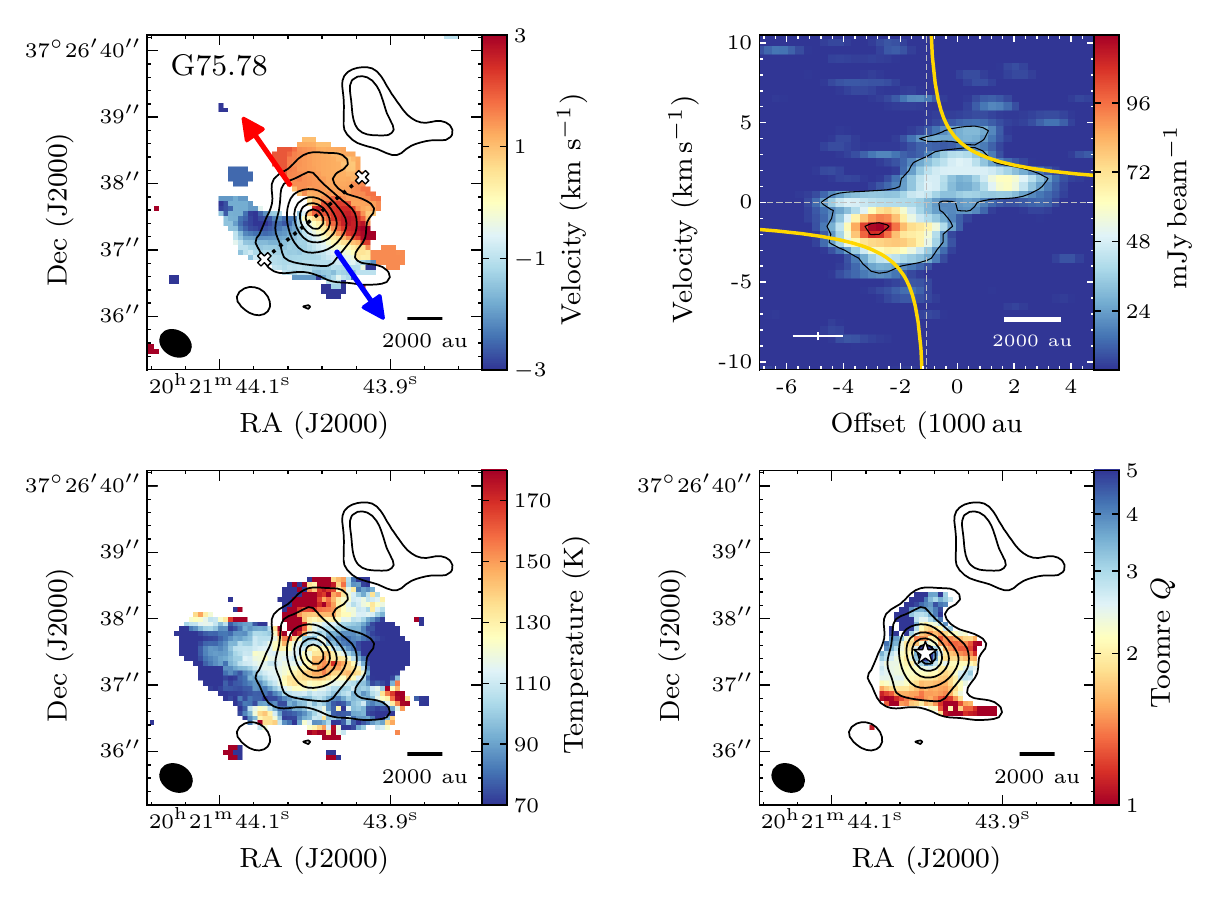}
    \caption{Summary plot showing the kinematics and derived properties for one source in the CORE sample (G75.78). {\em Top left panel:} intensity-weighted mean velocity (first moment) map of \mck{5} in colour with 1.37~mm continuum contours. The blue and red arrows correspond to the estimated directions of bipolar blueshifted and redshifted molecular outflows, respectively. The dotted line indicates the position of the strongest velocity gradient tracing the disk, i.e. perpendicular to the rotation axis. The edges of the assumed disk extent are marked with an $\times$. {\em Top right panel:} PV plot of \mck{5} along the cut in the direction of rotation as depicted by the dotted line in the top left panel. The contours correspond to the 6$\sigma$ level increasing in steps of 24$\sigma$. Yellow lines show the Keplerian rotation curves for an enclosed mass of 19~\mo. The cross in the bottom right corner corresponds to the spatial and spectral resolutions. {\em Bottom left panel:} Rotational temperature maps obtained by fitting \mckr{0}{6} and \mcisokr{0}{3} lines with \emph{XCLASS} in colour and 1.37~mm continuum contours. {\em Bottom right panel:} \tq\ map assuming a protostar is located at the position of the continuum peak as depicted by a star and accounting for the self-gravity of the disk, with 1.37~mm continuum contours. Regions outside of 6$\sigma$ continuum contours are masked out. The synthesised beam and a scale bar are shown in the bottom of the panels.}
    \label{f: summary}
\end{figure*}


\section{Summary and conclusions}\label{s: sample_summary}

In this work, we present the gas kinematics for a sample of 20 high-mass star-forming regions observed with the NOEMA interferometer at a resolution of $\sim$0.4\arcsec\ as part of the CORE survey. While the sample was chosen uniformly to target the early phase of high-mass star formation, there exists a diversity in age and chemical richness. We focused our analysis on 15 sources for which the emission of the dense gas tracer \mc\ was detected with the goal of searching for and characterising the properties of disk-like structures. Fig.~\ref{f: summary} shows the main results that can be obtained for a given region with respect to the kinematics and derived physical properties. The following is a summary of our findings:

\begin{itemize}

\item Using the $K$-ladder rotational transitions of the dense gas tracer \mck{K}, we found velocity gradients across 13 cores perpendicular to the directions of bipolar molecular outflows, making them excellent disk candidates. The directions of molecular outflows were determined from a combination of tracers ($^{12}\mathrm{CO}$, $^{13}\mathrm{CO}$, $\mathrm{C^{18}O}$, and SO), both from single-dish IRAM 30-m observations as well as combined NOEMA and single-dish observations, included in the CORE survey. 

\item The rotating structures vary in size from $\sim$1800 to $\sim$8500~au, measured from the extent of \mck{3} emission across the direction of the strongest velocity gradient. 

\item From the ratio of free-fall to rotational timescales as a function of gas mass, we established that the 13 sources in this sample are in fact rotationally supported disk candidates and not transient toroids.

\item The velocity profiles of some disk candidates resemble differential rotation with high-velocity components closest to the central object while others mimic rigid-body-like rotation. The latter may be a consequence of a lack of spatial resolution. We fitted the $5-7\sigma$ edge of the PV diagram for each disk with a Keplerian profile. The estimated dynamical masses, which for a Keplerian disk is the mass of the central object, are in the range $\sim10-25$~\mo.

\item Modelling the level population of \mckr{0}{6} lines under LTE conditions using \textit{XCLASS}, we provide temperature maps of the regions. Temperatures are on average warm (70--210~K, depending on the target) with a diversity in distributions. Some are warm in the center, cooler in the outskirts while others are more uniform in temperature. In some cases the ejection points of the outflows coincide with regions of warmer temperature where linewidths are also broader. This hints at the possibility that these regions have been heated by the outflows and associated shocks and/or regions that are carved out by the molecular outflows.

\item We calculated the gas mass of the disk-like structures assuming gas and dust temperatures are in equilibrium and optically thin dust emission. Gas masses range from 1~\mo\ to as high as 30~\mo, depending on the region. 

\item Radial profiles of the specific angular momentum for the best disk candidates span a range of 1--2 orders of magnitude, on average around $\sim10^{-3}$~\kms\,pc, and follow a $j \propto r^{1.7}$ relation. This scaling is consistent with synthetic observations of poorly resolved massive disk simulations where the envelope and disk components become completely blended. The radial profiles tapers off at $\sim$1000~au, suggesting that rotationally supported disks must exist below this scale. For the closest sources in our sample which are resolved at better linear resolutions, this limit was estimated at 300--400~au and 500--700~au for Cep~A and IRAS21078, respectively. 

\item Assuming the candidate disks to be in Keplerian rotation about a star with the estimated protostellar mass and accounting for the self-gravity of the disk by including its gas mass, we created maps of the Toomre~$Q$ parameter to determine their stability against axisymmetric gravitational collapse. Most disk candidates (11 out of 13) are Toomre unstable at the scales probed by our observations because of their high masses. In particular, disks with masses greater than $\sim10-20\%$ of the mass of their host (proto)stars are subject to fragmentation through gravitational instabilities. We find that more luminous YSOs tend to have disks that are more massive and as a result more prone to fragmentation. These findings remain valid even when accounting for the thickness of the disks. 

\end{itemize}

In conclusion, we find that disk-mediated accretion may be a common mechanism for high-mass star formation, similar to the formation scenario of their lower mass counterparts. We further discover that most high-mass disk candidates in the CORE sample (11 out of 13) are prone to fragmentation due to gravitational instabilities early in their formation due to their high disk to stellar mass ratio. This has profound implications for the formation of the most massive stars, as it modifies the accretion evolution of high-mass protostars. Considering the fact that almost all high-mass stars are found in binary or multiple systems, disk fragmentation seems to be an important mechanism in the formation of these systems.

\begin{acknowledgements}
The authors would like to thank the referee, Adam Ginsburg, whose comments helped improve the clarity and impact of the paper. AA and HB acknowledge support from the European Research Council under the European Community's Horizon 2020 framework program (2014-2020) via the ERC Consolidator Grant `From Cloud to Star Formation (CSF)' (project number 648505). RK acknowledges financial support via the Emmy Noether and Heisenberg Research Grants funded by the German Research Foundation (DFG) under grant no.~KU 2849/3 and 2849/9. RGM acknowledges support from UNAM-PAPIIT project IN108822. AP acknowledges financial support from the UNAM-PAPIIT IN111421 grant, the Sistema Nacional de Investigadores of CONACyT. RGM and AP also acknowledge support from the CONACyT project number 86372 of the `Ciencia de Frontera 2019' program, entitled `Citlalc\'oatl: A multiscale study at the new frontier of the formation and early evolution of stars and planetary systems', M\'exico.
\end{acknowledgements}

\bibliographystyle{aa} 
\bibliography{Ahmadi_core_kinematics} 

\begin{appendix}
\onecolumn
\section{Moment maps}\label{a: mom_panels_ch3cn_k3}

\begin{figure*}[h!]
    \centering
    \includegraphics[width=0.76\hsize]{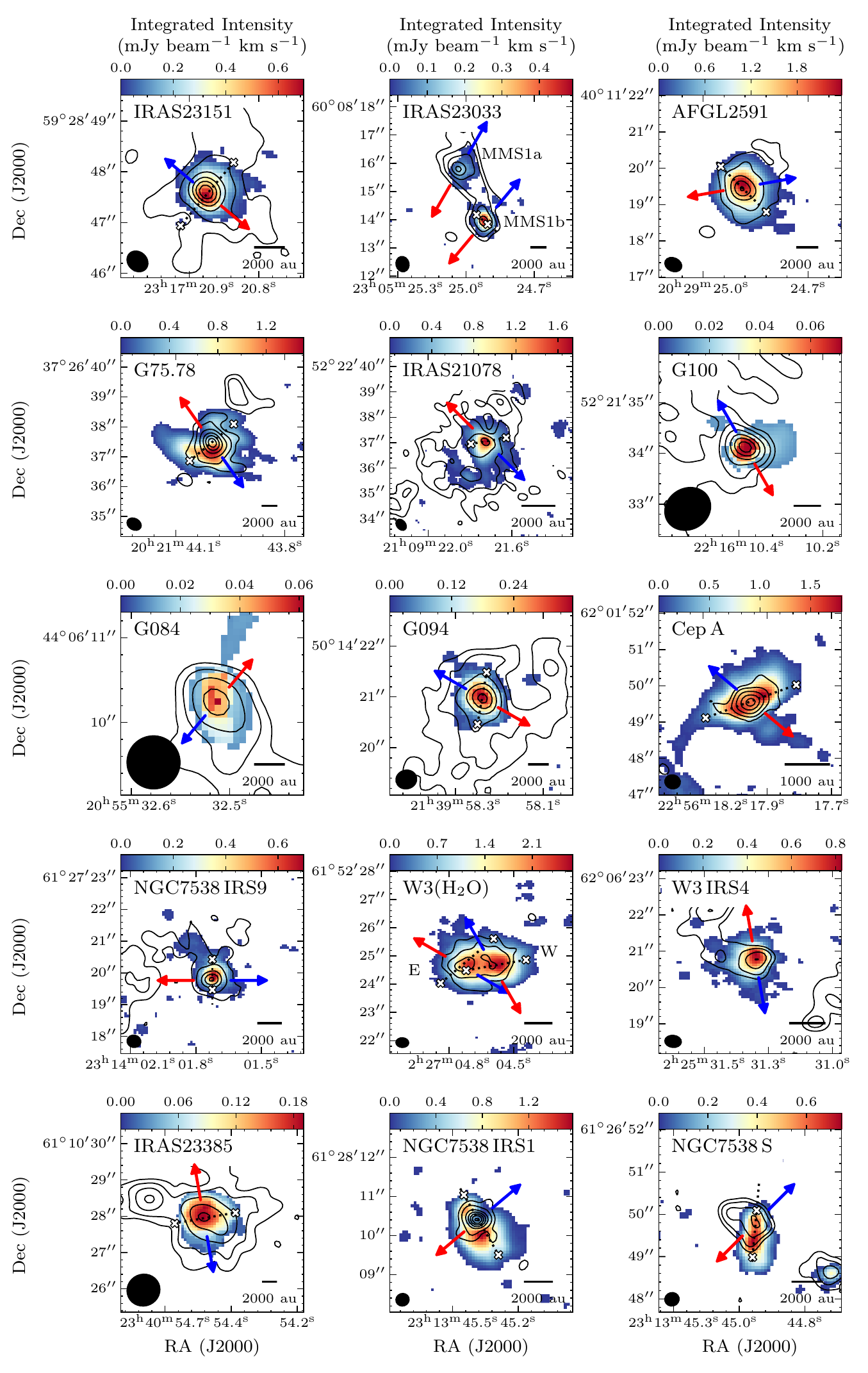}
    \caption{Integrated intensity (zeroth moment) maps of \mck{3} showing the dense gas distribution for 15 of the 20 sources in the CORE survey. The contours and features are as described in Fig.~\ref{f: mom_velo_ch3cn_k3}.}
    \label{af: mom0_panels_ch3cn_k3}
\end{figure*}

\begin{figure*}[hb!]
    \centering
    \includegraphics[width=0.82\hsize]{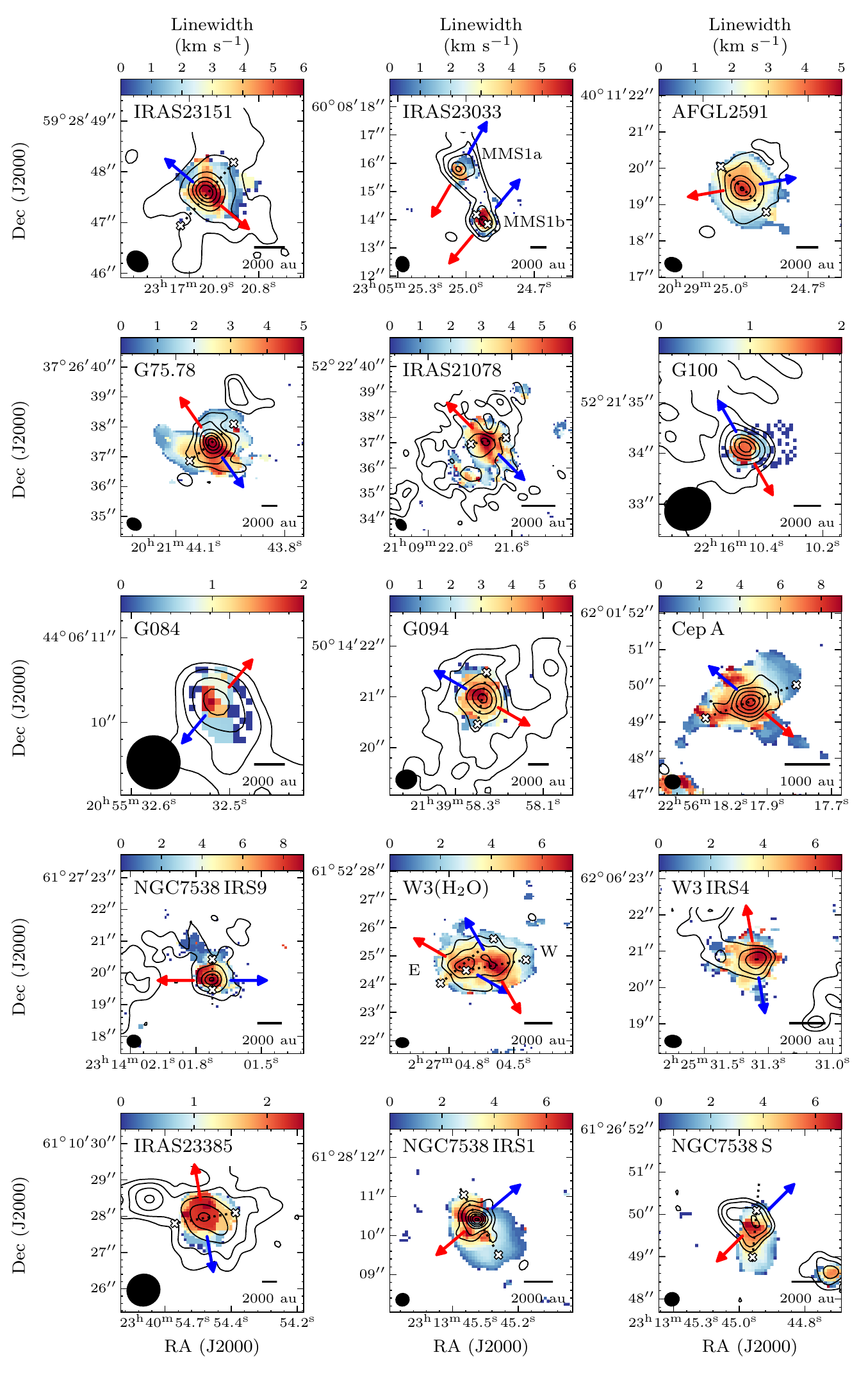}
    \caption{Intensity-weighted velocity dispersion (second moment) maps of \mck{3} showing the dense gas kinematics for 15 of the 20 sources in the CORE survey. The contours and features are as described in Fig.~\ref{f: mom_velo_ch3cn_k3}.}
    \label{af: mom2_panels_ch3cn_k3}
\end{figure*}

In this appendix, we present zeroth and second moment maps of \mck{3} in figures~\ref{af: mom0_panels_ch3cn_k3} and \ref{af: mom2_panels_ch3cn_k3}, respectively.

\clearpage
\section{Decomposed velocity maps}\label{a: decomp_maps}
\begin{figure*}[hb!]
    \centering
    \includegraphics[width=0.73\hsize]{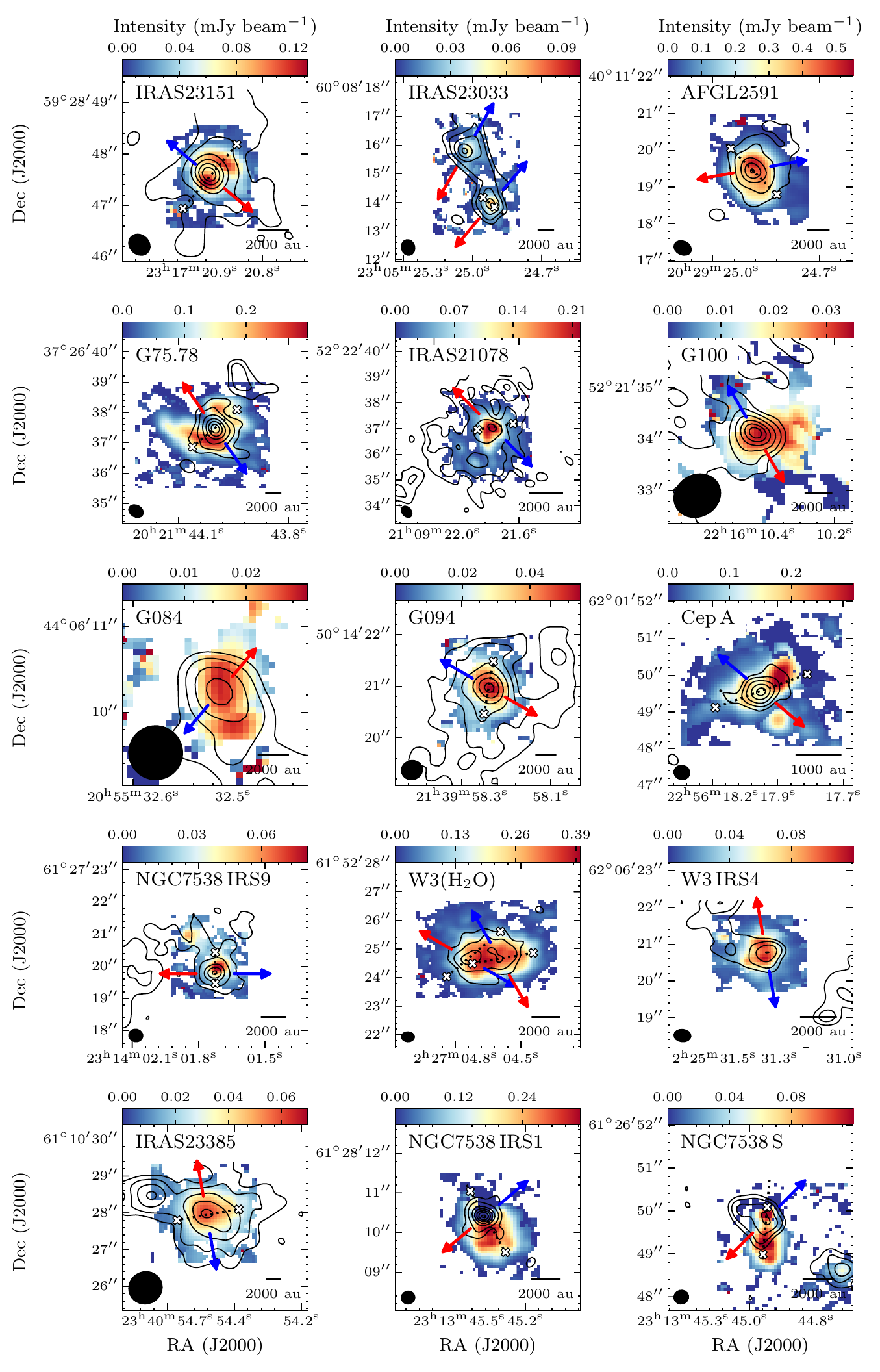}
    \caption{Peak intensity (amplitude) maps of \mck{3} showing the dense gas kinematics for 15 of the 20 sources in the CORE survey obtained by fitting Gaussian profiles to their spectra. The contours and features are as described in Fig.~\ref{f: mom_velo_ch3cn_k3}.}
    \label{af: amp_decomp}
\end{figure*}

\begin{figure*}[hb!]
    \centering
    \includegraphics[width=0.82\hsize]{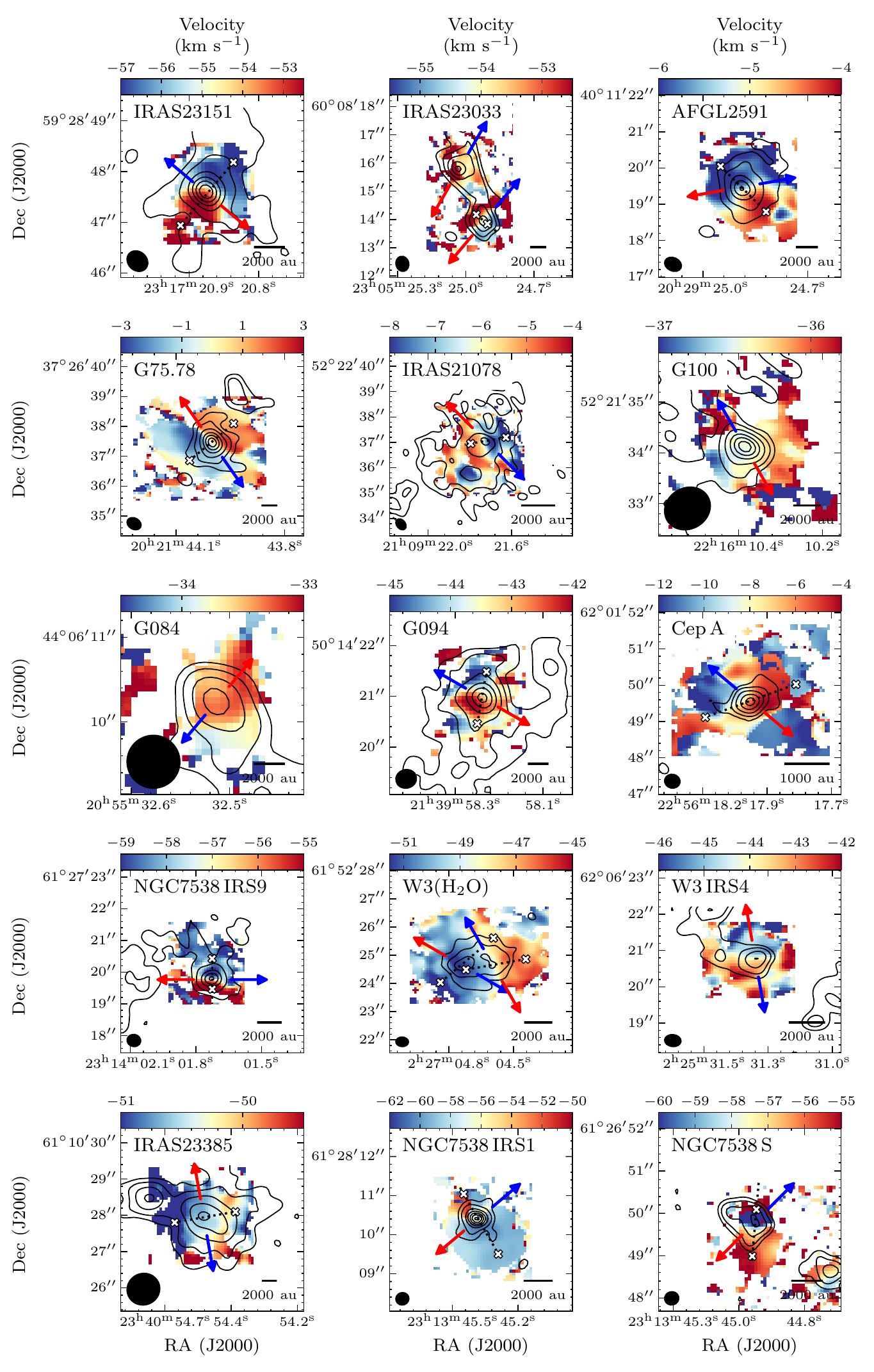}
    \caption{Peak velocity maps of \mck{3} showing the dense gas kinematics for 15 of the 20 sources in the CORE survey obtained by fitting Gaussian profiles to their spectra. The contours and features are as described in Fig.~\ref{f: mom_velo_ch3cn_k3}.}
    \label{af: velo_decomp}
\end{figure*}

\begin{figure*}[hb!]
    \centering
    \includegraphics[width=0.82\hsize]{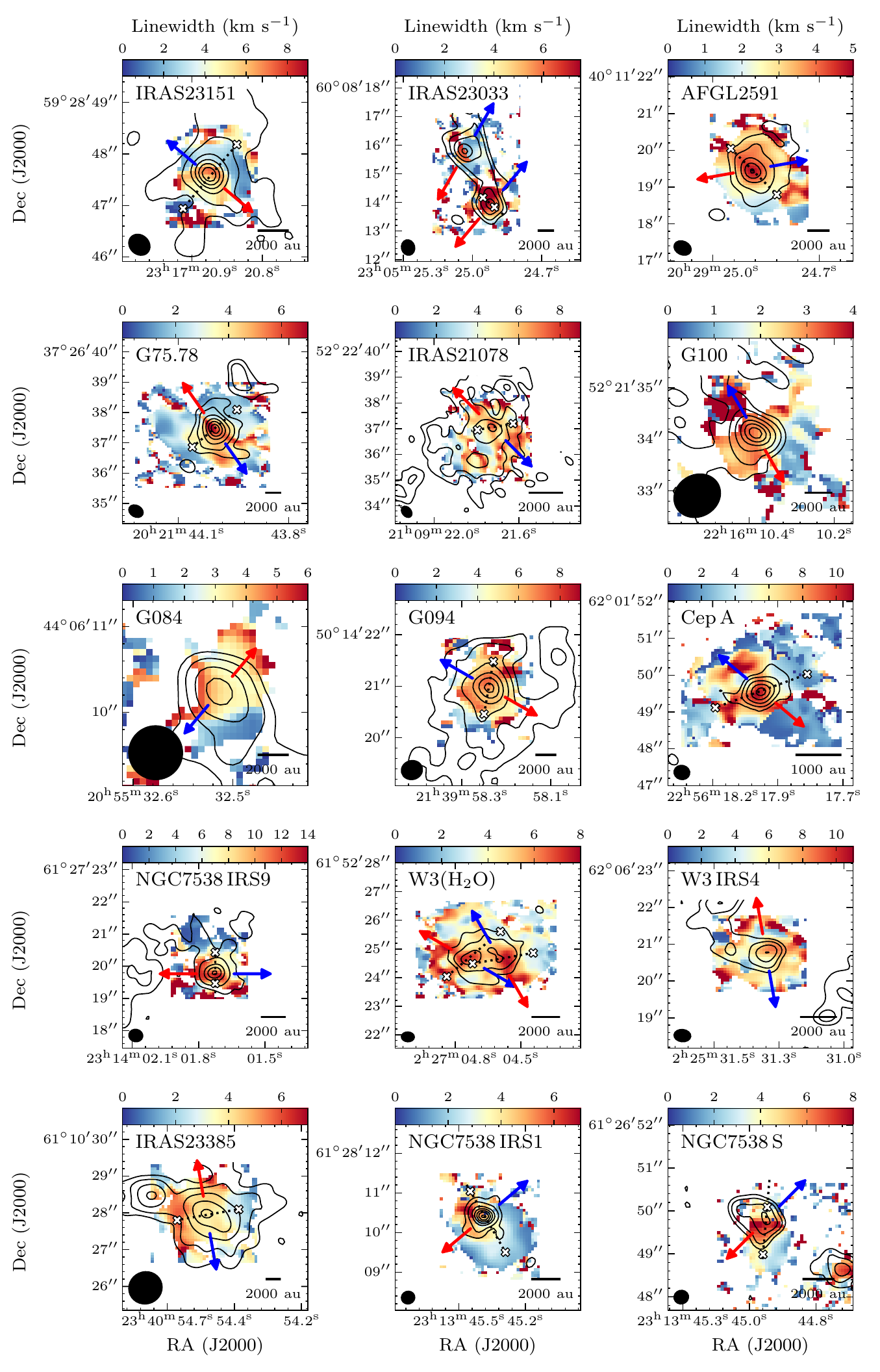}
    \caption{Linewidth (FWHM) maps of \mck{3} showing the dense gas kinematics for 15 of the 20 sources in the CORE survey obtained by fitting Gaussian profiles to their spectra. The contours and features are as described in Fig.~\ref{f: mom_velo_ch3cn_k3}.}
    \label{af: width_decomp}
\end{figure*}

We have fitted Gaussian profiles to the spectra of all sources using the \textit{Spectral-Cube} tool within the \textit{PySpecKit} package in Python (\citeads{2011ascl.soft09001G}; \citeads{2019zndo...2573901G}) and present the decomposed peak intensity, peak velocity, and full-width-at-half-maximum (FWHM) linewidth maps in this section (Figs.~\ref{af: amp_decomp}~--~\ref{af: width_decomp}). Fitting one Gaussian profile assumes that there is only one velocity component at each position.


\clearpage
\section{Molecular outflows}\label{a: outflow_panels} 
\begin{figure*}[h!]
    \centering
    \includegraphics[width=0.75\hsize]{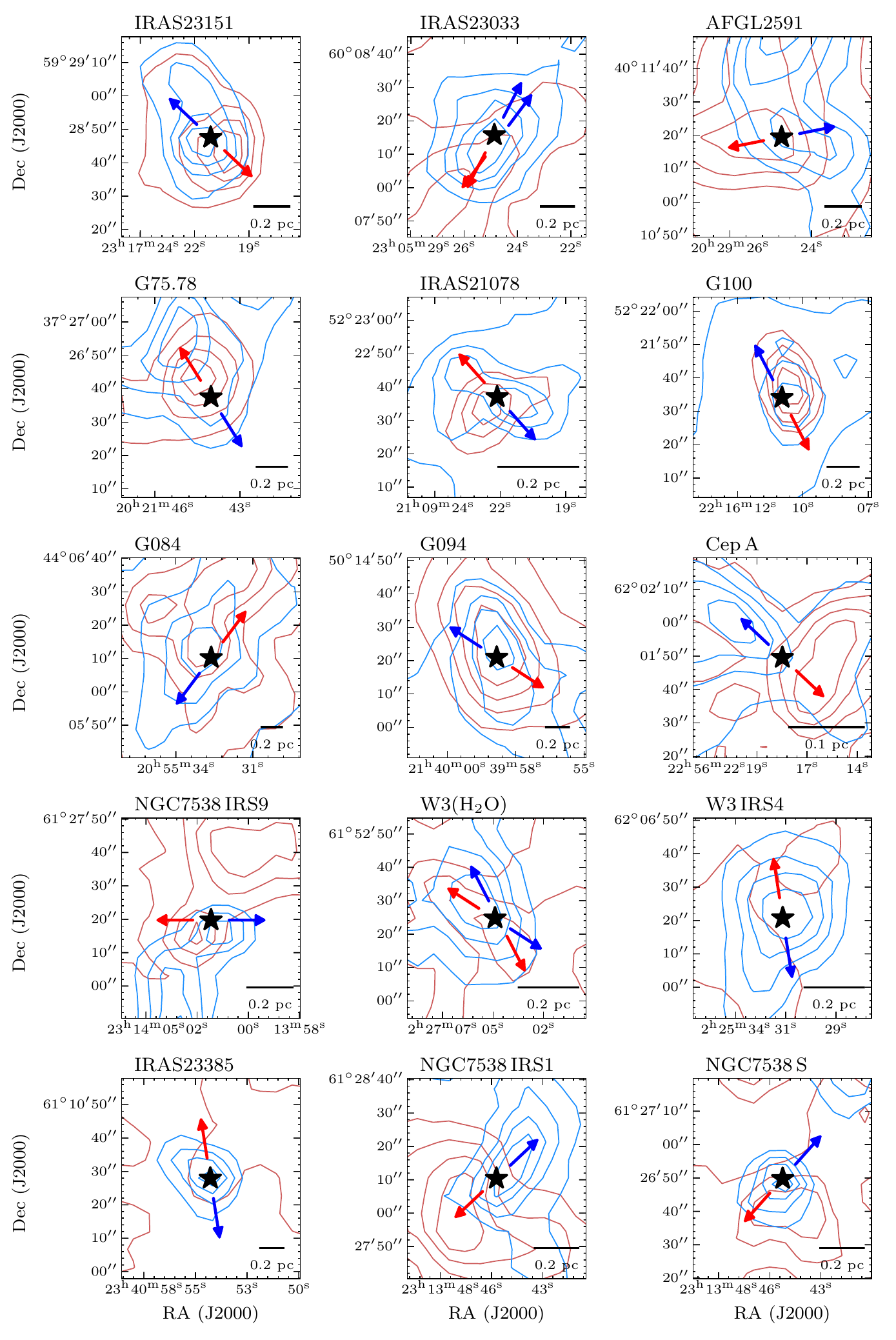}
    \caption{Intensity maps of CO\,(2--1) emission from IRAM 30-m telescope integrated over the blueshifted and redshifted wings of emission, showing the outflow structure. The position of the strongest source in the field is depicted by a star. The blue and red arrows correspond to the estimated directions of bipolar blueshifted and redshifted molecular outflows, respectively. A scale-bar is shown in the bottom right corner of each panel. The map size of the IRAM 30-m observations is 1.5\arcmin\ by 1.5\arcmin\ with a half-power beam width of $\sim$11\arcsec\ at this frequency.}
    \label{af: outflows_12co_sd}
\end{figure*}

\begin{figure*}[h!]
    \centering
    \includegraphics[width=0.82\hsize]{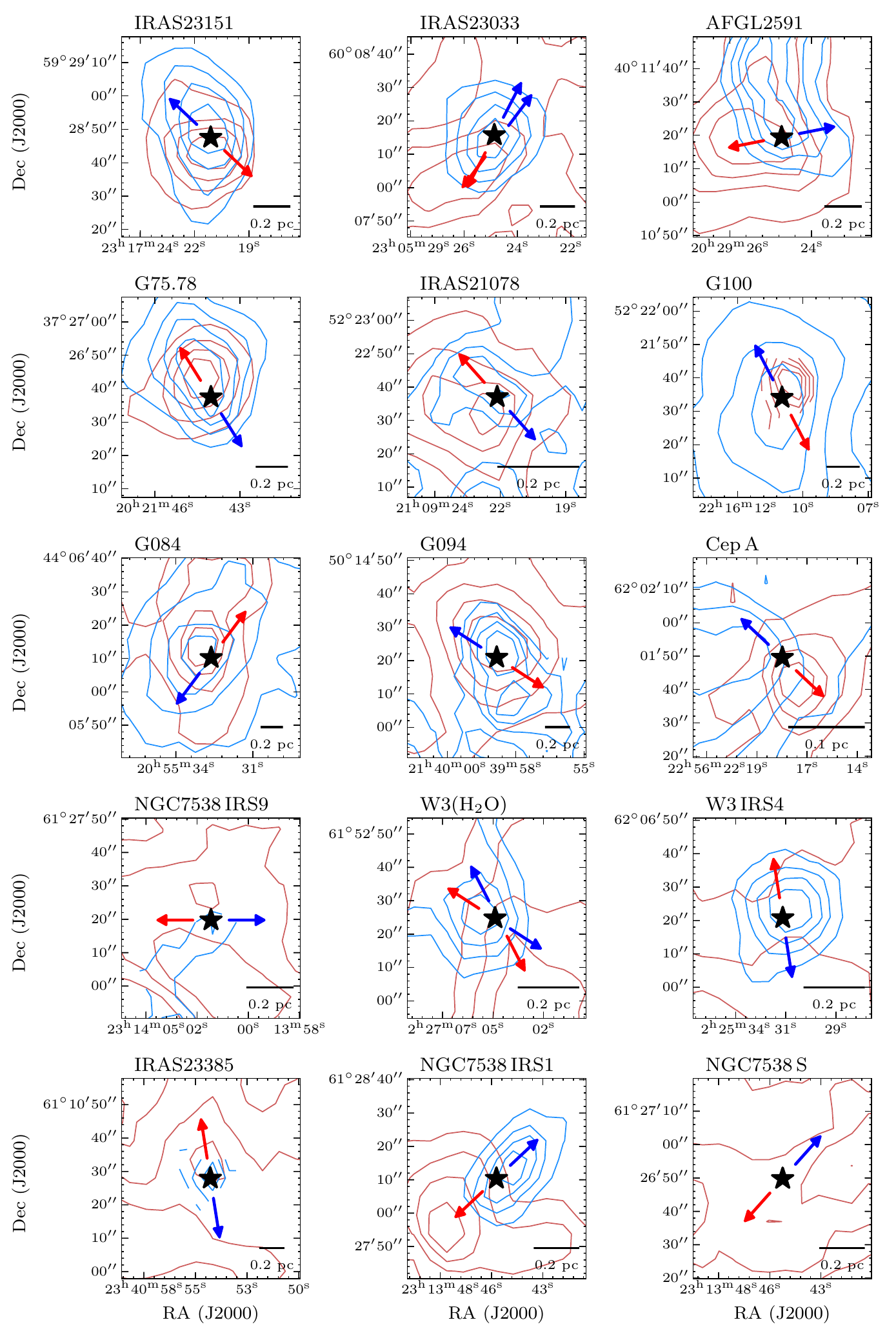}
    \caption{Intensity maps of $^{13}\mathrm{CO}\,(2-1)$ emission from IRAM 30-m telescope integrated over the blueshifted and redshifted wings of emission, showing the outflow structure. The position of the strongest source in the field is depicted by a star. The blue and red arrows correspond to the estimated directions of bipolar blueshifted and redshifted molecular outflows, respectively. A scale-bar is shown in the bottom right corner of each panel. The map size of the IRAM 30-m observations is 1.5\arcmin\ by 1.5\arcmin\ with a half-power beam width of $\sim$11\arcsec\ at this frequency.}
    \label{af: outflows_13co_sd}
\end{figure*}

\begin{figure*}[h!]
    \centering
    \includegraphics[width=0.82\hsize]{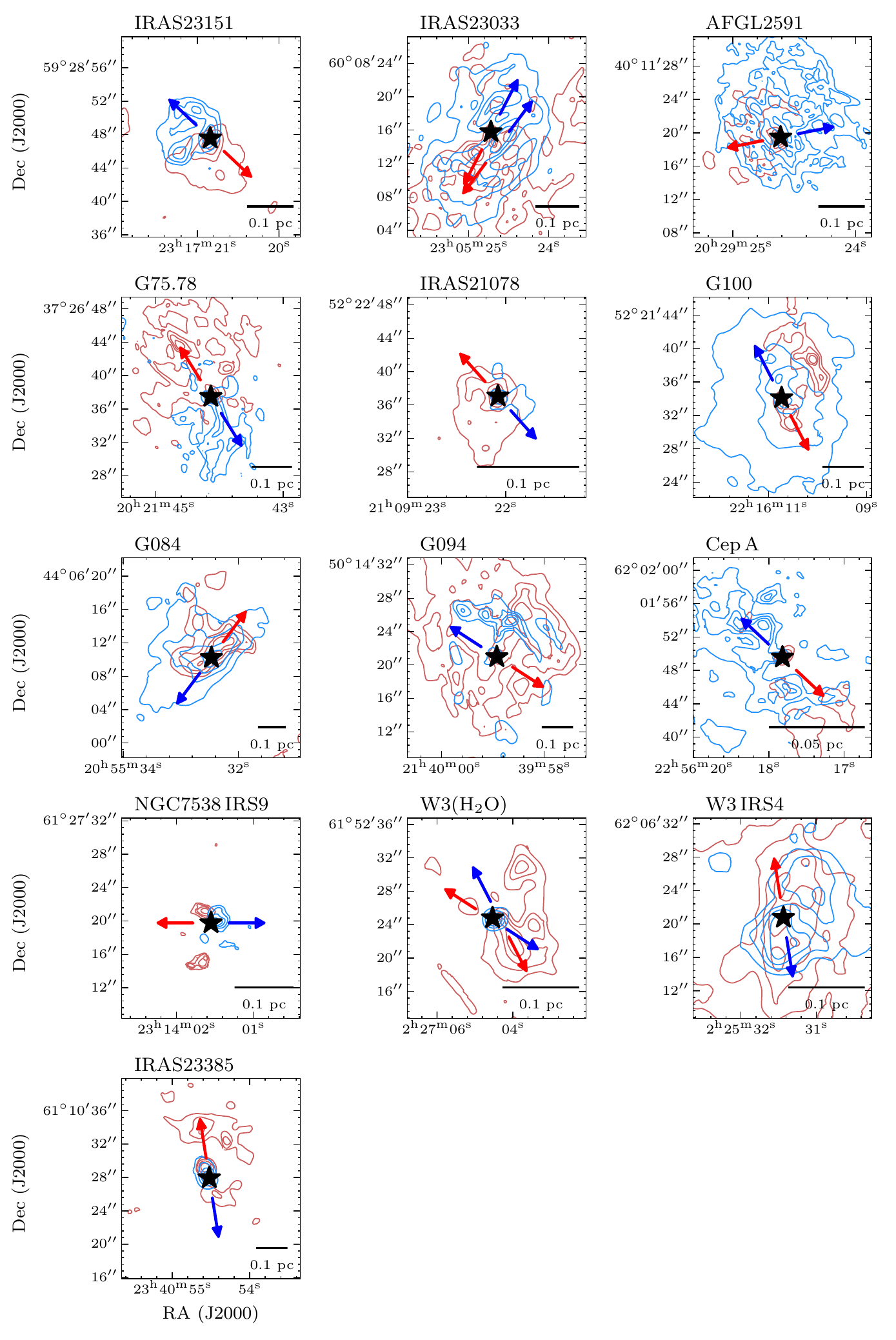}
    \caption{Intensity maps of $^{13}\mathrm{CO}\,(2-1)$ emission from merged NOEMA and IRAM 30-m data integrated over the blueshifted and redshifted wings of emission, showing the outflow structure. The position of the strongest source in the field is depicted by a star. The blue and red arrows correspond to the estimated directions of bipolar blueshifted and redshifted molecular outflows, respectively. A scale-bar is shown in the bottom right corner of each panel. We did not merge the data for the pilot sources as we did not observe this target with the most compact NOEMA configuration (D-array), hence the missing panels.}\label{af: outflows_13co_merged}
\end{figure*}

\begin{figure*}
    \centering
    \includegraphics[width=0.82\hsize]{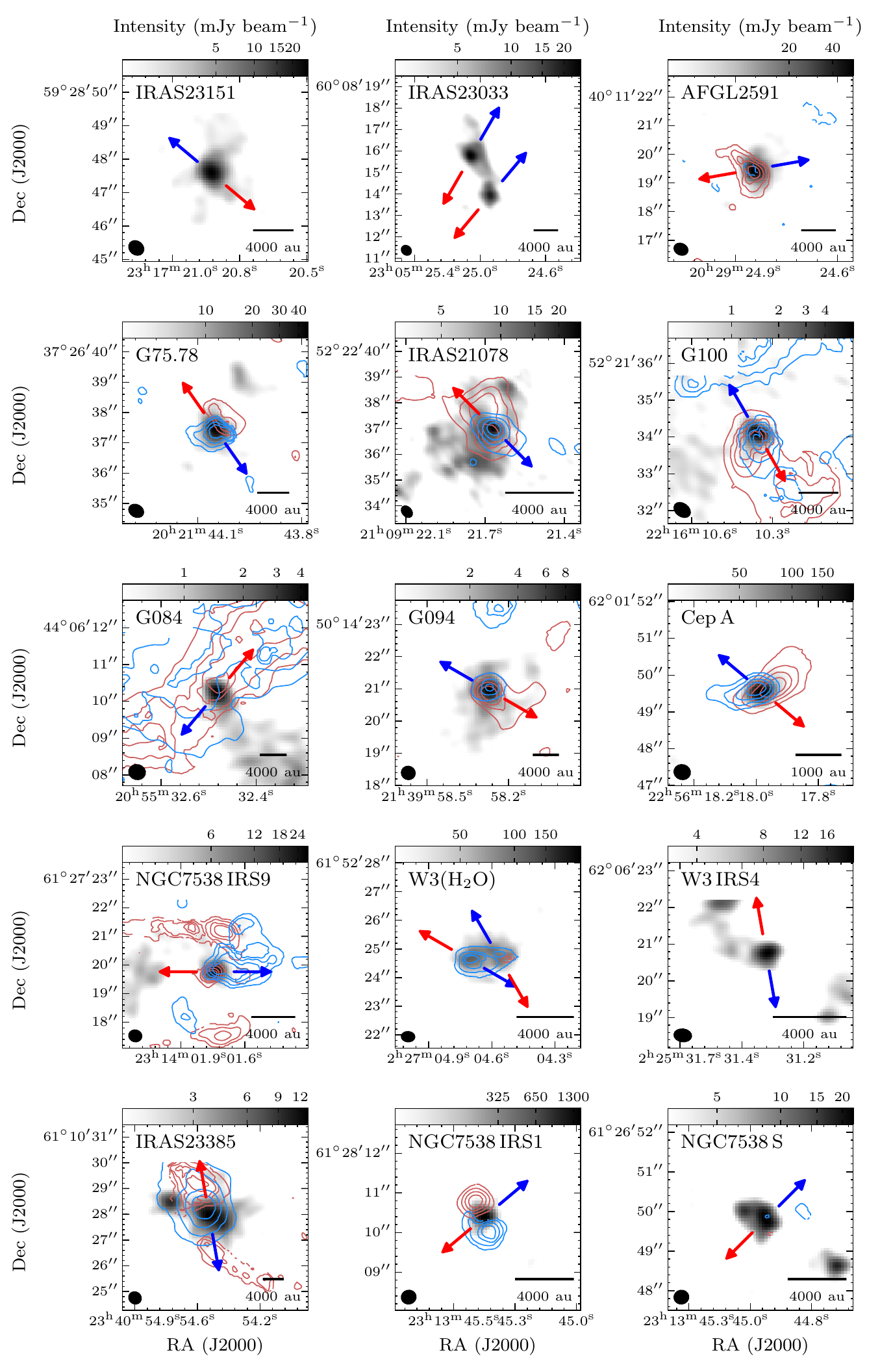}
    \caption{The greyscale corresponds to the 1.37~mm continuum while the blue and red contours correspond to the NOEMA intensity maps of $^{13}\mathrm{CO}\,(2-1)$ integrated over the blueshifted and redshifted wings of emission, tracing either outflows or disk winds.The blue and red arrows correspond to the estimated directions of bipolar blueshifted and redshifted molecular outflows, respectively. A scale-bar is shown in the bottom right corner of each panel. Note that most of the emission is filtered out by the interferometer.}
    \label{af: outflows_13co_noema}
\end{figure*}

\begin{figure*}[h!]
    \centering
    \includegraphics[width=0.82\hsize]{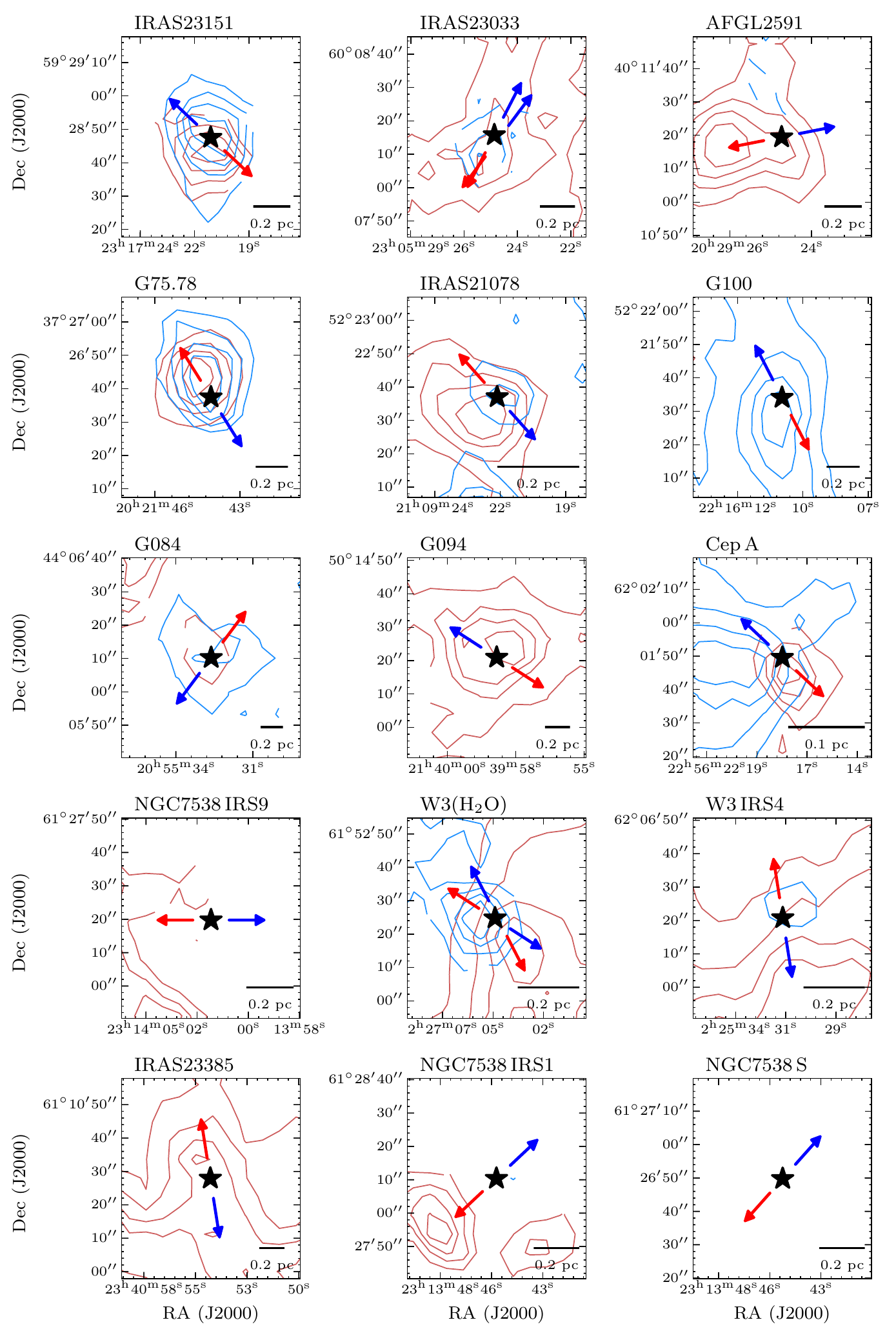}
    \caption{Intensity maps of $\mathrm{C^{18}O}\,(2-1)$ emission from IRAM 30-m telescope integrated over the blueshifted and redshifted wings of emission, showing the outflow structure. The position of the strongest source in the field is depicted by a star. The blue and red arrows correspond to the estimated directions of bipolar blueshifted and redshifted molecular outflows, respectively. A scale-bar is shown in the bottom right corner of each panel. The map size of the IRAM 30-m observations is 1.5\arcmin\ by 1.5\arcmin\ with a half-power beam width of $\sim$11\arcsec\ at this frequency. $\mathrm{C^{18}O}$ is not a good outflow tracer for all sources, hence the lack of contours in some targets.}
    \label{af: outflows_c18o_sd}
\end{figure*}

\begin{figure*}[h!]
    \centering
    \includegraphics[width=0.82\hsize]{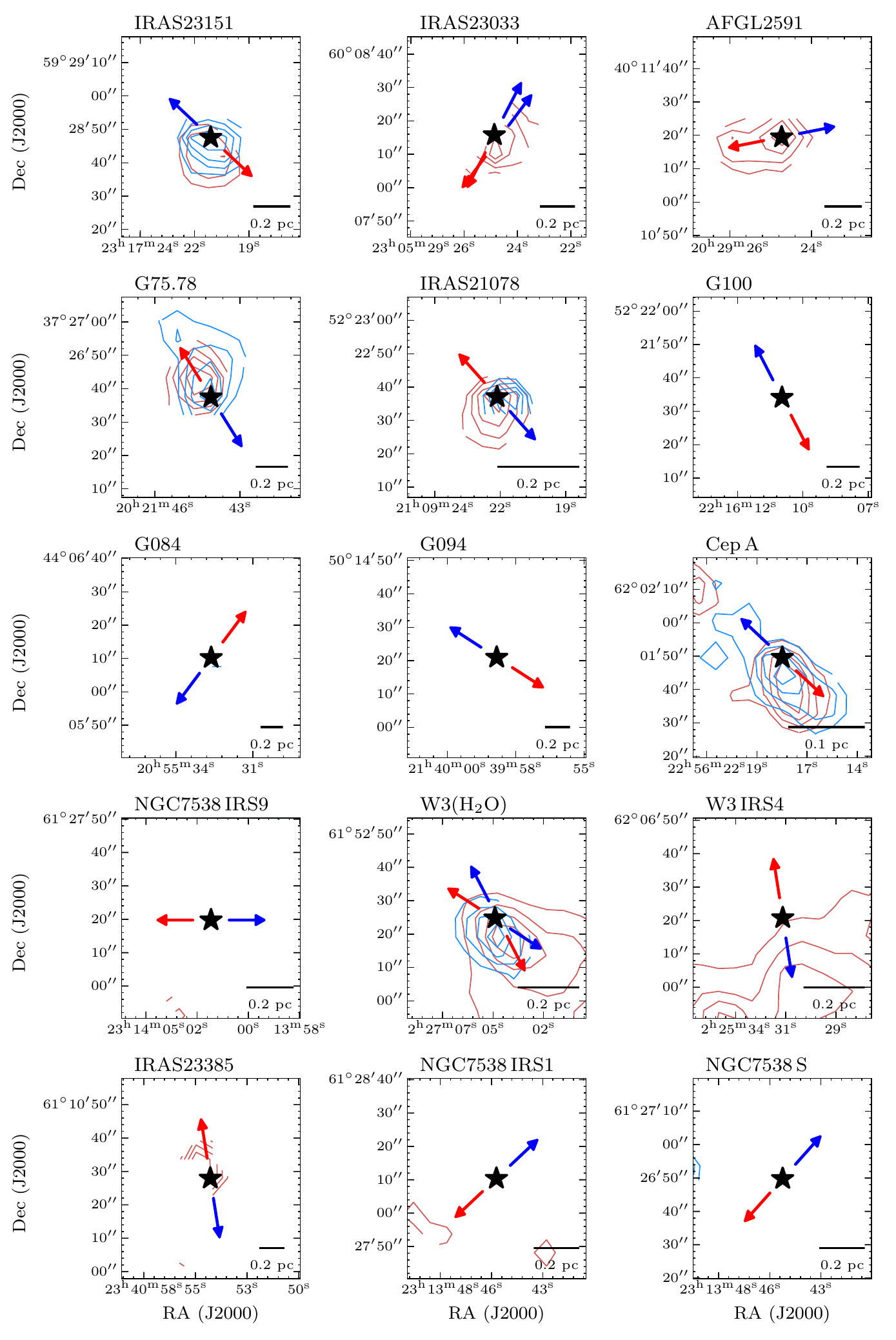}
    \caption{Intensity maps of $\mathrm{SO}\,(6_5-5_4)$ emission from IRAM 30-m telescope integrated over the blueshifted and redshifted wings of emission, showing the outflow structure. The position of the strongest source in the field is depicted by a star. The blue and red arrows correspond to the estimated directions of bipolar blueshifted and redshifted molecular outflows, respectively. A scale-bar is shown in the bottom right corner of each panel. The map size of the IRAM 30-m observations is 1.5\arcmin\ by 1.5\arcmin\ with a half-power beam width of $\sim$11\arcsec\ at this frequency.}
    \label{af: outflows_so_sd}
\end{figure*}

\begin{figure}[h!]
    \centering
    \includegraphics[width=\hsize]{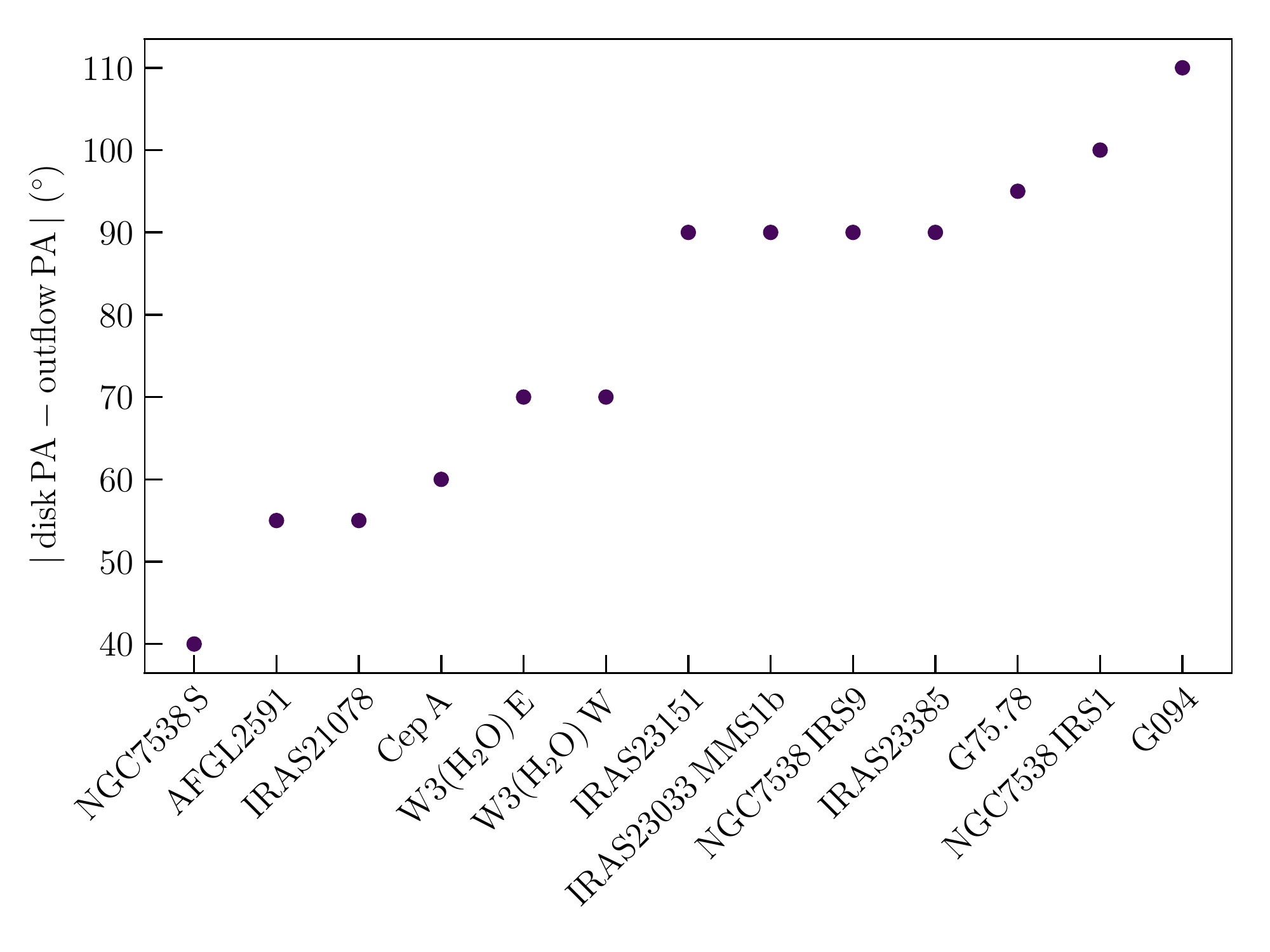}
    \caption{The distribution of absolute difference between the disk position angles and the assumed outflow position angles for our sample of 13 disk candidates in the CORE survey.}
    \label{af: outflows_disk_pa}
\end{figure}

In this section, the distribution of outflow and shock tracing transitions are presented in Figs~\ref{af: outflows_12co_sd}~--~\ref{af: outflows_so_sd}. The absolute difference between the assumed disk and outflow position angles is shown in Fig.~\ref{af: outflows_disk_pa}.

\FloatBarrier

\section{NGC7538\,IRS1}\label{a: ngc7538_irs1}
\subsection{Rotational temperature}\label{a: ngc7538_irs1_temp}
In NGC7538\,IRS1, the molecular lines are seen in absorption against the strong continuum source with many velocity components due to the complex dynamics in this source. To properly model the temperature structure across this disk candidate, we used \textit{XCLASS} to model the molecular line emission existing outside the continuum to the south-west where the spectra are not affected by absorption features. We then scaled the modelled rotational temperature from the edge of the continuum structure ($\sim$135~K) towards the continuum peak following the temperature power-law distribution $T\propto r^{-0.4}$, derived by \citetads{2021A&A...648A..66G} from the CORE survey. The resulting rotational temperature map is shown in Fig.~\ref{f: CORE_Trot}.
\subsection{Free-free contribution}\label{a: ngc7538_irs1_freefree}
NGC7538\,IRS1 is the only source in our sample of disk candidates with significant free-free contribution in the millimetre \citepads{2017A&A...605A..61B}. For this source, we smoothed the continuum emission map of \citetads{2017A&A...605A..61B} observed with the Karl G. Jansky Very Large Array (VLA) at 24.7~GHz (1.2~cm) to the resolution of our observations. We then scaled the emission of this smoothed map by $\nu^{-0.1}$, the expected frequency-dependence of free-free emission, and subtracted it from our millimetre continuum observations. The continuum flux densities before and after the subtraction are provided in the body of Table~\ref{t: gas_masses} and as a note under the table, respectively. The contribution is $\sim13\%$. 

\section{XCLASS maps}\label{a: xclass_panels}

\begin{figure*}[h!]
    \centering
    \includegraphics[width=0.89\hsize]{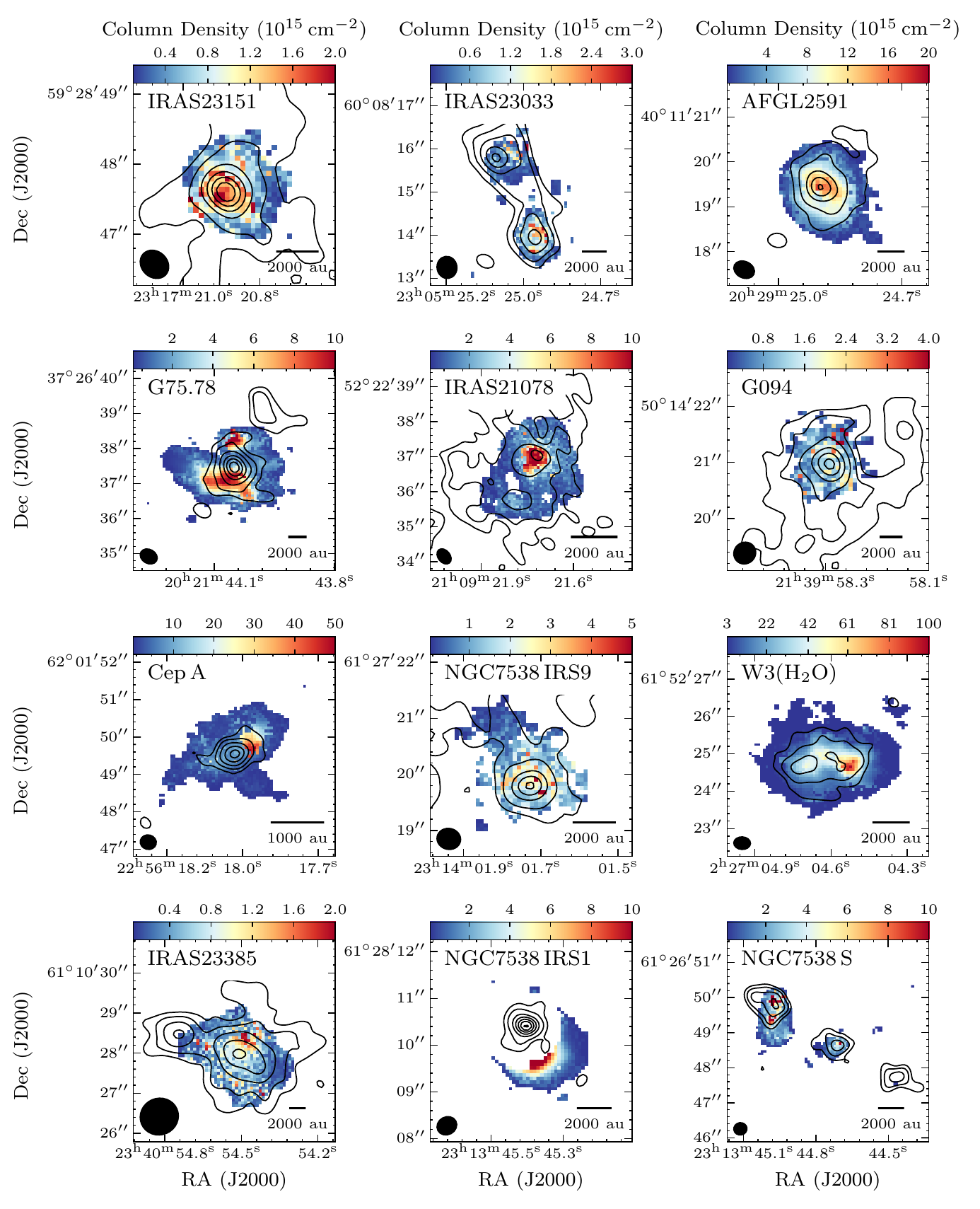}
    \caption{\mc\ column density maps obtained by fitting \mckr{0}{6} and \mcisokr{0}{3} lines with \emph{XCLASS} for the best disk candidates in the CORE survey. The contours correspond to the 1.37~mm continuum as described in Fig.~\ref{f: mom_velo_h2co}. For NGC7538\,IRS1, only the region outside the continuum to the south-west is modelled by \emph{XCLASS} as the molecular lines are seen in absorption against the strong continuum source with many velocity components due to the complex dynamics in this source. The synthesised beam is shown in the bottom left corner and a scale bar in the bottom right corner of each panel.}
    \label{af: XCLASS_Ntot}
\end{figure*}

\begin{figure*}
    \centering
    \includegraphics[width=0.97\hsize]{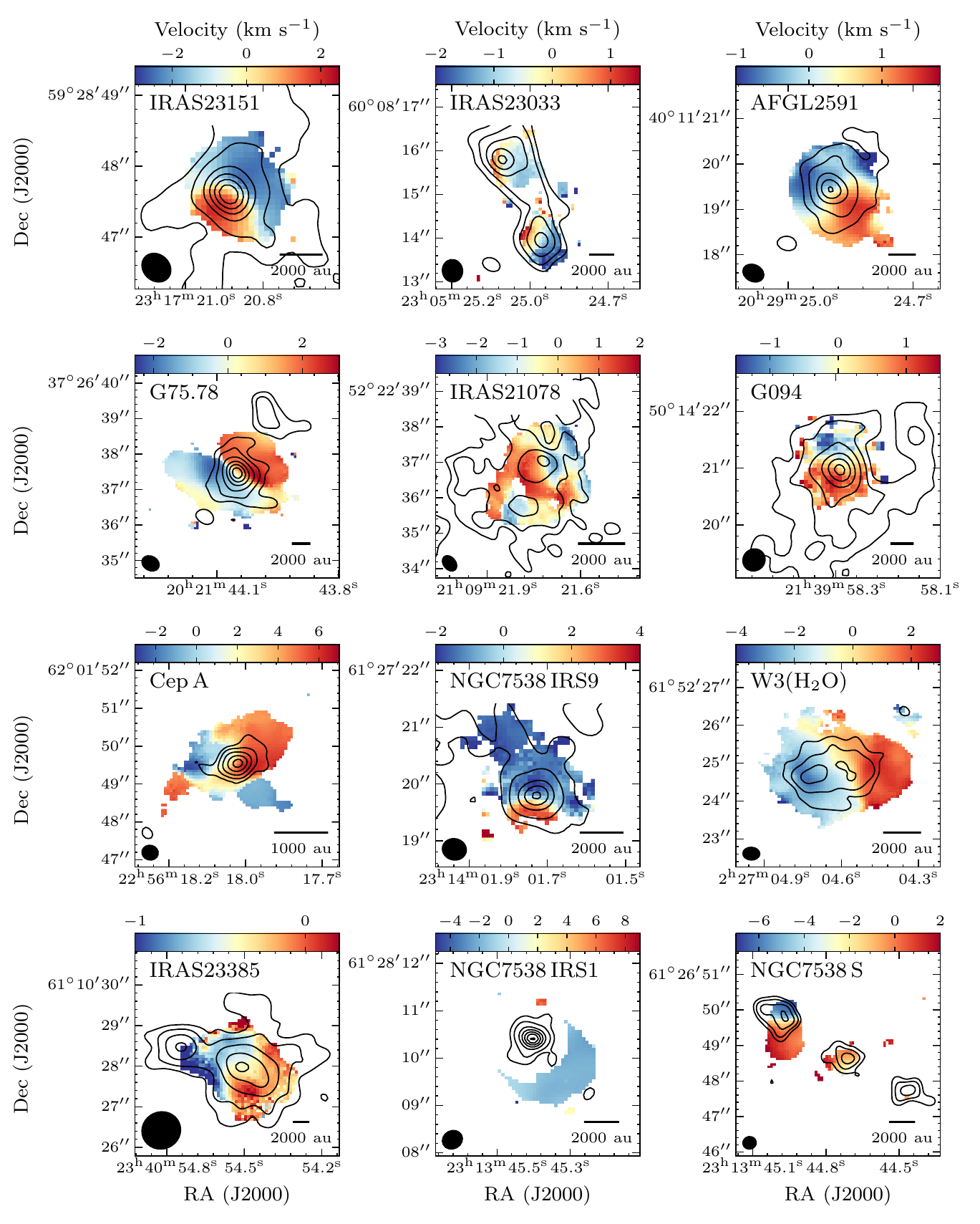}
    \caption{Maps of offset velocity with respect to the systemic velocity obtained by fitting \mckr{0}{6}  and \mcisokr{0}{3} lines with \emph{XCLASS} for the best disk candidates in the CORE survey. The contours correspond to the 1.37~mm continuum as described in Fig.~\ref{f: mom_velo_h2co}. For NGC7538\,IRS1, only the region outside the continuum to the south-west is modelled by \emph{XCLASS} as the molecular lines are seen in absorption against the strong continuum source with many velocity components due to the complex dynamics in this source. The synthesised beam is shown in the bottom left corner and a scale bar in the bottom right corner of each panel.}
    \label{af: XCLASS_Voff}
\end{figure*}

\begin{figure*}
    \centering
    \includegraphics[width=0.97\hsize]{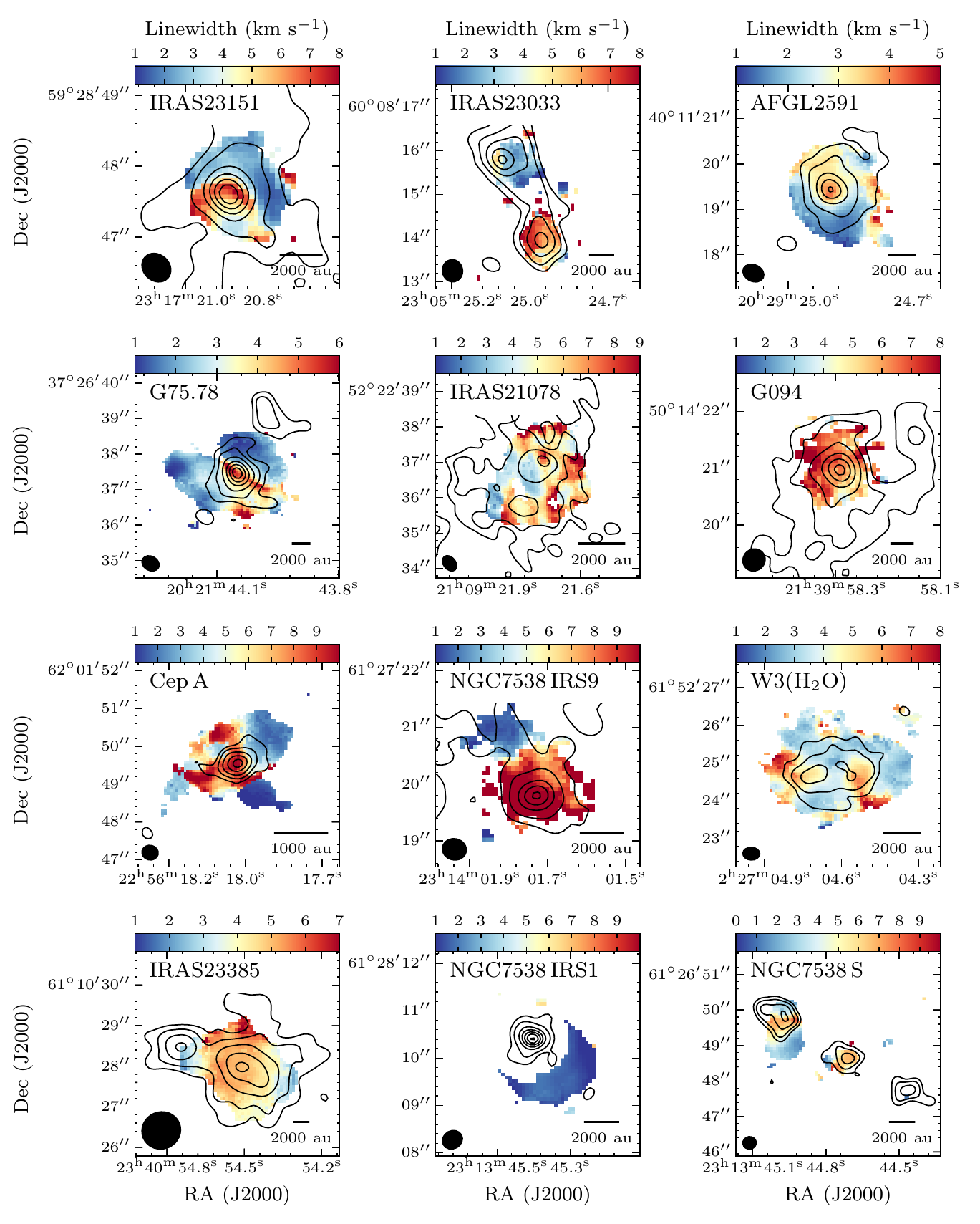}
    \caption[Maps of linewidth obtained by fitting \mc\ transitions with \emph{XCLASS} for 17 cores in the CORE survey]{Maps of linewidth obtained by fitting \mckr{0}{6}  and \mcisokr{0}{3} lines with \emph{XCLASS} for the best disk candidates in the CORE survey. The contours correspond to the 1.37~mm continuum as described in Fig.~\ref{f: mom_velo_h2co}. For NGC7538\,IRS1, only the region outside the continuum to the south-west is modelled by \emph{XCLASS} as the molecular lines are seen in absorption against the strong continuum source with many velocity components due to the complex dynamics in this source. The synthesised beam is shown in the bottom left corner and a scale bar in the bottom right corner of each panel.}
    \label{af: XCLASS_Vwidth}
\end{figure*}

\begin{figure*}
    \centering
    \includegraphics[width=0.97\hsize]{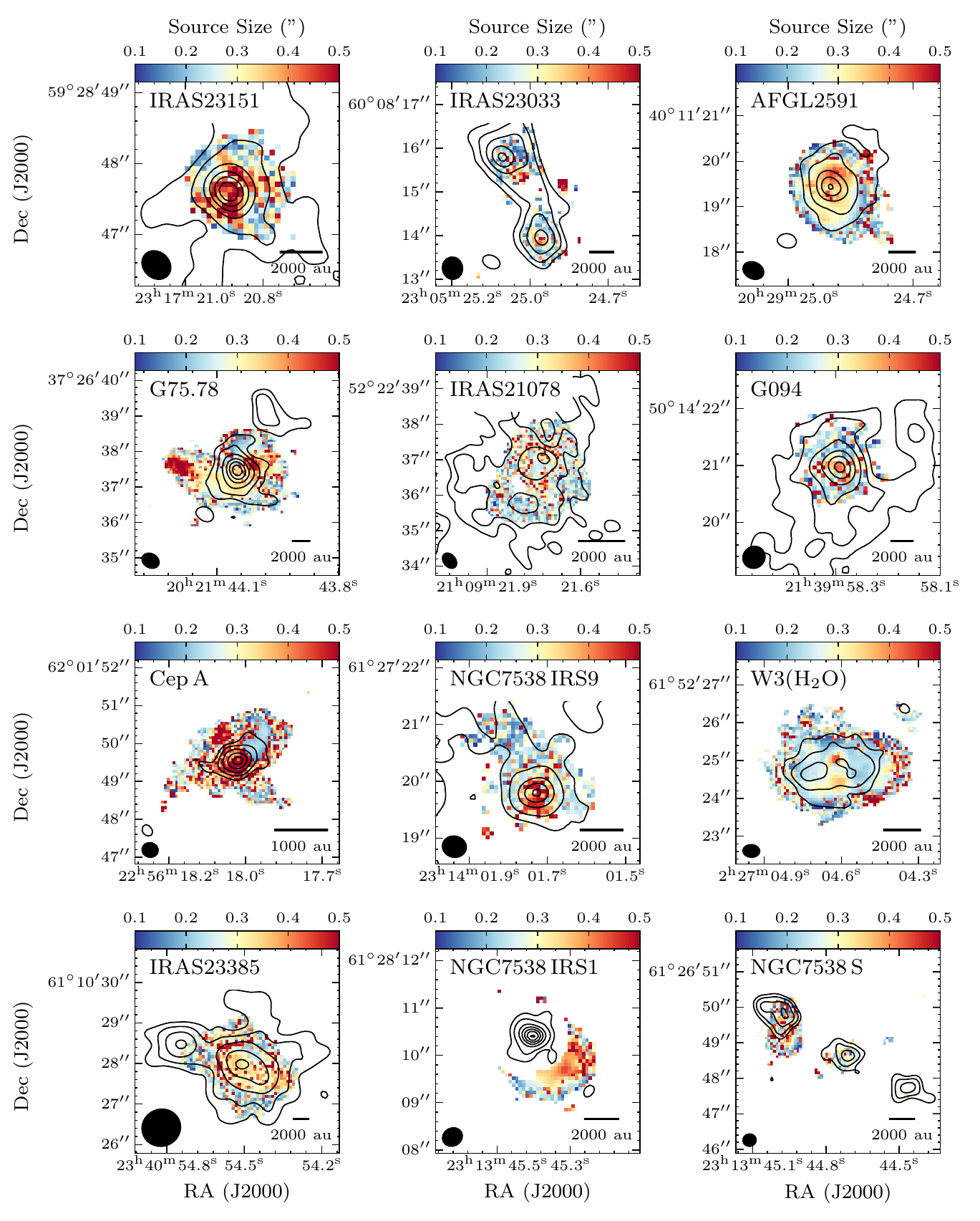}
    \caption{Maps of source size parameter obtained by fitting \mckr{0}{6}  and \mcisokr{0}{3} lines with \emph{XCLASS} for the best disk candidates in the CORE survey. The contours correspond to the 1.37~mm continuum as described in Fig.~\ref{f: mom_velo_h2co}. For NGC7538\,IRS1, only the region outside the continuum to the south-west is modelled by \emph{XCLASS} as the molecular lines are seen in absorption against the strong continuum source with many velocity components due to the complex dynamics in this source. The synthesised beam is shown in the bottom left corner and a scale bar in the bottom right corner of each panel.}
    \label{af: XCLASS_source_size}
\end{figure*}

\begin{figure*}
    \centering
    \includegraphics[width=0.79\hsize]{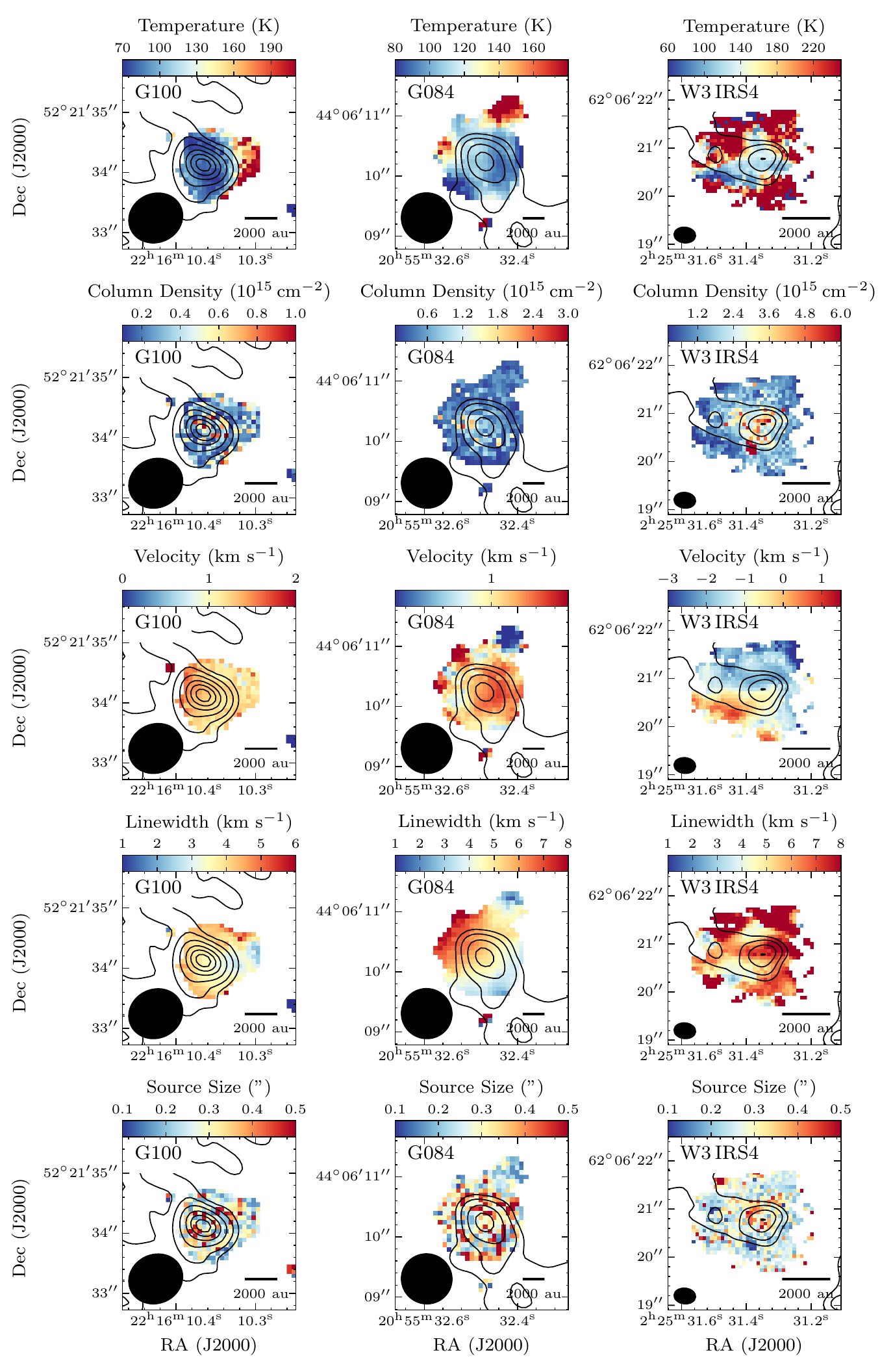}
    \caption{Maps of rotational temperature (top row), \mc\ column density (second row), offset velocity (third row), linewidth (fourth row), and source size parameter (bottom row) obtained by fitting \mckr{0}{6}  and \mcisokr{0}{3} lines with \emph{XCLASS} for the remaining three sources (columns) in the CORE survey. The contours correspond to the 1.37~mm continuum as described in Fig.~\ref{f: mom_velo_h2co}. The blue and red arrows correspond to the estimated directions of bipolar blueshifted and redshifted molecular outflows, respectively. The synthesised beam is shown in the bottom left corner and a scale bar in the bottom right corner of each panel.}
    \label{af: XCLASS_rest}
\end{figure*}

\textit{XCLASS} output maps such as the column density, velocity offset, linewidth, and source size maps for the best disk candidates in the CORE survey are shown in Figs.~\ref{af: XCLASS_Ntot}~--~\ref{af: XCLASS_source_size}. The \textit{XCLASS} output maps for the remaining three sources that had \mc\ emission but that were not good disk candidates are presented in Fig.~\ref{af: XCLASS_rest}.

\clearpage
\section{Specific angular momentum}\label{a: j_sims}

In order to better understand the observational findings, we analyse the expected specific angular momentum profiles for the numerical simulations presented in \citetads{2020A&A...644A..41O} to study the effects of inclination and angular resolution on the specific angular momentum, $j$. The 3D radiation-hydrodynamic simulation under investigation follows the collapse of a massive core with 200~\mo\ in gas and dust, which results in the formation and fragmentation of a massive disk. At this snapshot, the star has gained 10~\mo, being fed from a fragmented disk with a mass of 8~\mo. It is important to note that the simulation presented here does not include magnetic fields, which are an important ingredient for disk formation. It also does not include an inhomogeneous, large-scale accretion flow from which the disk is fed, and an initial axially symmetric rotation profile of $\Omega(r) \propto r^{-0.75}$ was adopted. 

\begin{figure*}[h!]
    \centering
    \includegraphics[width=\hsize]{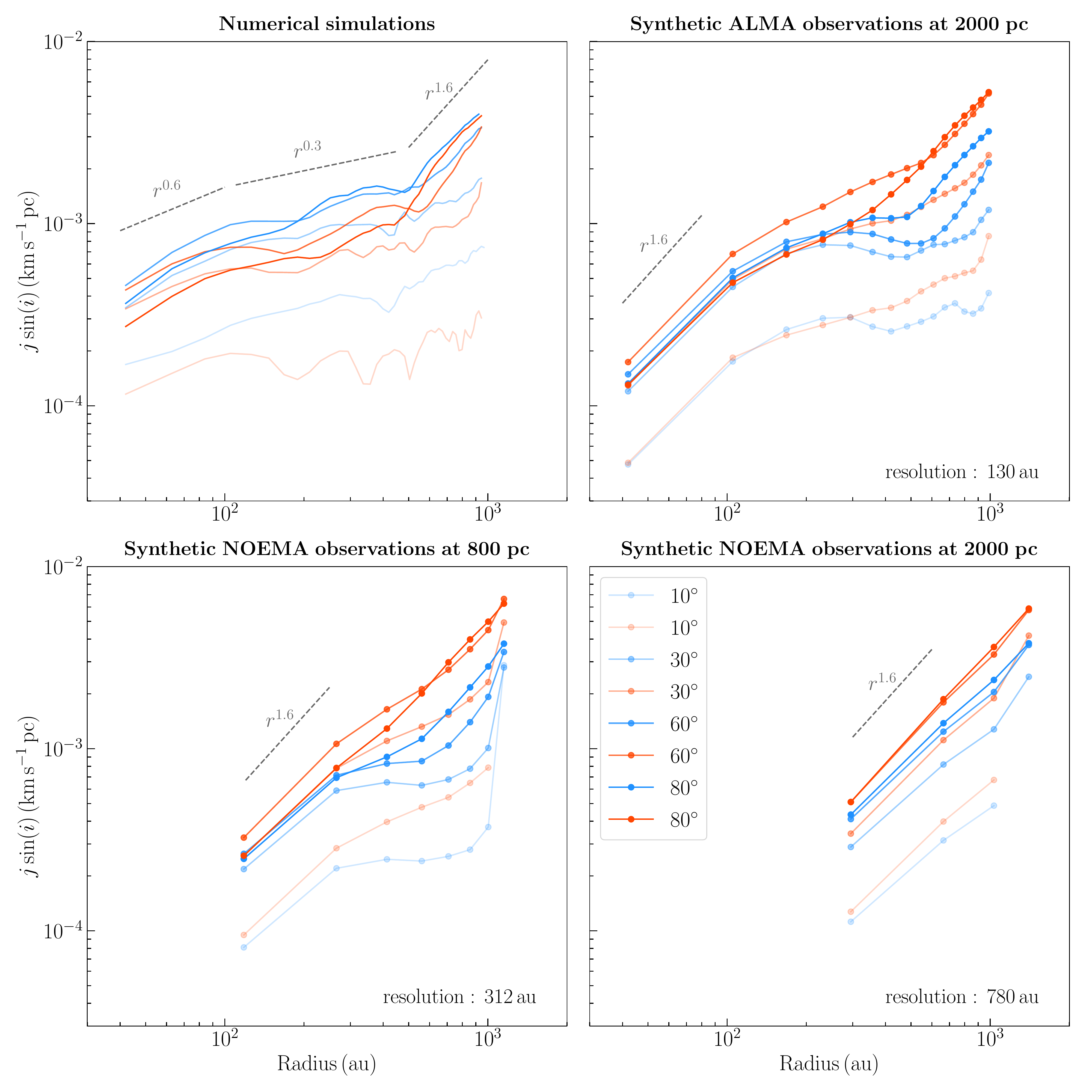}
    \caption{Specific angular momentum radial profiles (see Eq.~\ref{eq: j_obs}) for a high-resolution numerical simulation of a high-mass core having formed a massive disk that is feeding the central protostar, inclined to 10\degr, 30\degr, 60\degr, and 80\degr \emph{(top left)}, their synthetic ALMA observations at 2000~pc \emph{(top right)}, and synthetic NOEMA observations at 800~pc \emph{(bottom left)} and 2000~pc \emph{(bottom right)} (see \citeads{2019A&A...632A..50A}). The blue and red colours correspond to the blueshifted and redshifted sides, respectively, and the hues correspond to the different inclinations labeled in the bottom-right panel. Note that we have not corrected for the known inclination in order to compare the simulated observations to our NOEMA observations with unknown inclinations. The distance between each neighbouring point is a half beam spacing. The grey dashed lines show various slopes to guide the eye.}
    \label{f: sims_j}
\end{figure*}

We presented synthetic observations of this system in \citetads{2019A&A...632A..50A} and here calculate the observed $j~\mathrm{sin}(i)$ according to Eq.~\ref{eq: j_obs} using the first moment maps of \mck{3} at an inclination angle of 10\degr, 30\degr, 60\degr, and 80\degr from the plane of the sky along a cut in the E-W direction for the redshifted and blueshifted sides of the disk (see Fig.~5 of \citeads{2019A&A...632A..50A}). Figure~\ref{f: sims_j} shows the local specific angular momentum radial profiles for the simulations at the different inclinations as well as for a set of synthetic observations with successively lower resolution (as noted in the panels). Note that we have purposefully not corrected for the known inclination in order to compare the findings with our NOEMA observations. Similar to the observational findings for low-mass star formation, we find three distinct regimes in the $j$ radial profile. Least-squared power-law fit parameters to these three distinct regimes in the specific angular momentum plots for the numerical simulations are listed in Table~\ref{t: SIMS_j_slopes}. The inner disk region ($r<200$~au) actively accreting material onto the central protostar with $j\propto r^{0.6}$ close to Keplerian rotation ($j\propto r^{0.5}$), an intermediate region where $j$ is quite flat but not exactly constant, and the region beyond 700~au is rather consistent with the $j \propto r^{1.6}$ relation that had been found for the rotation of low-mass dense cores and Class 0 envelopes (\eg~\citeads{2020A&A...637A..92G}). The intermediate regime has a significant amount of substructure as four fragments reside in this range of radii at this snapshot, accreting part of the disk material. The discontinuity at $\sim$700~au roughly corresponds to the position where the rotating core/envelope begins to dynamically collapse. We find that the simulation with more of a face-on view has almost an order of magnitude lower specific angular momentum with flatter radial profiles and for which the distinction between the different regimes is unclear.

As the angular resolution worsens, the substructure in the $j$ radial profile becomes smoother with the power-law relations for the different regimes becoming more continuous and not as clear to disentangle. The specific angular momentum of the blueshifted side seems to have a more pronounced kink than the redshifted side, best visible in the synthetic observations at 800~pc. This is due to the existence of a fragment at this position, accreting some of the disk material and complicating the angular momentum profile at this position. At the scale of synthetic NOEMA observations at 2~kpc, the envelope and disk components become completely blended, with only one power-law relation seen across all scales, with $j \propto r^{1.6}$. Interestingly, there still exists a small shallowing of the slope (a kink) in the $j$ profiles. 

\begin{table*}[h!]
\str{1.3}
\centering
\caption{Fit parameters for the different regions of the specific angular momentum radial profiles of the simulations shown in the top left panel of Fig.~\ref{f: sims_j}.}
\label{t: SIMS_j_slopes}
\begin{tabular}{lcccc}
\hline \hline
 &  \multicolumn{4}{c}{Inner ($r<100$~au)} \\
    \cline{2-5} Inclination & $a_\mathrm{blue}$ & $a_\mathrm{red}$ & $b_\mathrm{blue}$ & $b_\mathrm{red}$ \\
\hline
10\degr & $0.56\pm0.06$ & $0.56\pm0.06$ & $-3.16\pm0.06$ & $-3.16\pm0.06$ \\
30\degr & $0.54\pm0.07$ & $0.54\pm0.07$  & $-2.71\pm0.07$ & $-2.71\pm0.07$\\
60\degr & $0.56\pm0.09$& $0.56\pm0.09$ & $-2.6\pm0.1$ & $-2.6\pm0.1$\\
80\degr & $0.75\pm0.08$ & $0.75\pm0.08$ & $-2.51\pm0.08$ & $-2.51\pm0.08$ \\
\hline
 &  \multicolumn{4}{c}{Middle ($100~\mathrm{au}<r<400~\mathrm{au}$)} \\
    \cline{2-5} Inclination & $a_\mathrm{blue}$ & $a_\mathrm{red}$ & $b_\mathrm{blue}$ & $b_\mathrm{red}$ \\
\hline
10\degr & $0.14\pm0.05$ & $-0.01\pm0.08$ & $-3.36\pm0.03$ & $-3.77\pm0.05$\\
30\degr & $0.15\pm0.04$ & $0.26\pm0.04$ & $-2.96\pm0.02$ & $-3.04\pm0.02$ \\
60\degr & $0.35\pm0.03$ & $0.54\pm0.05$ & $-2.69\pm0.02$ & $-2.7\pm0.03$ \\
80\degr & $0.52\pm0.04$ & $0.48\pm0.03$ & $-2.58\pm0.02$ & $-2.83\pm0.02$ \\
\hline
 &  \multicolumn{4}{c}{Outer ($r>400$~au)} \\
    \cline{2-5} Inclination & $a_\mathrm{blue}$ & $a_\mathrm{red}$ & $b_\mathrm{blue}$ & $b_\mathrm{red}$ \\
\hline
10\degr & $0.78\pm0.03$ & $0.6\pm0.1$ & $-3.12\pm0.01$ & $-3.53\pm0.02$ \\
30\degr & $0.74\pm0.05$ & $1.0\pm0.1$ & $-2.76\pm0.01$ & $-2.85\pm0.02$ \\
60\degr & $1.30\pm0.05$ & $1.75\pm0.09$ & $-2.45\pm0.01$ & $-2.45\pm0.01$ \\
80\degr & $1.45\pm0.03$ & $1.75\pm0.05$ & $-2.34\pm0.01$ & $-2.34\pm0.01$ \\
\hline
\end{tabular}
\tablefoot{The heading parameters $a$ and $b$ satisfy the following relation: $j\,\mathrm{sin}(i)=10^b\,(\frac{r}{1000~\mathrm{au}})^a~\mathrm{km\,s^{-1}\,pc}$, for the blueshifted and redshifted sides.}
\end{table*}

\end{appendix}

\end{document}